\documentclass[twocolumn,superscriptaddress,amsmath,amssymb,aps,prx,nofootinbib]{revtex4-2}

\usepackage{wasysym}
\usepackage{tikz}
\usepackage{hyperref}
\usepackage{natbib}
\usepackage{dcolumn}
\usepackage{bm}
\usepackage{oplotsymbl}
\usepackage{graphicx}
\graphicspath{ {./images/} }
\usepackage{physics}
\usepackage[english]{babel}
\usepackage{xcolor}
\usepackage[normalem]{ulem}
\usepackage{mathtools}
\usepackage{tabularx}
\usepackage{array}
\usepackage{booktabs}
\usepackage{sansmath}
\usepackage{bm}
\usepackage{yhmath}
\usepackage{caption}
\usepackage{ragged2e}
\usepackage{mathrsfs}  
\usepackage{subcaption}

\newcommand\TikCircle[1][2.5]{\tikz[baseline=-2]{\draw(0,0)circle[radius=1mm];}}
\newcommand\BigTikCircle[1][2.5]{\tikz[baseline=-2]{\draw(0,0)circle[radius=1.5mm];}}

\makeatletter
\newcommand\incircbin{\mathpalette\@incircbin}
\newcommand\@incircbin[2]{\mathbin{\ooalign{\hidewidth$#1#2$\hidewidth\crcr$#1\TikCircle$}}}

\newcommand\bigincircbin{\mathpalette\@bigincircbin}
\newcommand\@bigincircbin[2]{\mathbin{\ooalign{\hidewidth$#1#2$\hidewidth\crcr$#1\BigTikCircle$}}}

\newcommand{\osum}{\incircbin{\sum}}

\makeatother

\newcommand\TikCircleSup[1][2.5]{\tikz[baseline=-1.5]{\draw(0,0)circle[radius=0.75mm];}}
\newcommand{\sumlr}[1]{\sum^{}_{#1} {\vphantom{\sum}}'}

\makeatletter
\newcommand\incircbinsup{\mathpalette\@incircbinsup}
\newcommand\@incircbinsup[2]{\mathbin{\ooalign{\hidewidth$#1#2$\hidewidth\crcr$#1\TikCircleSup$}}}

\makeatother

\newcommand{\tet}{
  \mathchoice
    {\includegraphics[height=1ex]{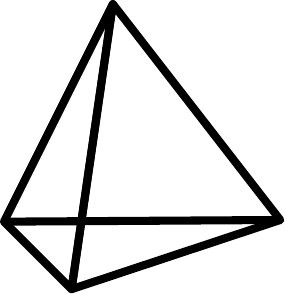}} 
    {\includegraphics[height=1.7ex]{tetrahedron}} 
    {\includegraphics[height=1.5ex]{tetrahedron}} 
    {\includegraphics[height=.5ex]{tetrahedron}} 
}

\newcommand{\hexA}{\hspace{-0.1cm}\raisebox{-0.1cm}{
\begin{tikzpicture}
    \newdimen\r
    \newdimen\rad
    \r = 0.15cm
    \rad = 1pt
    \draw[rotate=30] (-0.15cm:\r) \foreach \x in {60,120,...,360} {  -- (\x:\r) };
   \foreach \x in {-30,90,210} { \filldraw [black] (\x:\r)  circle (\rad);}
   \foreach \x in {30,150,270} { \filldraw [draw=black,fill=white] (\x:\r)  circle (\rad);}
\end{tikzpicture}}
}

\newcommand{\hexB}{\hspace{-0.1cm}\raisebox{-0.1cm}{
\begin{tikzpicture}
    \newdimen\r
    \newdimen\rad
    \r = 0.15cm
    \rad = 1pt
   \draw[rotate=30] (0:\r) \foreach \x in {60,120,...,360} {  -- (\x:\r) };
   \foreach \x in {30,150,270} { \filldraw [black] (\x:\r)  circle (\rad);}
   \foreach \x in {-30,90,210} { \filldraw [draw=black,fill=white] (\x:\r)  circle (\rad);}
\end{tikzpicture}}
}

\newcommand{\dual}[1]{\boldsymbol{\mathsf{#1}}}

\newcommand{\newtext}[1]{\textcolor{black}{#1}}

\newcommand{\capt}[1]{\caption[.]{\justifying{#1}}}
\begin{document}

\title{Quantum spin ice in three-dimensional Rydberg atom arrays}

\author{Jeet Shah}
\affiliation{Joint Quantum Institute, Department of Physics, University of Maryland,
College Park, MD 20742, USA}
\affiliation{Joint Center for Quantum Information and Computer Science, NIST/University of Maryland,
College Park, MD 20742, USA}
\author{Gautam Nambiar}
\affiliation{Joint Quantum Institute, Department of Physics, University of Maryland,
College Park, MD 20742, USA}
\author{Alexey V. Gorshkov}
\affiliation{Joint Quantum Institute, Department of Physics, University of Maryland,
College Park, MD 20742, USA}
\affiliation{Joint Center for Quantum Information and Computer Science, NIST/University of Maryland,
College Park, MD 20742, USA}
\author{Victor Galitski}
\affiliation{Joint Quantum Institute, Department of Physics, University of Maryland,
College Park, MD 20742, USA}
\affiliation{Center for Computational Quantum Physics, The Flatiron Institute, New York, NY 10010, USA}
\date{\today}

\begin{abstract}

  Quantum spin liquids are exotic phases of matter whose low-energy physics is described as the deconfined phase of an emergent gauge theory. With recent theory proposals and an experiment showing preliminary signs of $\mathbb{Z}_2$ topological order [G. Semeghini et al., Science \textbf{374}, 1242 (2021)], Rydberg atom arrays have emerged as a promising platform to realize a quantum spin liquid. In this work, we propose a way to realize a $U(1)$ quantum spin liquid in three spatial dimensions, described by the deconfined phase of $U(1)$ gauge theory in a pyrochlore lattice Rydberg atom array. We study the ground state phase diagram of the proposed Rydberg system as a function of experimentally relevant parameters. Within our calculation, we find that by tuning the Rabi frequency, one can access both the confinement-deconfinement transition driven by a proliferation of ``magnetic" monopoles and the Higgs transition driven by a proliferation of ``electric" charges of the emergent gauge theory. We suggest experimental probes for distinguishing the deconfined phase from ordered phases. This work serves as a proposal to access a confinement-deconfinement transition in three spatial dimensions on a Rydberg-based quantum simulator. 
  
\end{abstract}

\maketitle

\section{Introduction}
When the classical part of a many-body Hamiltonian is frustrated, quantum fluctuations can break the degeneracy in interesting ways. An exotic form of such breaking was pointed out by Anderson~\cite{anderson1973resonating} where the ground state is a  superposition of several almost-degenerate states, and the excitations are ``fractional"\cite{savary2017}. Broadly, a common feature tying together such systems called quantum spin liquids is that, at low energies, they can be described as lying in a deconfined phase of an emergent gauge theory. The fractional excitations are the ``charge''-like and ``flux/monopole''-like excitations of this gauge theory. When these fractional excitations get confined, they cease to be important for the low-energy physics, and the system becomes ordered.  From this point of view, transitions from a spin liquid to conventional ordered phases are understood as a confinement-deconfinement transition, driven by a proliferation of ``flux/monopole''-like excitations, or a Higgs transition, driven by a proliferation of ``charge"-like excitations \cite{senthil2004deconfined,hermele2005algebraic,song2019unifying,nambiar2022monopole, chen2017symmetry, nussinov2007high}. Gauge theories and their phase transitions are of fundamental importance in physics~\cite{marciano1978quantum,meyer1996phase,kogut1983lattice,o2000gauge}. The prospect of this physics emerging in many-body systems provides an important motivation for studying quantum spin liquids. They are also interesting due to their possible role in the physics of strongly correlated materials~\cite{lee2006doping} and possible application in quantum computing~\cite{kitaev2003fault,nayak2008non}. 

Traditionally, the main search space for spin liquids has comprised of solid state systems. While consistent progress has been made~\cite{savary2017,knolle2019field}, conclusive evidence for spin liquids is still lacking in these systems. One reason is that the same feature that makes spin liquids interesting---being characterized by non-local order parameters---also makes them hard to detect. Meanwhile, over the past decade, Rydberg atom arrays have emerged as a promising platform for engineering interacting Hamiltonians~\cite{weimer2010rydberg,ebadi2021quantum,wu2021concise,saffman2010,lukin2001,labuhn2016,jaksch2000,browaeys2016,barredo2016anatom,beguin2013,barredo2018,semeghini2021, bernien2017probing,bluvstein2022quantum,guardado-sanchez2018probing,keesling2019quantum,levine2019parallel,lienhard2018observing,madjarov2020highfidelity,omran2019generation,schauss2015crystallization,song2021quantum}. Rydberg states have large principal quantum number $n$ $(\sim 20-100)$, and the van der Waals interaction between them scales as $n^{11}$. The strong tunable interactions, along with the ability to customize the lattice of atoms, locally control qubits, and take wavefunction snapshots, make Rydberg atom arrays a competitive platform to explore quantum many-body physics. Following theory proposals~\cite{verresen2021,samajdar2021}, promising signs of $\mathbb{Z}_2$ topological order have been observed experimentally on this platform~\cite{semeghini2021}. This has sparked a lot of activity over the past few years in the general direction of proposing ways to realize exotic states on quantum devices using analogue quantum simulation~\cite{myerson2022construction,slagle2022quantum, giudice2022trimer,zhou2022quantum, tarabunga2023classification, kornjaca2023trimer,ohler2023quantum}, digital quantum simulation~\cite{liu2021methods}, and projective measurements~\cite{tantivasadakarn2022shortest,verresen2021efficiently}. \newtext{However, all of these proposals have been for two-dimensional Rydberg atom arrays.} 

\begin{figure*}[t]
  \centering
  \includegraphics[width=0.95\textwidth]{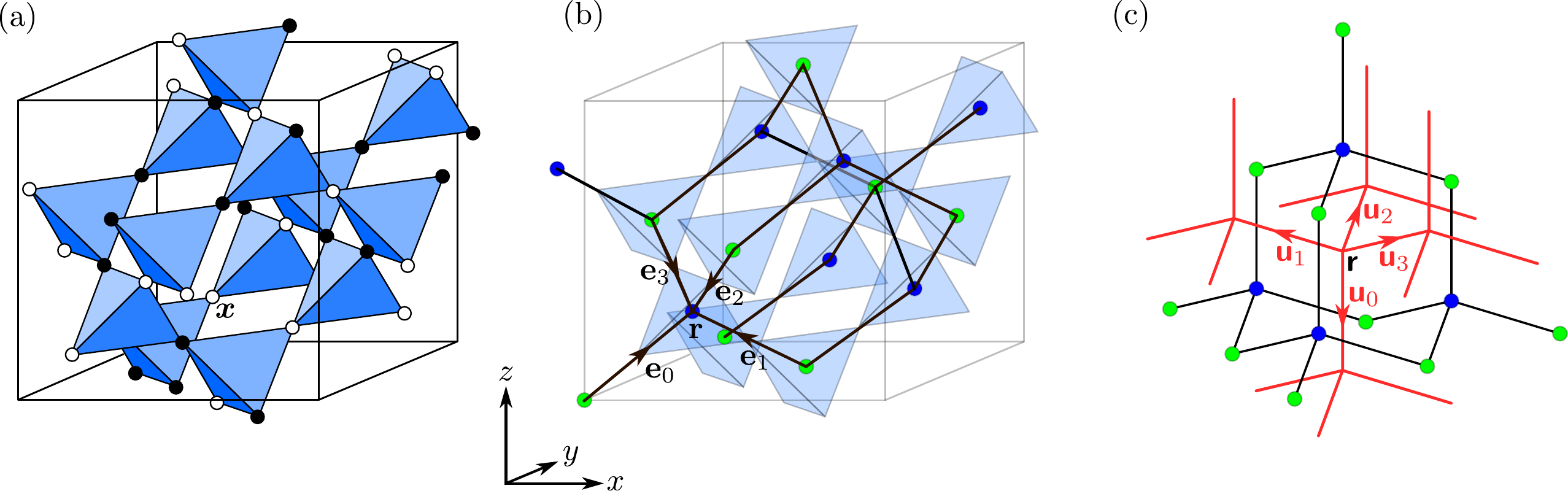}
  \capt{(a) The pyrochlore lattice. White circles denote atoms in the ground state, while black circles denote atoms in the Rydberg state. The configuration shown satisfies $n_{\tet} = 2$ on each tetrahedron. The label $\bm{x}$ is used to denote the sites of the pyrochlore lattice. (b) The diamond lattice. It is the bipartite lattice formed by the centers of the tetrahedra marked by green (A sublattice) and blue (B sublattice) dots. $\vb{e}_\mu$ for $\mu \in \{ 0,1,2,3 \}$ label the vectors joining an A site to its neighboring B sites. The label $\mathbf{r}$ is used to denote the sites of the diamond lattice. (c) The red links are the edges of the lattice dual to the diamond lattice shown in (b). This lattice is also a diamond lattice, and we refer to it as the ``dual diamond lattice" in this paper to distinguish it from the ``diamond lattice" in (b). The sites of the dual diamond lattice are centers of the ``polyhedra" formed by four puckered hexagons of the diamond lattice. $\dual{u}_\mu$ for $\mu \in \{ 0,1,2,3 \}$  label the vectors joining an A site to its neighboring B sites on the dual diamond lattice. The label $\dual{r}$ [notice the difference in the font as compared to $\mathbf{r}$ in (b)] is used to denote the sites of the dual diamond lattice.}
  \label{fig:pyrochlore_diamond}
\end{figure*}

Our work is a proposal for realizing a $U(1)$ quantum spin liquid, described by the deconfined phase of a compact $U(1)$ gauge theory on three-dimensional Rydberg atom  arrays,  with an eye towards accessing the confinement-deconfinement transition. 
\newtext{With our proposal, we intend to push the search for a $U(1)$ quantum spin liquid, which has traditionally remained limited to solid state systems, in the direction of three-dimensional Rydberg atom arrays.
The connection between Rydberg-atom arrays and Abelian gauge theories in one dimension and two dimensions has been studied in depth in the literature before~\cite{cheng2024emergent, ohler2022self, surace2020lattice, cheng2023gauge}; however, this connection in three dimensions has remained unexplored. It is known that gauge theories in three dimensions can have a significantly different behavior than those in two dimensions. This is illustrated by the compact $U(1)$ gauge theory.}
It was shown by Polyakov~\cite{polyakovCompactGaugeFields1975,polyakov1977} that compact $U(1)$ gauge theory in 2+1 dimensions is always in the confined phase in the thermodynamic limit due to a proliferation of monopole events. Therefore we turn to 3+1 dimensions, where Polyakov argued~\cite{polyakov1977} for the existence of both deconfined and confined phases separated by a transition driven by monopole excitations. The deconfined phase consists of gapless ``photons'', gapped ``monopoles'' and gapped ``charge'' excitations. In the early 2000s, lattice models of spins~\cite{hermele2004} and dimers~\cite{huse2003} on corner-sharing polyhedra were constructed that were strongly argued to realize this phase---a $U(1)$ spin liquid, using perturbation theory, solvable limits~\cite{hermele2004} and later Quantum Monte Carlo simulations~\cite{banerjee2008,shannon2012}. Our work is based on a spin model with easy-axis antiferromagnetic interactions introduced by Hermele et al.~\cite{hermele2004} on the pyrochlore lattice consisting of corner-sharing tetrahedra (see Fig.~\ref{fig:pyrochlore_diamond}). 
The classical Ising limit of this model is the widely studied classical spin ice~\cite{bramwell2020history,gingras2011spin,udagawa2021spin,skjaervo2020advances,nisoli2013colloquium}, which has a large residual entropy at low temperatures similar to water-ice~\cite{pauling1935}. This is because the ground states form an exponentially degenerate set of states obeying the ``ice rule'' (see Sec.~\ref{sec:proposal}). The quantum model in Ref.~\cite{hermele2004} has also been a subject of intense study in the context of pyrochlore materials like $\text{Yb}_2\text{Ti}_2\text{O}_7$ and $\text{Er}_2\text{Ti}_2\text{O}_7$ as potential quantum spin ice (another name for the $U(1)$ spin liquid) candidates~\cite{gingras2014quantum}. 

It was observed in Ref.~\cite{tewari2006emergence} that the Hamiltonian in Ref.~\cite{hermele2004} can be viewed as that of hard-core bosons hopping on an optical lattice with nearest-neighbor repulsion, thus extending its relevance to the cold atom setting. Ref.~\cite{glaetzle2014quantum} studied a similar model of hard-core bosons \textit{hopping} on a two-dimensional checkerboard lattice. In Ref.~\cite{glaetzle2014quantum}, the atom's internal state was largely the ground state, but a dressing with Rydberg states was used to engineer interactions between atoms. Later, Ref.~\cite{celi2020emerging} showed that dimer models in two dimensions can be implemented on configurable Rydberg arrays---where the atoms themselves are stationary but can internally be either in a ground state or in a Rydberg state. In this setting, the atoms are driven with a laser (or a pair of lasers making a two-photon transition) that is detuned from the ground to Rydberg transition. The Rydberg interactions and the detuning define a (frustrated) ``classical'' energy landscape. The laser driving induces quantum fluctuations controlled by the Rabi frequency, leading (perturbatively) to dimer moves or ring exchange terms that are required to deconfine a gauge theory. The proposal~\cite{verresen2021} and experiment~\cite{semeghini2021} mentioned above worked in the same setting. Our work is also based on this setting in which the atom array is configured in a 3D pyrochlore lattice.

In Sec.~\ref{sec:proposal}, we explain our proposal. We show that within a window of laser detunings, the classical landscape is identical to the set of ice rule obeying states. Our Hamiltonian, when restricted to nearest-neighbor interactions, is equivalent to the transverse-field Ising model on the pyrochlore lattice. In the limit of small Rabi frequencies, it is perturbatively equivalent to the model in~\cite{hermele2004}, which was argued to have a spin liquid ground state. Away from the perturbative limit, there is numerical evidence for a spin liquid phase~\cite{emonts2018monte}. However, once we include the long-range $1/r^6$ interactions beyond nearest-neighbor, the classical landscape is no longer degenerate, and it is a priori unclear if the spin liquid survives as the ground state. We attempt to answer this in Sec.~\ref{sec:meanfield} by comparing the energy of an ansatz wave function of the spin liquid with that of an ordered state. 
Within our approximation, we find that by dialing up the Rabi frequency, for fixed detuning and interaction strength, one goes through a confinement-deconfinement transition from an ice rule obeying ferromagnetic state into a deconfined spin liquid phase. Then, by further increasing the Rabi frequency, one goes through a Higgs transition from the spin liquid to a transverse-field-polarized state (see Sec.~\ref{sec:Higgs}). 
\newtext{Thus both the deconfinement-confinement and the Higgs transitions of the compact $U(1)$ gauge theory can be accessed by changing the Rabi frequency in our model.}
While the analysis till this point focuses on the ground state, in Sec.~\ref{sec:dynamics}, we comment on the role played by dynamical state-preparation in deciding the nature of the state prepared in experiment. 
In Sec.~\ref{sec:diagnostics}, we present correlation functions that distinguish the spin liquid from the confined phases, and provide experimental protocols for measuring them.
\newtext{We explain the behavior of the correlation functions in each phase of the phase diagram. We also provide protocols to measure them.}
Finally, in Sec.~\ref{sec:discussion}, we present general discussions and conclusions.

\section{Proposal to realize a $U(1)$ quantum spin liquid using Rydberg atoms}\label{sec:proposal}

In this section, we describe our proposal to realize a $U(1)$ Quantum Spin Liquid (QSL) in Rydberg atom arrays. 
Consider a 3D Rydberg array in which the atoms are positioned on the sites of the pyrochlore lattice [see Fig.~\ref{fig:pyrochlore_diamond}(a)]. Each of the atoms can either be  in the ground state $\ket{g}$ or in the  Rydberg state $\ket{r}$. In the rotating wave approximation and in a rotating frame, the Hamiltonian is
\begin{equation}
\begin{aligned}
  \hat{H}_{\text{ryd}} =& -\delta \sum_i \hat{n}_i + \frac{V}{2} \sum_{i \neq j} \left( \frac{a}{|\bm{x}_i -\bm{x}_j|} \right)^6 \hat{n}_i \hat{n}_j\\
  &+\frac{\Omega}{2} \sum_i (\hat{b}_i + \hat{b}_i^\dagger ),
  \end{aligned}
\end{equation}
where $\hat{b}_i = \ket{g_i}\bra{r_i}$, $\hat{n}_i = \hat{b}_i^\dagger \hat{b}_i$, $\Omega$ is the Rabi frequency, $\delta$ is the laser detuning, $V$ is the nearest-neighbor van der Waals interaction strength,  and $a$ is the distance between two neighboring atoms. The summation $\sum_{i \neq j}$ is over distinct sites $i$ and $j$ of the pyrochlore lattice (each pair is being counted twice), and  $\sum_i$ is over sites $i$. Below, we briefly describe the pyrochlore lattice.

The pyrochlore lattice is a face-centred cubic (FCC) lattice with a four-site basis formed by the four vertices of an up-pointing tetrahedron. (Since each lattice site belongs to one up-pointing tetrahedron and one down-pointing tetrahedron, the down-pointing tetrahedra are formed automatically once we create the up-pointing tetrahedra.) In Cartesian coordinates, the primitive vectors of the FCC lattice are 
\begin{equation}
\begin{aligned}
    \vb{a}_1&=\sqrt{2}a(0,1,1),\\
    \vb{a}_2&=\sqrt{2}a(1,0,1),\\
    \vb{a}_3&=\sqrt{2}a(1,1,0).
\end{aligned}
\end{equation}
The pyrochlore lattice sites are physically located at $\vb{r}+\vb{e}_\mu/2$ [and labeled $(\vb{r},\mu)$], where $\vb{r}$ is an FCC lattice vector, and the vectors $\vb{e}_\mu$ for $\mu \in \{ 0,1,2,3 \}$ are defined as [see Fig.~\ref{fig:pyrochlore_diamond}(b)]
\begin{equation}
\begin{aligned}
    \vb{e}_0&=\frac{a}{\sqrt{2}}(1,1,1) = \frac{1}{4}(\vb{a}_1+\vb{a}_2 + \vb{a}_3),\\
    \vb{e}_1&=\frac{a}{\sqrt{2}}(1,-1,-1),\\
    \vb{e}_2&=\frac{a}{\sqrt{2}}(-1,1,-1),\\
    \vb{e}_3&=\frac{a}{\sqrt{2}}(-1,-1,1).
\end{aligned}
\end{equation}

We map the two levels of the atoms to spins-1/2s: $\ket{g} \rightarrow \ket{\downarrow}$, $\ket{r} \rightarrow \ket{\uparrow}$, $\hat{n}_i \rightarrow \hat{S}_i^z +1/2$  and $\hat{b}_i + \hat{b}_i^\dagger \rightarrow 2 \hat{S}_i^x$. The term $\hat{n}_i \hat{n}_j$ therefore maps to an $ \hat{S}_i^z \hat{S}_j^z$ interaction in addition to a Zeeman term $\hat{S}_i^z$. Written in terms of spins, the Hamiltonian, up to an additive constant, is
\begin{equation} \label{eq:spin_hamiltonian_long_range}
\begin{aligned}
\hat{H}_{\text{ryd}} = &- h \sum_i \hat{S}_i^z + \frac{V}{2} \sum_{i \neq j} \left( \frac{a}{|\bm{x}_i - \bm{x}_j| }\right)^6 \hat{S}_i^z \hat{S}_j^z  \\
& + \Omega \sum_i \hat{S}_i^x ,
\end{aligned}
\end{equation}
where
\begin{equation} \label{eq:h_def}
\begin{aligned}
  h = \delta -\frac{V}{2}   \sum^{}_{i \neq 0}  \left( \frac{a}{|\bm{x}_i - \bm{x}_0|} \right)^6, 
  \end{aligned}
\end{equation}
and is independent of the choice of $\bm{x}_0$ for an infinite lattice.
Evaluating this sum numerically for the pyrochlore lattice, we obtain  $h = \delta - 3.46 V$. It is useful to separate the total Hamiltonian, Eq.~\eqref{eq:spin_hamiltonian_long_range}, into three parts, $\hat{H}_{\text{ryd}}=\hat{H}_0 + \hat{H}_{\Omega}+\hat{H}_{\text{LR}}$, where
\begin{equation}\label{eq:hamsplit}
    \begin{aligned}
      \hat{H}_0=& \frac{V}{2}\sum_{\langle i,j \rangle }\hat{S}_i^z \hat{S}_j^z - h \sum_i \hat{S}_i^z,\\
        \hat{H}_{\Omega}=&\Omega \sum_i \hat{S}_i^x, \text{ and }
        \hat{H}_{\text{LR}}= \frac{V}{2} \sumlr{i \neq j} \left( \frac{a}{|\bm{x}_i - \bm{x}_j| }\right)^6 \hat{S}_i^z \hat{S}_j^z,  
    \end{aligned}
\end{equation}
where $\sum_{\langle i,j \rangle }$ is over nearest-neighbor pairs and $\sum'_{i \neq j}$ in $\hat{H}_{\text{LR}}$ is over the remaining pairs that are not nearest-neighbor (in both $\sum$ and $\sum'$, each pair is counted twice). 
Since the interaction drops very rapidly with distance, we will drop $\hat{H}_{\text{LR}}$ for the rest of this section because doing so allows us to connect to some previously known results~\cite{hermele2004,shannon2012,Sikora2009}. We will study the effect of the long-range van der Waals interaction $\hat{H}_{\text{LR}}$ in Sec.~\ref{sec:meanfield}. 

\begin{figure}[t]
    \centering
    \includegraphics[width=0.4\textwidth]{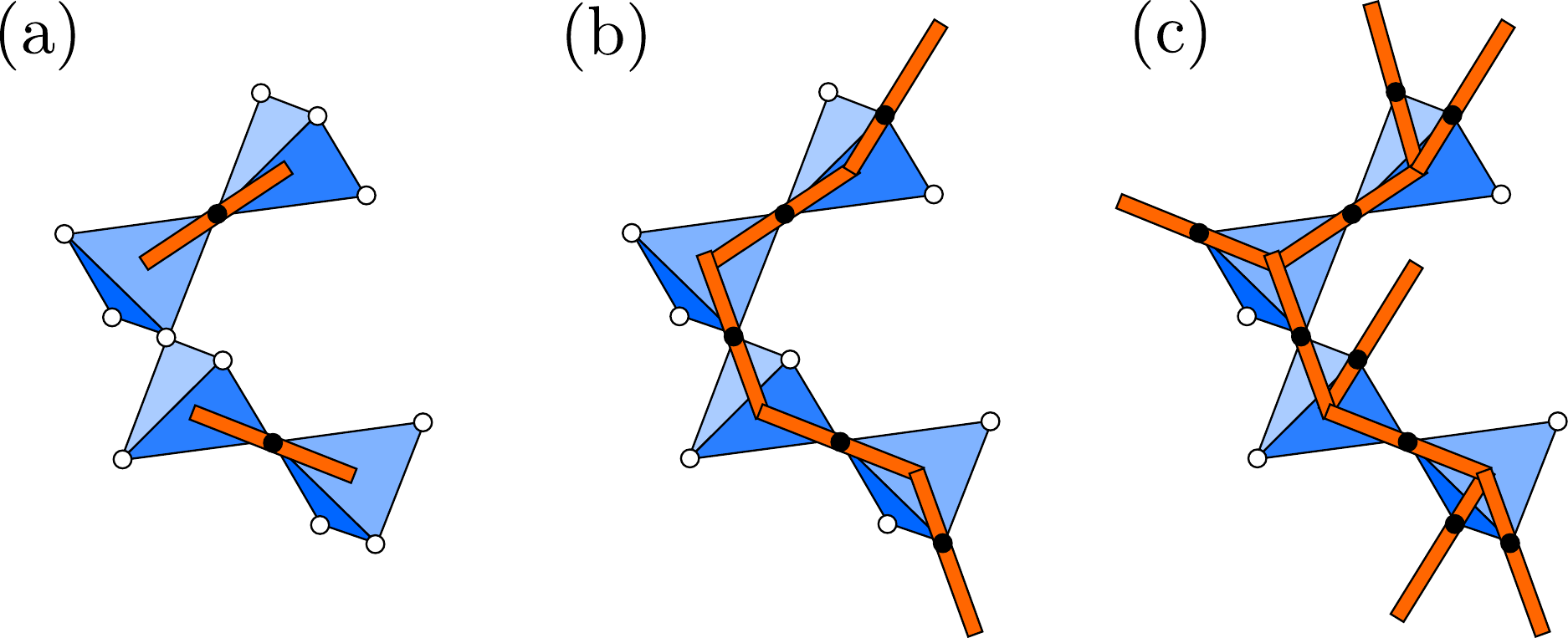}
    \capt{Mapping between Rydberg array configurations and dimer configurations. A Rydberg atom (black dot) is mapped to the presence of a dimer (orange bar), while a ground state atom (white dot) is mapped to the absence of a dimer. (a), (b), and (c) show example dimer configurations corresponding to $n_{\tet} = 1$, $2$, and $3$, respectively. In each case, $n_{\tet}$ many dimers touch the center of each tetrahedron (the centers of the tetrahedra form the diamond lattice).}
    \label{fig:dimer_constraints}
\end{figure}

Since the pyrochlore lattice is made of corner-sharing tetrahedra, we see that $\hat{H}_0$ can be written up to an additive constant as (for convenience, in the expression below, we switch back to the hard-core boson notation)
\begin{equation}
    \hat{H}_0 = \frac{V}{2} \sum_{\tet_{\mathbf{r}}} \left( \hat{n}_{\tet_{\mathbf{r}}} -\rho \right)^2, \label{eq:classical_h}
\end{equation}
where the sum is over all tetrahedra, $\rho= \frac{1}{2}\left( 4+\frac{h}{V} \right)= \frac{1}{2} \left( 0.54 +\frac{\delta}{V} \right)$, and $\hat{n}_{\tet_{\mathbf{r}}} = \sum_{i\in {\tet_{\mathbf{r}}}} \hat{n}_i$ denotes the total number of atoms in the excited state on a given tetrahedron $\tet_{\mathbf{r}}$. Minimizing $\hat{H}_0$ to obtain the classical ground state imposes a constraint on $n_{\tet}$ for each tetrahedron depending on the value of $\rho$:
\begin{equation}\label{eq:constraint}
    n_{\tet} = 
    \begin{cases}
      0 & \text{ if  }\rho< 1/2, \\
        \text{floor}\left(\rho+\frac{1}{2} \right) & \text{ if }1/2 < \rho < 7/2,\\
        4 & \text{ if } 7/2 < \rho. 
    \end{cases}
\end{equation}
\newtext{The classical ground state is unique for $n_{\tet}=0$ and $n_{\tet}=4$, while it is exponentially degenerate (in system size) for $n_{\tet} = 1,2,3$}. 
The number of states satisfying the constraint $n_{\tet} = 2$, is approximately $(3/2)^{N_{\text{tetrahedra}}}$ (where $N_{\text{tetrahedra}}$ is the number of tetrahedra)~\cite{anderson1956ordering}. This is based on an argument similar to the one given by Pauling to explain the residual entropy of water-ice at zero temperature~\cite{pauling1935}. From now on, we will refer to the condition $n_{\tet}=2$ as the ``ice rule". An ice rule obeying configuration is shown in Fig.~\ref{fig:pyrochlore_diamond}(a). 
In these non-trivial cases, the configurations with fixed $n_{\tet}$ can be mapped to configurations of dimers on the bipartite diamond lattice formed by the centers of tetrahedra of the pyrochlore lattice [Fig.~\ref{fig:pyrochlore_diamond}(b)], with exactly $n_{\tet}$ many dimers touching each diamond site (see Fig.~\ref{fig:dimer_constraints}). The A and B sites of the diamond lattice are located at $\vb{n}$ and $\vb{n}+\vb{e}_0$, respectively, where $\vb{n}$ is an FCC lattice vector. For later use in this paper, we also show the lattice dual to this diamond lattice in Fig.~\ref{fig:pyrochlore_diamond}(c) (also a diamond lattice, which we call the ``dual diamond lattice"). 
An atom in the Rydberg state on site $i$ is mapped  to a dimer on the corresponding link of the diamond lattice, while an atom in the ground state is mapped to no dimer. Such dimer models have been studied extensively in both two and three dimensions \cite{rokhsar1988,hermele2004,balasubramanian2022exact,balasubramanian2022classical}.

In the limit $\Omega \ll V$, $\hat{H}_\Omega$ leads to quantum fluctuations that break the exponential degeneracy of the low-energy manifold. We will study this effect perturbatively in the following section (Sec.~\ref{sec:perturbation_theory}). 
Classically, the energy gap between the degenerate ground state space and the lowest excited states corresponding to two tetrahedra violating Eq.~\eqref{eq:constraint} by either $+1$ or $-1$ is $2V \times \text{min}\left(\{\rho +1/2 \},1-\{\rho +1/2\}\right)$. Here, $\{x\}\equiv x - \text{floor}(x)$ is the fractional part of $x$. 
It should be noted that, in the borderline cases when $\rho = m + 1/2$ with $m\in \{0,1,2,3\}$, the energy gap closes and our perturbative analysis cannot be used. We assume going forward that $\rho$ is away from these borderline values. 

\subsection{Perturbation theory} \label{sec:perturbation_theory}

We work in the limit $\Omega \ll V$ and treat $\hat{H}_\Omega$ as a perturbation over $\hat{H}_0$, ignoring for now $\hat{H}_{LR}$ whose effects will be considered later in Sec.~\ref{sec:meanfield}. We calculate the effective Hamiltonian within the ground state manifold of $\hat{H}_0$ using the Schrieffer-Wolff formulation of perturbation theory. For simplicity, we present the calculation of the effective Hamiltonian only for  $n_{\tet} = 2$ here. The only difference between these three cases will be the Hilbert space on which the Hamiltonian acts.
Calculating, at $k^\text{th}$ order in perturbation theory,  the matrix element of the effective Hamiltonian between two states $\ket{n} $ and $\ket{m}$ lying in the degenerate manifold  involves starting from $\ket{m}$, applying the perturbation $k$ times, and reaching the state $\ket{n}$. Since $\hat{H}_\Omega$ changes the particle number by $\pm 1$, the corrections at all odd orders are zero.

\begin{figure}[t]
    \centering
    \includegraphics[width=0.45\textwidth]{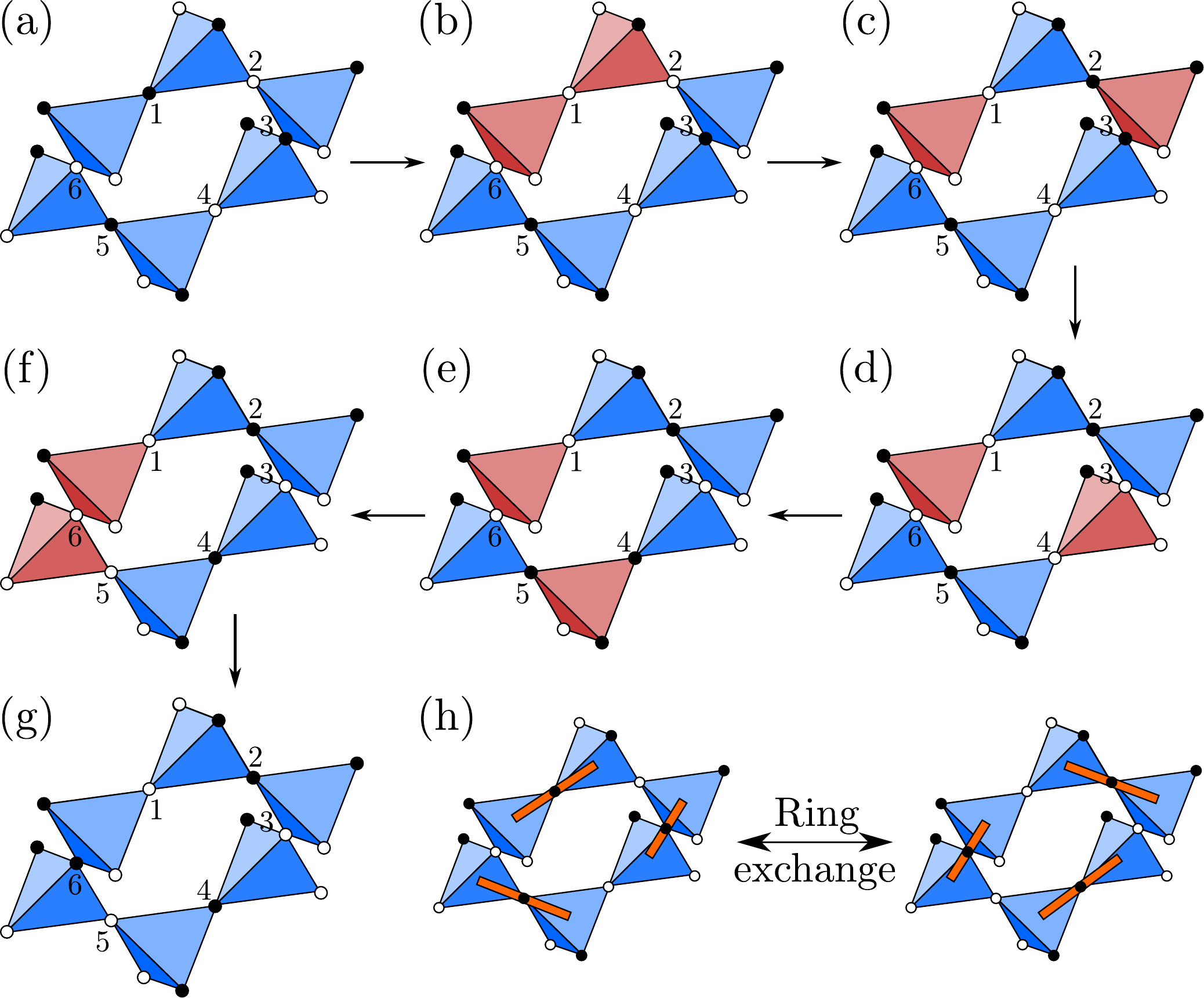}
    \capt{(a) and (b) constitute a virtual process at second order in perturbation theory in $\Omega/V$. Starting from (a) which is a configuration that satisfies $n_{\tet} = 2$ on all sites, $\hat{b}_1 + \hat{b}_1^\dagger$ is applied giving (b). To complete the second order process, $\hat{b}_1 + \hat{b}_1^\dagger$ is applied to (b)  giving back (a). Tetrahedra for which  $n_{\tet} \neq 2$ are shaded in red. Sub-figures (a)--(g) constitute a sixth-order process in the perturbation theory that contributes to the ring exchange term in the effective Hamiltonian, Eq.\ \eqref{eq:ringex}. Starting from (a), the perturbation $\hat{b}_i+\hat{b}_i^\dagger$ is applied sequentially on sites $i=1,2,\ldots,6$. At the end of the six steps, a configuration with $n_{\tet} = 2$ is obtained as shown in (g). Note that the configuration of the atoms on the hexagon is flipped in (g) as compared to (a) thereby producing the effect of a ring exchange. Other sixth-order processes where the perturbation is not applied sequentially also contribute to Eq.\ \eqref{eq:ringex}, but are not shown here. (h) Ring exchange process which appears in the effective Hamiltonian Eq.~\eqref{eq:ringex}. A flippable configuration is mapped to the complimentary flippable configuration. }
    \label{fig:perturbation_th}
\end{figure}

Acting with $\frac{\Omega}{2} (\hat{b}_i + \hat{b}_i^\dagger )$ on an ice rule obeying state creates two excited tetrahedra (whose common site is $i$), which violate the constraint $n_{\tet} = 2$. Therefore, the only second-order process that takes us back to the ice manifold (the degenerate manifold of the ice rule obeying states) is the one in which two excited tetrahedra are created and annihilated, as illustrated in  Figs.~\ref{fig:perturbation_th}(a) and (b). 

\begin{figure}[t]
    \centering
    \includegraphics[width=0.3\textwidth]{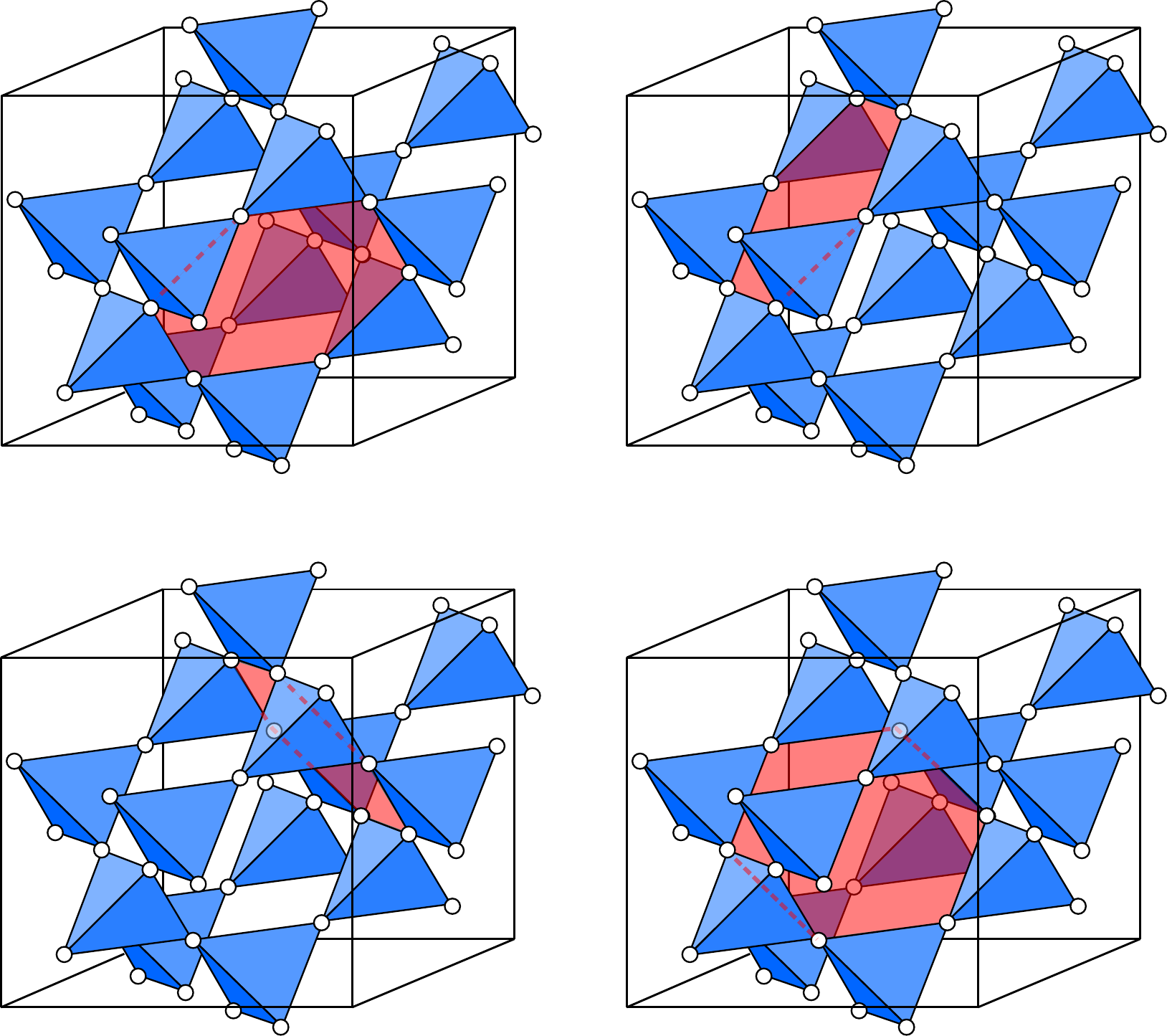}
    \capt{Shaded in red are the four nonequivalent hexagonal plaquettes of the pyrochlore lattice.}
    \label{fig:pyrochlore_plaquettes}
\end{figure}

\begin{figure}[t]
    \centering
\includegraphics[width=0.7\columnwidth]{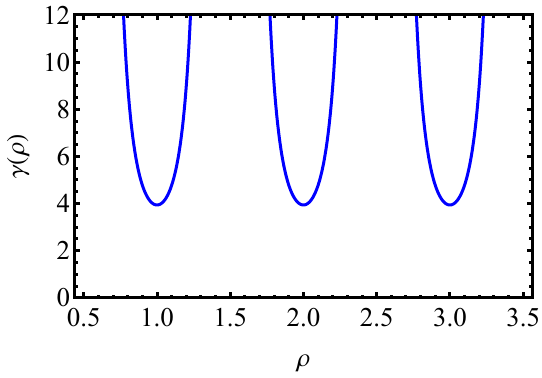}
\capt{Plot showing the variation of $\gamma(\rho)$ (which is the proportionality constant in $J_{\text{ring}} (\rho) = \gamma(\rho) \Omega^6/V^5$) as a function of $\rho$. For $\rho =0.5,1.5,2.5,$ and $3.5$, the energy gap between the low-energy  and the high-energy sectors closes and $\gamma(\rho)$ diverges. }
    \label{fig:jring}
\end{figure}

Since such processes are present for all the states of the ice manifold, they contribute only a constant energy shift and can be ignored. The same is true for the fourth-order processes. Now, the pyrochlore lattice has hexagonal plaquettes, some of which are shown in Fig.~\ref{fig:pyrochlore_plaquettes}.
This allows for non-trivial processes to exist at sixth order. In fact, non-trivial ring exchange over hexagonal plaquettes of the pyrochlore lattice is obtained by the process shown in Figs.~\ref{fig:perturbation_th}(a)--(g)
(some sixth-order processes also result in a constant energy shift which we neglect). A flippable configuration---one in which atoms on a hexagonal plaquette are alternately in the  ground and Rydberg states---is mapped to the complementary flippable configuration by the ring exchange process as illustrated in Fig.~\ref{fig:perturbation_th}(h).
Thus, the effective Hamiltonian consists of ring exchange terms:
\begin{equation}\label{eq:ringex}
  \hat{H}_{\text{eff}} = -J_{\text{ring}} (\rho) \sum_{\hexagon} \ket{\hexA}\bra{\hexB} + \text{H.c.},
\end{equation}
where $J_{\text{ring}} (\rho)= \gamma(\rho) \Omega^6/V^5$, the sum is over all hexagonal plaquettes of the pyrochlore lattice, and $\gamma(\rho)$ is a dimensionless number obtained by summing over virtual processes and is plotted as a function of $\rho$ in Fig.~\ref{fig:jring} . 
We note that, when $\rho$ is an integer,  the value of $\gamma(\rho)$ is $63/16$ and is the same as the one appearing in Refs.~\cite{savary2017disorder,chern2010}. Although the effective Hamiltonian was derived here assuming $n_{\tet} = 2$, the effective Hamiltonian we obtain for $n_{\tet} = 1,3$ is also given by Eq.~\eqref{eq:ringex}. 

In terms of dimers on the diamond lattice, the effective Hamiltonian Eq.~\eqref{eq:ringex} corresponds to a kinetic energy of the dimers. It is well known that dimer models can be made exactly solvable by adding a potential energy $V_{RK}$ for the dimers and tuning to a special point $V_{RK}=J_{\text{ring}}$ called the Rokhsar-Kivelson (RK) point~\cite{rokhsar1988}. The Hamiltonian with such a potential energy term takes the form
\begin{align}\label{eq:Hgeneral}
    \hat{H}_{\text{dimer}} =& -J_{\text{ring}}(\rho) \sum_{\hexagon} \ket{\hexA}\bra{\hexB} + \text{H.c.} \\ \nonumber
    &+ V_{RK} \sum_{\hexagon} \ket{\hexA} \bra{\hexA} + \ket{\hexB} \bra{\hexB}.
\end{align}

The Rydberg system we are interested  in [Eq.~\eqref{eq:ringex}] is obtained from Eq.~\eqref{eq:Hgeneral} by setting $V_{RK}=0$. 

\subsection{$U(1)$ quantum spin liquid---relation to Hermele-Fisher-Balents~\cite{hermele2004}} \label{sec:pyrochlore_photons}

The Hamiltonian in Eq.~\eqref{eq:Hgeneral} was also derived by Hermele, Fisher and Balents in Ref.~\cite{hermele2004} starting from the Heisenberg model on the pyrochlore lattice and taking the easy-axis limit where the Hamiltonian is
\begin{equation}
\label{eq:easyaxis}
\hat{H}_\text{easy-axis}= \frac{1}{2}\sum_{\langle i,j \rangle} \left[ J_z  \hat{S}^z_i \hat{S}^z_j + J_\perp \left(\hat{S}^x_i \hat{S}^x_j + \hat{S}^y_i \hat{S}^y_j\right) \right],
\end{equation}
where $J_z \gg J_{\perp} > 0$. When $J_{\perp} = 0$, the ground state is exponentially degenerate with $S^z_{\tet}=0 $ on each tetrahedron, which is equivalent to $n_{\tet} = 2$. The $J_{\perp}$ term was treated as a perturbation over the $J_z$ term, and at third order, a ring exchange term identical to Eq.~\eqref{eq:ringex} was obtained. Written in terms of the spins, the ring-exchange term is
\begin{equation} \label{eq:hermele_h_eff}
  \hat{H}_{\text{eff}}  = -J_{\text{ring}}\sum_{\hexagon} \hat{S}_1^+ \hat{S}_2^- \hat{S}_3^+ \hat{S}_4^- \hat{S}_5^+ \hat{S}_6^- + \text{H.c.},
\end{equation}
where the sum is over hexagonal plaquettes of the pyrochlore lattice. The RK potential term was added by hand in Ref.~\cite{hermele2004} giving Eq.~\eqref{eq:Hgeneral}.

Hermele et al.~then go to the quantum rotor variables $n_{\vb{r}\vb{r}'} \in \mathbb{Z}$ and $\theta_{\vb{r}\vb{r}'} \in [-\pi,\pi)$, which live on the links $\vb{r}\vb{r}'$ of the diamond lattice (equivalently, sites of the pyrochlore lattice) and satisfy the canonical commutation relations $[\hat{n}_{\vb{r}\vb{r}'} ,\hat{\theta}_{\vb{r}\vb{r}'} ] = i$:
\begin{equation} \label{eq:rotor_mapping}
  \hat{S}^z \rightarrow \hat{n} - \frac{1}{2},  \quad  \hat{S}^\pm  \rightarrow e^{ \pm i\hat{\theta}}.
\end{equation}
The constraint $n = 0 \text{ or }1$ is imposed by adding a term to the Hamiltonian that energetically penalizes states violating this constraint: 
\newtext{
\begin{align} \label{eq:Heff_gauge_theory}
  \hat{H}_{\text{eff}} & =   \frac{U}{2} \sum_{\langle \vb{r}\vb{r}' \rangle} \left( \hat{n}_{\vb{r}\vb{r'}} -\frac{1}{2} \right)^2 \\ \nonumber
&  - 2 J_{\text{ring}} \sum_{\hexagon_p} \cos\left(\hat{\theta}_{\vb{r}_1\vb{r}_2} -\hat{\theta}_{\vb{r}_2 \vb{r}_3}+\hat{\theta}_{\vb{r}_3\vb{r}_4}- \hat{\theta}_{\vb{r}_4\vb{r}_5}\right.\\
& \quad \quad \quad \quad \quad \quad \quad \left.+\hat{\theta}_{\vb{r}_5\vb{r}_6}-\hat{\theta}_{\vb{r}_6\vb{r}_1}\right),
\end{align}
where the first sum is over all the links $\langle\vb{r}\vb{r}'\rangle$ of the diamond lattice. The second sum is over the puckered hexagonal plaquettes of the diamond lattice $\hexagon_p$  whose vertices are $\vb{r}_1, \vb{r}_2, \ldots ,\vb{r}_6$.}
In the limit $U \rightarrow \infty$, Eq.~\eqref{eq:Heff_gauge_theory} reduces to the effective Hamiltonian Eq.~\eqref{eq:hermele_h_eff}.

The local constraint, $S_{\tet_{\vb{r}}}^z = 0$ for each tetrahedron, gives a gauge structure to the effective Hamiltonian where the gauge transformations are generated by $\hat{S}_{\tet_{\vb{r}}}^z$.
The presence of this local symmetry motivated Hermele et al.~to write Eq.~\eqref{eq:Heff_gauge_theory} as a lattice $U(1)$ gauge theory. The electric field and the vector potential were defined as
\begin{equation} \label{eq:EM_mapping}
  \hat{e}_{\vb{r}\vb{r}'} = \pm\left(  \hat{n}_{\vb{r}\vb{r}'} - \frac{1}{2} \right), \quad \hat{a}_{\vb{r}\vb{r}'} = \pm \hat{\theta}_{\vb{r}\vb{r}'}.
\end{equation}
The positive (negative) sign is chosen if $\vb{r}$ belongs to $A$ ($B$) sublattice of the diamond lattice. The Hamiltonian written in terms of the electric field and the vector potential takes the form of a compact $U(1)$ lattice gauge theory~\cite{wilsonConfinementQuarks1974,polyakovCompactGaugeFields1975}:
\begin{equation} 
  \hat{H}_{\text{eff}} = \frac{U}{2} \sum_{\langle \vb{r},\vb{r}' \rangle } \hat{e}_{\vb{r}\vb{r}'}^2 - 2 J_{\text{ring}} \sum_{\hexagon} \cos\left( (\text{curl } \hat{a})_{\hexagon}\right),
\end{equation}
where the second summation is over hexagonal plaquettes of the diamond lattice and 
\begin{equation}
  \left( \text{curl }\hat{a}\right)_{\hexagon} = \bigincircbin{\sum_{\vb{r},\vb{r}' \in \hexagon}}  \;\; \;\hat{a}_{\vb{r} \vb{r}'} ,\medskip \equiv \hat{B}_{\dual{r}, \mu}
\end{equation}
where $\osum_{\;\vb{r},\vb{r}' \in \hexagon}$ is a sum along the directed bonds of a hexagonal plaquette of the diamond lattice. 
\newtext{ Such a plaquette can be uniquely defined by $(\dual{r}, \mu)$, where $\dual{r}$ belongs to the dual lattice and $\dual{u}_\mu$ for $\mu \in\left\{1,2,3,4\right\}$ gives the plaquette orientation [as defined in Fig.~\ref{fig:pyrochlore_diamond}(c)], and $\hat{B}_{\dual{r}, \mu}$ is the magnetic field operator at that plaquette.} 
The adjective ``compact'' refers to the vector potential $\hat{a}_{\vb{r}\vb{r}'}$ being an angular variable. There is an important difference between the above gauge theory and the compact $U(1)$ gauge theory studied by Polyakov~\cite{polyakovCompactGaugeFields1975, polyakov1977,polyakovThermalPropertiesGauge1978}---the gauge theory obtained by Hermele et al.~is an odd gauge theory, i.e.,~electric fields are half-integers, $e_{\vb{r}\vb{r}'} \in \mathbb{Z}+1/2$, while the gauge theory studied by Polyakov 
was an even gauge theory, i.e.,~the electric fields were integers, $e_{\vb{r}\vb{r}'} \in \mathbb{Z}$. Because of this difference, the phases of the two theories differ. 
\newtext{For readers familiar with the Schwinger model, we point out that the even and odd compact $U(1)$ gauge theories are reminiscent of gauge theories with a $\theta$-term in $1+1$ dimensions at $\theta=0$ and $\theta=\pi$, respectively.} 

The phases of a gauge theory can be characterized by the interaction between two externally added opposite electric charges separated by a distance $R$. If the potential between charges goes to zero (or increases as at most $\log{R}$ in $2+1$D) as $R \rightarrow \infty$, then the gauge theory is in the deconfined phase. On the other hand, if the potential increases linearly with $R$ or faster, then these opposite charges cannot be separated, and the gauge theory is in the confined phase. In the limit $U\rightarrow \infty$, the even gauge theory was shown to be in the confined phase in Refs.~\cite{polyakovCompactGaugeFields1975, wilsonConfinementQuarks1974}, while the odd gauge theory can be in either the confined phase or the deconfined phase~\cite{hermele2004}. This can be understood intuitively as follows.

In the even gauge theory, in the limit $U \rightarrow \infty$, the electric fields are forced to be 0, $e_{\vb{r}\vb{r}'} = 0$, to minimize the energy in the absence of any external charges. However, in the presence of two opposite external charges,  the Gauss's law requires that the electric field can no longer be zero everywhere. The spreading of the electric field is, however, penalized by the term $\frac{U}{2} \sum_{\langle \vb{r},\vb{r'} \rangle } \hat{e}_{\vb{r}\vb{r}'}^2$. This forces the electric field to be nonzero  only in  a narrow tube joining the two charges, leading to a linearly rising potential between the two charges. Thus, in the limit $U \rightarrow \infty$,  the even gauge theory is in a confined phase, and there is no deconfined phase in this limit. This confinement of charges has been shown in Refs.~\cite{polyakovCompactGaugeFields1975,polyakov1977,wilsonConfinementQuarks1974,kogutIntroductionLatticeGauge1979}.

On the other hand, in an odd gauge theory, in the limit $U \rightarrow \infty$, the electric field can take two values, $e_{\mathbf{r}\mathbf{r}'} = \pm 1/2$. This results in frustration, i.e.,~allows for many configurations of the electric field, so that the ground state in this limit is non-trivial. When two external charges are introduced, the electric field is not necessarily confined in a string between the charges, but can spread in space similar to the familiar Coulomb-law field lines of a non-compact $U(1)$ gauge theory. This suggests that it is possible for the odd gauge theory to be in the deconfined phase even in the $U\rightarrow \infty$ limit. In fact, the odd gauge theory on the pyrochlore lattice \eqref{eq:Heff_gauge_theory} is indeed in the deconfined phase in the $U \rightarrow \infty$ limit~\cite{banerjee2008, shannon2012, Sikora2009}.

Hermele et al.~have shown that the dimer model with the Hamiltonian Eq.~\eqref{eq:Hgeneral} is described by the deconfined phase of the underlying compact $U(1)$ gauge theory close to the RK point (for $V_{RK}$ smaller than $J_\text{ring}$ but close to $J_\text{ring}$). This phase is the $U(1)$ quantum spin liquid. It has three types of emergent excitations---gapless  photons, gapped magnetic monopoles and gapped fractionalized electric charges, also called as spinons. The spinons are the tetrahedra which violate the constraint on $n_{\tet}$, Eq.~\eqref{eq:constraint}.

\subsection{Previous numerical work} \label{sec:previous_numerical_work}

In this section, we summarize some of the known work on the dimer model with the Hamiltonian Eq.~\eqref{eq:Hgeneral} and on the nearest-neighbor transverse-field Ising model on the pyrochlore lattice.

\begin{figure}[t]
    \centering
    \includegraphics[width=.95\columnwidth]{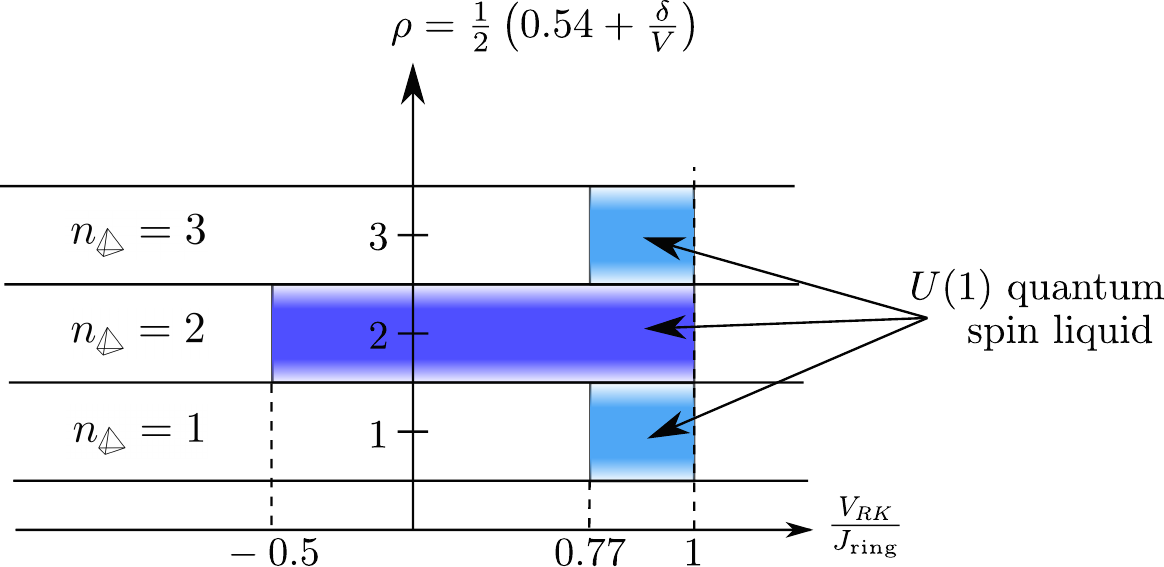}
    \capt{For $\rho \in \left( 3/2,5/2 \right)$, corresponding to $n_{\tet} = 2$, the system is in the $U(1)$ spin liquid phase at $V_{RK} = 0$~\cite{shannon2012}. On the other hand, for $\rho \in \left( 1/2,3/2 \right)$ and $\rho \in \left( 5/2,7/2 \right)$, corresponding to $n_{\tet} = 1$ and $3$,  respectively, the system is in an ordered phase at $V_{RK} =0$~\cite{Sikora2009}. Note that for 
     $\rho = 1/2$, $3/2$, and $5/2$,  the perturbation theory described in Sec.~\ref{sec:perturbation_theory} does not apply, and we cannot comment on the phase of the system.}
    \label{fig:different_dimer_constraints}
\end{figure}

Using quantum Monte Carlo simulations, Refs.~\cite{shannon2012} and~\cite{Sikora2009} studied  the range of $V_{RK}$ [see Eq.~\eqref{eq:Hgeneral}] over which the $U(1)$ spin liquid exists. They  found that the spin liquid is present in the range  $-0.5 J_{\text{ring}}< V_{RK} < J_{\text{ring}}$ for the dimer model with $n_{\tet} = 2$ and in the range $0.77 J_{\text{ring}} < V_{RK} < J_{\text{ring}}$ for the dimer model with $n_{\tet}=1$. The dimer model with $n_{\tet} = 3$ is equivalent to the one with $n_{\tet} = 1$ by a particle-hole transformation.  These numerical results are summarized in Fig.~\ref{fig:different_dimer_constraints}. 

  While a theory proposal to realize the RK potential exists~\cite{celi2020emerging}, the RK potential is a six-body term for the pyrochlore lattice and is difficult to engineer experimentally. Thus, we focus on the case where $V_{RK}=0$. 
From Fig.~\ref{fig:different_dimer_constraints}, we see that to obtain a spin liquid phase for $V_{RK} = 0$, one must have $n_{\tet} = 2$, which corresponds to $3/2 <\rho < 5/2$. 
  In the cases $n_{\tet} = 1$ and $3$, the system is in an ordered state when $V_{RK} = 0$.  
  Hence, in conclusion, assuming the long-range interactions $\hat{H}_{LR}$ can be ignored, we expect that, in the limit $\Omega \ll V$, the Rydberg system will be in a $U(1)$ quantum spin liquid phase for $3/2 < \rho < 5/2$.

  When $\rho = 2$, or equivalently $h = 0$, and the long-range interactions $\hat{H}_{LR}$ are ignored, the Hamiltonian of the system $\hat{H}_0 + \hat{H}_{\Omega}$ in Eq.\ (\ref{eq:hamsplit}) is the transverse field Ising model on the pyrochlore lattice.  For $\Omega \ll V$, we know from the perturbative analysis of Sec.~\ref{sec:perturbation_theory} and Ref.~\cite{Sikora2009} that the system is in the $U(1)$ quantum spin liquid phase. For large $\Omega/V$, where  perturbation theory cannot be applied, Ref.~\cite{emonts2018monte} found using quantum Monte Carlo calculations that the $U(1)$ spin liquid exists in the region $\Omega < 0.55(5) V$, while, for $\Omega > 0.55(5) V$, the system is in a transverse-field-polarized (TFP) phase, which extends to $\Omega/V \rightarrow \infty$ where the ground state is polarized in the $x$-direction. This transition was also studied in Ref.~\cite{rochner2016spin} using perturbation theory, where a transition was found at $\Omega\approx 0.6 V$. 

  The effects of adding a third nearest-neighbor interaction, $V_{\text{3NN}}$, to the dimer model were considered in Ref.~\cite{pace2021}. It was found that the quantum spin liquid transitioned into an ordered state (antiferromagnet~\cite{chen2016magnetic}) at $V_{\text{3NN}} \approx J_{\text{ring}}$. 
  Thus non-nearest-neighbor interactions can destabilize the quantum spin liquid.
  In fact, in a 2D model with neutral atoms located on the bonds of a kagome lattice (same as the sites of a ruby lattice), a spin liquid ground state was found if the interactions were short-ranged using DMRG on cylinders~\cite{verresen2021,semeghini2021}. However, with the full long-range van der Waals interactions, the spin liquid ceased to be the ground state~\cite{verresen2021,semeghini2021}. 
  \newtext{While these works suggest that long-range interactions could destabilize the quantum spin liquid and favor an ordered state, it is not always the case as was discovered in Ref.~\cite{isakov2005why}, where the degeneracy of the ice manifold was preserved despite the introduction of a dipolar-like long-range interaction. Thus, in our work, it is important to study the effects of the van der Waals interaction more closely.}
  In the following section, we will study the phase diagram of Hamiltonian~\eqref{eq:hamsplit} in the presence of long-range interactions, using approximate methods. 

\section{Phase Diagram---Approximate Methods}\label{sec:meanfield}
The goal of this section is to study the ground state phase diagram of Hamiltonian \eqref{eq:spin_hamiltonian_long_range} for $\delta = 3.46 V$ (which corresponds to $\rho = 2$) including long-range interactions $\hat{H}_{\text{LR}}$.

\subsection{Confinement-deconfinement transition---Monte Carlo assisted perturbation theory}\label{sec:mcpert}
Consider the full Hamiltonian $\hat{H}=\hat{H}_0 + \hat{H}_\Omega + \hat{H}_{\text{LR}}$ from Eq.~\eqref{eq:hamsplit} in the case $\rho = 2$ [see Eq.~\eqref{eq:classical_h}]:
\begin{equation}
    \begin{aligned}
        \hat{H}_0=& \frac{V}{2} \sum_{\tet_{\mathbf{r}}}\left( \sum_{i\in \tet_{\mathbf{r}}} \hat{S}^z_i  \right)^2,\\
        \hat{H}_{\Omega}=&\Omega \sum_i \hat{S}_i^x, \text{ and } \hat{H}_{\text{LR}}= \frac{V}{2} \sumlr{i \neq j} \left( \frac{a}{|\bm{x}_i - \bm{x}_j| }\right)^6 \hat{S}_i^z \hat{S}_j^z.  
    \end{aligned}
\end{equation}
The long-range interaction $\hat{H}_{\text{LR}}$ splits the exponential degeneracy of the ice manifold, and selects one configuration diagonal in the $\hat{S}^z$ basis as the ground state of $\hat{H}_0 + \hat{H}_{\text{LR}}$, which we call the ``ordered state''. On the other hand, $\hat{H}_\Omega$ prefers superpositions of ice rule obeying states, the $U(1)$ quantum spin liquid (QSL) being one such superposition. Further, we also note that quantum fluctuations around the ``ordered state" due to $\hat{H}_\Omega$ may also lead to a change in its energy relative to the QSL. 
It is this competition between kinetic energy and long-range interactions that we will study in this section.

We first show that the ground state in the classical limit $\Omega=0$ is the zero-momentum state satisfying the ice rule which we call the ``ice ferromagnet''. We assume that, as one increases $\Omega$, there is no phase transition to a different ordered state before the putative transition to a QSL. In order to determine whether a QSL phase exists and, if yes, at what $\Omega$ the transition to the QSL occurs, one needs to compare the energies of ansatz wavefunctions of the QSL and the ordered state. When $\Omega\neq 0$, such wavefunctions would necessarily involve configurations that violate the ice rule. We incorporate the effect of nonzero $\Omega$ on the wavefunction using perturbation theory. Our strategy is as follows. We treat $\hat{H}_1 \equiv \hat{H}_\Omega + \hat{H}_{\text{LR}}$, i.e.,~both the laser driving term and the long-range interactions, as a perturbation to $\hat{H}_0$ (unlike Sec.~\ref{sec:perturbation_theory}, where we dropped $\hat{H}_{\text{LR}}$). We perturbatively find an effective Hamiltonian $\hat{H}_{\text{eff}}$ acting on the low-energy ice manifold. We then compare the expectation value of $\hat{H}_{\text{eff}}$ in candidate wavefunctions that live entirely in this low-energy space. 
Since a QSL wavefunction is a linear superposition of exponentially (in system size) many ice rule obeying states, we calculate $\expval{\hat{H}_{\text{eff}}}$ numerically using classical Monte Carlo sampling.

\subsubsection{Expression for $\hat{H}_{\text{eff}}$} \label{sec:expression_for_Heff}
We perform a Schrieffer-Wolff transformation 
\begin{equation}\label{eq:scwtrans}
    \hat{\tilde{H}}= \hat{U}_S \hat{H} \hat{U}_S^{\dagger} = \hat{U}_S \left(\hat{H}_0 + \hat{H}_\Omega + \hat{H}_{\text{LR}}\right)\hat{U}_S^{\dagger},
\end{equation}
for a unitary $\hat{U}_S=e^{\hat{S}}$, where $\hat{S}$ is an anti-hermitian operator  
chosen to make $\hat{\tilde{H}}$ block-diagonal in the (degenerate) eigenbasis of $\hat{H}_0$, i.e.,~
\begin{equation}\label{eq:effhamdef}
\hat{\tilde{H}}=\hat{\mathcal{P}}\hat{\tilde{H}}\hat{\mathcal{P}} + (1-\hat{\mathcal{P}})\hat{\tilde{H}}(1-\hat{\mathcal{P}}), 
\end{equation}
where $\hat{\mathcal{P}}$ projects onto the ice manifold. In the remainder of this paper, we will restrict ourselves to the low-energy sector and therefore only consider the $\hat{H}_{\text{eff}} \equiv\hat{\mathcal{P}}\hat{\Tilde{H}}\hat{\mathcal{P}}$ term above. We calculate $\hat{H}_{\text{eff}}$ perturbatively in $\hat{H}_1 = \hat{H}_\Omega +\hat{H}_{\text{LR}}$ (see Appendix B of Ref.~\cite{slagle2017} for general expressions of $\hat{H}_\text{eff}$). As we saw in Sec.~\ref{sec:perturbation_theory}, if we consider only $\hat{H}_\Omega$ as the perturbation, then the first non-trivial term appearing in $\hat{H}_{\text{eff}}$ is $ -J_{\text{ring}} \sum_{\hexagon} \ket{\hexA}\bra{\hexB} + \text{H.c.}$, where 
\begin{equation}
  J_{\text{ring}}= \frac{63}{16}\frac{\Omega^6}{V^5} + \Theta\left( \frac{\Omega^8}{V^7} \right).
\end{equation} 
Since we are performing perturbation theory in two operators $\hat{H}_{\Omega}$ and $\hat{H}_{LR}$, each of them comes with its own small parameter. Since the perturbative expansion will involve polynomials in these two small parameters, there is some arbitrariness in deciding how to compare the two parameters relative to each other and thus in how to truncate the expansion. In our calculation, we follow an operational scheme of keeping all the terms up to sixth order in $\hat{H}_\Omega+\hat{H}_{\text{LR}}$. Following this truncation scheme, we get (up to additive constants)
\begin{equation}\label{eq:Hpertlr}
\begin{aligned}
     \hat{H}_{\text{eff}}\approx&-J_{\text{ring}} \sum_{\hexagon} \ket{\hexA}\bra{\hexB} + \text{H.c.}\\
     &+\left(1-\frac{\Omega^2}{V^2}-\frac{61}{18}\frac{\Omega^4}{V^4}\right)\hat{H}_{\text{LR}}\\& - \frac{\Omega^2}{V}\left( \hat{W}^{(2)}_{\text{LR}}+\hat{W}^{(3)}_{\text{LR}}+\hat{W}^{(4)}_{\text{LR}}\right)\\
     &-\frac{\Omega^4}{V^3}\left(\frac{152}{27} \hat{W}^{(2)}_{\text{LR}}-\hat{L}^{(2)}_{\text{LR}}+\hat{M}^{(2)}_{\text{LR}}\right),
\end{aligned}
\end{equation}
where 
\begin{align}
    \hat{W}^{(2)}_{\text{LR}}&\equiv \frac{1}{4}\sum_{j} \sum_{\substack{k_1\neq j \\ k_2\neq j}}v_{j,k_1} v_{j, k_2}\hat{S}^z_{k_1}\hat{S}^z_{k_2},\\
    \hat{L}^{(2)}_{\text{LR}}&\equiv \frac{109}{432}\sum_{\substack{j_1 \neq k_1 \\ j_2 \neq k_2}}\delta_{\langle j_1, j_2 \rangle}  v_{j_1,k_1} v_{j_2, k_2} \hat{S}^z_{k_1}\hat{S}^z_{k_2}, \\
    \hat{M}^{(2)}_{\text{LR}}&\equiv \frac{20}{27} \sum_{\substack{j_1 \neq k_1 \\ j_2 \neq k_2}}\delta_{\langle j_1, j_2 \rangle}  v_{j_1,k_1} v_{j_2, k_2} \hat{S}^z_{j_1}\hat{S}^z_{k_1}\hat{S}^z_{j_2}\hat{S}^z_{k_2}, \\
    \hat{W}^{(3)}_{\text{LR}}&\equiv \frac{1}{2}\sum_{j}\sum_{\substack{k_1 \neq j \\ k_2 \neq j \\ k_3 \neq j}}v_{j,k_1}v_{j, k_2} v_{j, k_3}\hat{S}^z_{k_1}\hat{S}^z_{k_2}\hat{S}^z_{k_3}\hat{S}^z_{j},\\
    \hat{W}^{(4)}_{\text{LR}}&\equiv \frac{1}{4}\sum_{j}\sum_{\substack{k_1 \neq j \\ k_2 \neq j \\ k_3 \neq j \\k_4 \neq j}}v_{j,k_1}v_{j, k_2} v_{j, k_3}v_{j, k_4}\hat{S}^z_{k_1}\hat{S}^z_{k_2}\hat{S}^z_{k_3}\hat{S}^z_{k_4},\\
    \text{ and } &v_{i,j} \equiv \begin{cases}\frac{a^6}{\left|\bm{x}_i - \bm{x}_j\right|^6} \text{ if }\bm{x}_i, \bm{x}_j \text{ are not nearest neighbors,}\\0 \;\;\;\;\;\;\;\;\;\;\;\text{ otherwise.}\end{cases}
\end{align}
In the above equations, $\delta_{\langle i,j \rangle}$ enforces $i$ and $j$ to be nearest neighbors.

The expectation value of the Hamiltonian~\eqref{eq:hamsplit} in a given state $\ket{\Psi}$ is 
\begin{equation}
    \mel{\Psi}{\hat{H}}{\Psi}=\left(\bra{\Psi}\hat{U}_S^\dagger\right)\left(\hat{U}_S \hat{H} \hat{U}_S^\dagger \right)\left(\hat{U}_S \ket{\Psi} \right).
\end{equation}
Suppose $\hat{U}_S \ket{\Psi} $ (i.e.,~$\ket{\Psi}$ transformed by the Schrieffer-Wolff transformation) lies entirely in the ice manifold, then using Eq.~\eqref{eq:scwtrans}, we get
\begin{equation}
    \mel{\Psi}{\hat{H}}{\Psi}= \left(\bra{\Psi}\hat{U}_S^\dagger\right) \hat{H}_{\text{eff}}\left(\hat{U}_S \ket{\Psi} \right).
\end{equation}
For the ground state, $\ket{\Psi_g}$ of the full Hamiltonian $\hat{H}$, $\hat{U}_S \ket{\Psi_g}$ lies entirely in the ice manifold. Thus, we pick an ansatz wavefunction for $\hat{U}_S \ket{\Psi}$ that also lies entirely in the ice manifold and compute the expectation value of $\hat{H}_{\text{eff}}$ in our ansatz state to get the energy. Before describing our ansatz states in Sec.~\ref{sec:ansatz_wf}, we first consider the limit $\Omega=0$ in the next section.

\subsubsection{Classical ground state of the long-range Hamiltonian}\label{sec:classical}

Here, we will find the ground state selected by long-range interactions in the limit $\Omega = 0$ where there are no quantum fluctuations. The Hamiltonian is $\hat{H}_{\text{cl}} = \hat{H}_0 +  \hat{H}_{LR}$. We find the ground state by going to the Fourier space. Since the pyrochlore lattice is an FCC lattice with a four-site basis, we use the notation $\hat{S}^z_{\mathbf{r},\mu}$ for spins where $\mathbf{r}$ is an FCC lattice vector and $\mu \in \{ 0, 1,2,3 \}$ labels the sites within the basis. The spin $\hat{S}^z_{\mathbf{r},\mu}$ is physically located at $\mathbf{r} + \mathbf{e}_\mu/2$ where $\mathbf{e}_\mu$ are the vectors joining a diamond A site to a neighboring diamond B site. (See Fig.~\ref{fig:pyrochlore_diamond}(b) for the precise definition.)
\newtext{Using the Luttinger-Tisza method~\cite{luttinger1946theory, lyons1960method,litvin1974luttinger}, we are able to determine the exact ground state of the classical Hamiltonian at $\Omega=0$. We explain this calculation below.}
As we are considering the classical limit in this section, we drop hats on quantities which would otherwise be operators. The Fourier transform of $S_{\mathbf{r},\mu}^z$ is
\begin{equation}
  S^z_{\mathbf{r},\mu} = \frac{1}{\sqrt{N_{\text{u.c.}}}} \sum_{\mathbf{k}} e^{i \mathbf{k} \cdot \mathbf{r}}S^z_{\mathbf{k},\mu},
\end{equation}
where $N_{\text{u.c.}}$ is the number of FCC unit cells. Substituting this in $H_{\text{cl}}$, we get 
\begin{equation}
  H_{\text{cl} } = \sum_{\mu,\nu,\mathbf{k}} V_{\mu\nu,\mathbf{k}} S^z_{\mathbf{k},\mu} S^z_{-\mathbf{k},\nu},
\end{equation}
where $\mathbf{k}$ is a vector in the Brillouin zone of the FCC lattice and $V_{\mu\nu,\mathbf{k}}$ is the Fourier transform of the van der Waals potential:
\begin{equation} \label{eq:vlr_ft}
V_{\mu\nu,\mathbf{k}} = \frac{V}{2 } \sum_{\mathbf{r}} e^{i \mathbf{k}\cdot \mathbf{r}} \left( \frac{a}{|\mathbf{r} + (\mathbf{e}_\mu - \mathbf{e}_\nu)/2 |} \right)^6.
\end{equation}
Diagonalizing the matrix $V_{\mu\nu,\mathbf{k}}$ for each $\mathbf{k}$ gives
\begin{equation} \label{eq:H_Omega_0_fourier}
  H_\text{cl} = \sum_{\mu, \mathbf{k}} \varepsilon_{\mathbf{k},\mu} |S^{'z}_{\mathbf{k},\mu}|^2,
\end{equation}
where $S^{'z}_{\mathbf{k},\mu}$ is related to $S^z_{\mathbf{k},\nu}$ through a multiplication by a unitary matrix $U_{\mu\nu,\mathbf{k}}$ which diagonalizes $V_{\mu\nu,\mathbf{k}}$: $S^{'z}_{\mathbf{k},\mu} = \sum_\nu U_{\mu\nu,\mathbf{k}} S^z_{\mathbf{k},\nu}$.
Recall that $S^z_{\mathbf{r},\mu}$ is either $+1/2$ or $-1/2$. This imposes the following constraint:
\begin{equation} \label{eq:sk2_constraint}
  \sum_{\mathbf{k},\mu} | S^{'z}_{\mathbf{k},\mu} |^2 = \sum_{\mathbf{k},\mu } |S^z_{\mathbf{k},\mu} |^2 =  \sum_{\mathbf{r},\mu} \left(  S^z_{\mathbf{r},\mu} \right)^2 = N_{\text{u.c.}} .
\end{equation}
From Eq.~\eqref{eq:H_Omega_0_fourier}, the energy can be interpreted as a weighted sum of $\varepsilon_{\mathbf{k},\mu}$ with the corresponding weights being $|S^{'z}_{\mathbf{k},\mu}|^2$. Because of the constraint in Eq.~\eqref{eq:sk2_constraint}, the energy is minimized by having the full weight on the smallest $\varepsilon_{\mathbf{k},\mu}$ and no weight on the rest of the $\varepsilon_{\mathbf{k},\mu}$. This holds provided that such a configuration of $S^{'z}_{\mathbf{k},\mu}$ in the momentum space corresponds to some configuration in the real space where $S^z_{\mathbf{r},\mu}$ are $\pm 1/2$.

\begin{figure}[t]
    \centering
    \includegraphics[width=0.25\textwidth]{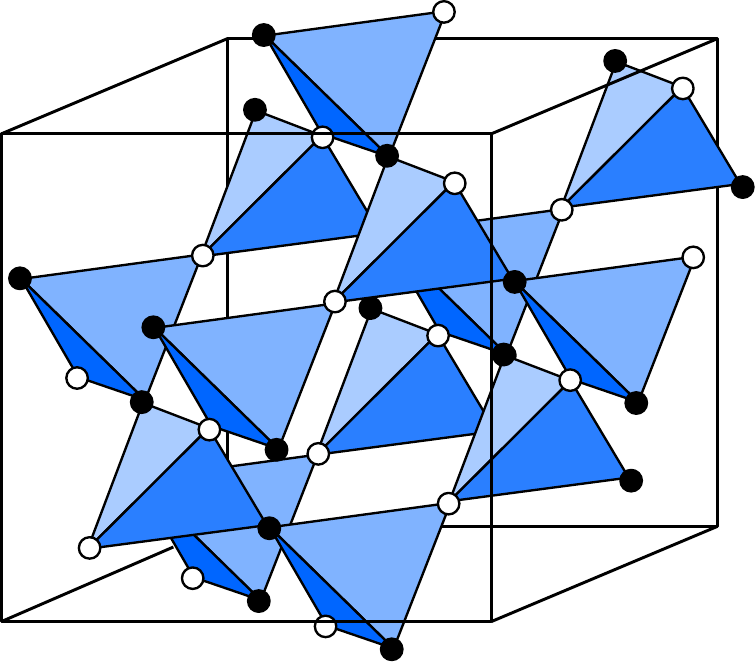}
    \capt{An ice ferromagnet state. It is an ice rule obeying state (i.e.,\ $n_{\tet} = 2$ on every tetrahedron) with $\mathbf{k} = \bm{0}$. All the up-pointing tetrahedra are copies of each other. The same is true for the down-pointing tetrahedra. There are six (${}^4C_2$) such states, and together they make up the ground subspace of $\hat{H}_{\text{cl}}$.}
    \label{fig:ice_fm}
\end{figure}

Calculating the Fourier transform of the long-range potential, Eq.~\eqref{eq:vlr_ft}, and its eigenvalues $\varepsilon_{\mathbf{k},\mu}$, we find that the minimum of $\varepsilon_{\mathbf{k},\mu}$ occurs for $\mathbf{k} =\boldsymbol{0}$ and is triply degenerate. In particular,
\begin{equation} \label{eq:Vk0_matrix}
  V_{\mu\nu,\mathbf{k}=\boldsymbol{0}} = 
  \begin{pmatrix}
    v_1 & v_2 & v_2 & v_2\\
    v_2 & v_1 & v_2 & v_2\\
    v_2 & v_2 & v_1 & v_2\\
    v_2 & v_2 & v_2 & v_1\\
  \end{pmatrix},
\end{equation}
where $v_1 = 0.113 V$ and $v_2 =1.12 V$. Its eigenvalues are $\varepsilon_{\boldsymbol{0},0} = 3.46V$ and $\varepsilon_{\boldsymbol{0},1}=\varepsilon_{\boldsymbol{0},2}=\varepsilon_{\boldsymbol{0},3} = -1.004 V$. The unitary that diagonalizes the above matrix also relates $S^{'z}_{\boldsymbol{0},\mu}$ to $S^{z}_{\boldsymbol{0},\nu}$ as 
\begin{equation} \label{eq:s_to_s_prime}
\renewcommand\arraystretch{1.3}
  \begin{pmatrix}
    S^{'z}_{\boldsymbol{0},0} \\ 
    S^{'z}_{\boldsymbol{0},1} \\ 
    S^{'z}_{\boldsymbol{0},2} \\ 
    S^{'z}_{\boldsymbol{0},3} \\ 
  \end{pmatrix} 
  = \frac{1}{2} 
  \begin{pmatrix}
    1 &  1 &  1 &  1 \\
    1 &  1 & -1 & -1 \\
    1 & -1 &  1 & -1 \\
    1 & -1 & -1 &  1 \\
  \end{pmatrix} 
  \begin{pmatrix}
    S^{z}_{\boldsymbol{0},0} \\ 
    S^{z}_{\boldsymbol{0},1} \\ 
    S^{z}_{\boldsymbol{0},2} \\ 
    S^{z}_{\boldsymbol{0},3} \\ 
  \end{pmatrix}.
\end{equation}
Since $\varepsilon_{\boldsymbol{0},1},\varepsilon_{\boldsymbol{0},2}$ and $\varepsilon_{\boldsymbol{0},3}$ are the minimum eigenvalues, the energy is minimized by having all the weight distributed between $S^{'z}_{\boldsymbol{0},1}, S^{'z}_{\boldsymbol{0},2}$ and $S^{'z}_{\boldsymbol{0},3}$ and no weight on the remaining $S^{'z}_{\mathbf{k},\mu}$, that is, $S^{'z}_{\mathbf{k}\neq\boldsymbol{0},\mu} = 0$ and $S^{'z}_{\boldsymbol{0},0} = 0$. 
There indeed exist states satisfying these two conditions. The first condition, $S^{'z}_{\mathbf{k}\neq\boldsymbol{0},\mu} = 0$, implies that the ground state is a $\mathbf{k} = \boldsymbol{0}$ state, while the second condition, $S^{'z}_{\boldsymbol{0},0} = 0$,  implies that the ground state satisfies the ice rule (so that the total spin, which is $S^{'z}_{\boldsymbol{0},0}$ is 0), see Eq.~\eqref{eq:s_to_s_prime}. There are six such states, and we refer to them as the ``ice ferromagnet'' or ``ice FM" states. One of these is shown in Fig.~\ref{fig:ice_fm}.
\newtext{We note that ice ferromagnet is one of the ``chain states" that were described in Ref.~\cite{mcclarty2015chain}.}
\newtext{We here point out an interesting question: if we add a next-nearest-neighbor $\hat{S}^z\hat{S}^z$ interaction to $\hat{H}_{\text{cl}}$, do the ground states of this new Hamiltonian satisfy the ice rule and are they also ``chain states" as described in Ref.~\cite{mcclarty2015chain}? We leave it for future work to answer this question.}

\subsubsection{Ansatz wavefunctions for the ordered state and for the quantum spin liquid} \label{sec:ansatz_wf}
We now assume that, as one increases $\Omega$ starting from $\Omega=0$, the ground state remains adiabatically connected to the ice ferromagnet derived in the previous section till the point where it undergoes the putative phase transition to the QSL. Therefore, our ansatz for the ordered state is 
\begin{equation}
    \ket{\Psi_{\text{ord}}}=\hat{U}^{\dagger}_S \ket{\Psi_{\text{IFM}}},
\end{equation}
where $\ket{\Psi_\text{IFM}}$, a product state in the $\hat{S}^z$ basis, is the $\vb{k}=\boldsymbol{0}$ ice ferromagnet defined Sec.~\ref{sec:classical}. This configuration is given by $S^z_{\mathbf{r},\mu}=\frac{1}{2}\varepsilon_\mu$ (independent of $\mathbf{r}$), where $\left(  \varepsilon_0 , \varepsilon_1 , \varepsilon_2 , \varepsilon_3 \right) \equiv \left( 1,1, -1,-1\right)$. We note that there are six such choices for $\varepsilon_\mu$ that satisfy the ice rule. We pick one such choice, but our calculations are not sensitive to which one we pick. 
$\ket{\Psi_{\text{IFM}}}$ lives entirely in the ice manifold. Left-multiplication by $\hat{U}^{\dagger}_S$ takes it back to the original Hilbert space with ice rule violations.

Our ansatz wave function for the spin liquid state is 
\begin{equation}
\label{eq:qsl_ansatz}
    \ket{\Psi_{\text{QSL}}}=\hat{U}^{\dagger}_S \ket{\Psi_{\text{RK}}},
\end{equation}
where $\ket{\Psi_{\text{RK}}}$ is a uniform superposition of all dimer coverings~\cite{rokhsar1988} of the diamond lattice (with $n_{\tet}=2$). $\ket{\Psi_{\text{RK}}}$ lives in the ice manifold. Like before, we left-multiply it by $\hat{U}^{\dagger}_S$ to take it back to the original Hilbert space. The justification for our choice is the following. $\ket{\Psi_{\text{RK}}}$ is the ground state of the dimer model at the RK point [see Eq.~\eqref{eq:Hgeneral}]. When the RK potential is zero, $\ket{\Psi_\text{RK}}$ has an energy expectation value of $-4N_{\text{u.c.}}J_{\text{ring}}\overline{n}_{\text{flip}}$, where $\overline{n}_{\text{flip}}$ is the average fraction of flippable hexagons in the RK wavefunction. We find numerically that $\overline{n}_{\text{flip}}=0.1757$ (also calculated in Ref.~\cite{hermele2004}). Therefore, the energy of $\ket{\Psi_{\text{RK}}}$ is $-0.7028 J_{\text{ring}} N_{\text{u.c.}}$ which is not too far from the ground state energy of the dimer model \eqref{eq:hermele_h_eff} found in Ref.~\cite{shannon2012} to be $-0.756 J_{\text{ring}}N_{\text{u.c.}}$. Even though $\ket{\Psi_{\text{RK}}}$ has slightly higher energy, it has the advantage of being simpler to sample by classical Monte Carlo. This explains our choice.

For comparison, we will also calculate the energy of a different ordered state $\ket{\Psi_{\text{ord}}'}=\hat{U}_S^\dagger \ket{\Psi_{\text{IAFM}}}$ that we call an ice antiferromagnet. Here $\ket{\Psi_{\text{IAFM}}}$ is an ice rule obeying state with ordering wave vector $\vb{k}=\pi \left(\vb{b}_1 + \vb{b}_2\right)$, where $\vb{b}_1$, $\vb{b}_2$ and $\vb{b}_3$ are primitive reciprocal lattice vectors of the FCC lattice satisfying $\vb{a}_i \cdot \vb{b}_j =\delta_{ij}$. This state is known elsewhere in literature as the $2\pi (001)$ state (this nomenclature uses an enlarged cubic unit cell of the FCC lattice) \cite{melko2001long,melko2004monte,chen2016magnetic}.

\subsubsection{Numerical results---energy expectation values and phase diagram}\label{sec:numerics}

\begin{figure}[t]
  \centering
  \includegraphics[width=0.95\columnwidth]{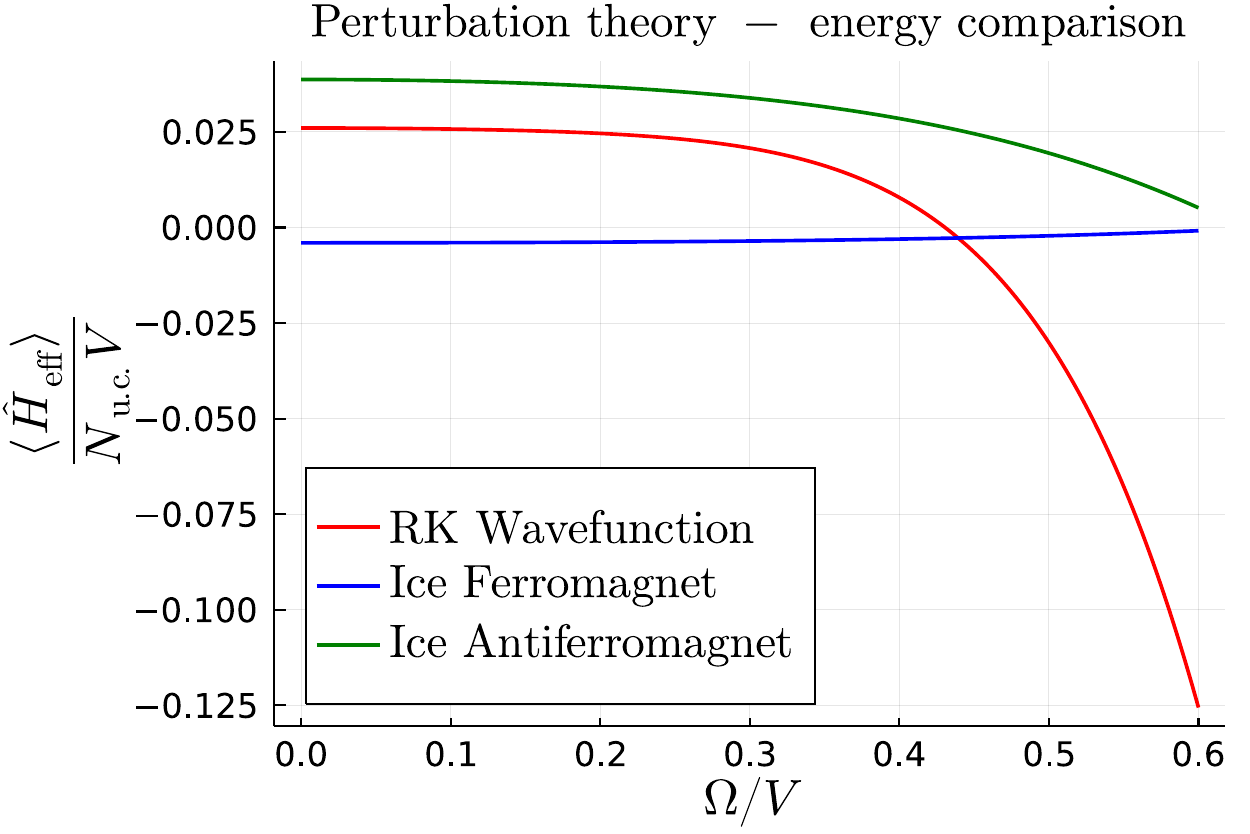}
  \capt{$\langle \hat{H}_\text{eff} \rangle $ in $\ket{\Psi_{\text{RK}}}$, $\ket{\Psi_{\text{IFM}}}$, and $\ket{\Psi_{\text{IAFM}}}$ calculated by inserting the values in Table~\ref{tab:numerics} in Eq.~\eqref{eq:Hpertlr}.}
  \label{fig:pertenergy}
\end{figure}

We now describe our computation of the expectation value of $\hat{H}_{\text{eff}}$ [see Eq.~\eqref{eq:Hpertlr}] in $\ket{\Psi_{\text{RK}}}$, $\ket{\Psi_{\text{IFM}}}$ and in $\ket{\Psi_{\text{IAFM}}}$. While the expectation value in $\ket{\Psi_{\text{IFM}}}$ and $\ket{\Psi_{\text{IAFM}}}$ can be computed straightforwardly, the expectation value in $\ket{\Psi_{\text{RK}}}$ requires classical Monte Carlo sampling. We use a system with $8 \times 8 \times 8$ unit cells (i.e.,~containing $2048$ pyrochlore sites) with periodic boundary conditions in the $\vb{a}_1$, $\vb{a}_2$, and $\vb{a}_3$ directions.  
We restrict our sampling to sectors in which the total electric flux piercing through any 2D torus cross-section (as defined in Sec.~IV~B of Ref.~\cite{hermele2004}) is 0. Our sampling is done using loop moves as described in Refs.~\cite{hermele2004,melko2001long,melko2004monte} -- in each Monte Carlo run, we perform $512\times 500,000$ loop moves. We calculate $\Bar{n}_{\text{flip}}$, $H_{\text{LR}}$, $W^{(2)}_{\text{LR}}$ and $L^{(2)}_{\text{LR}}$ after every 512 loop moves, i.e., we take 500,000 data points. We calculate $M^{(2)}_{\text{LR}}$, $W^{(3)}_{\text{LR}}$, and $W^{(4)}_{\text{LR}}$ after every $512\times 10,000$ loop moves, i.e. we take 50 data points. We repeat this procedure for 9 independent runs in order to calculate the uncertainties. Our results are summarized in Table~\ref{tab:numerics}. 
\begin{table*}[t]
  \setlength{\extrarowheight}{1.1ex}
  \begin{ruledtabular}
    \begin{tabular}{cccc}
         Operator & $\ket{\Psi_{\text{RK}}}$ & $\ket{\Psi_{\text{IFM}}}$ & $\ket{\Psi_{\text{IAFM}}}$ \\  \hline
         $\hat{R} $ & $0.70288(4)N_{\text{u.c.}}$ & $0$ & $0$\\ 
         $\hat{H}_{\text{LR}}$ & $ 2.6037(1) \times 10^{-2}N_{\text{u.c.}}$ & $-0.4002 \times 10^{-2} N_{\text{u.c.}}$ & $3.8722\times 10^{-2}N_{\text{u.c.}}$\\ 
         $\hat{W}^{(2)}_{\text{LR}}$ & $1.11778(1) \times 10^{-3} N_{\text{u.c.}}$ & $0.01642\times 10^{-3} N_{\text{u.c.}}$ & $1.4994 \times 10^{-3} N_{\text{u.c.}}$ \\
         $\hat{L}^{(2)}_{\text{LR}}$ & $ -2.7467(3)\times 10^{-4} N_{\text{u.c.}}$ & $-0.0829\times 10^{-4} N_{\text{u.c.}}$ & $ -7.5662\times 10^{-4} N_{\text{u.c.}}$\\
         $\hat{M}^{(2)}_{\text{LR}}$ & $ 2.96(3)\times 10^{-3} N_{\text{u.c.}}$ & $0.073\times 10^{-3} N_{\text{u.c.}}$ & $ 6.66\times 10^{-3} N_{\text{u.c.}}$\\
         $\hat{W}^{(3)}_{\text{LR}}$ & $ 5.25(4)\times 10^{-5} N_{\text{u.c.}}$ & $-0.00665\times 10^{-5} N_{\text{u.c.}}$ & $ 5.81\times 10^{-5} N_{\text{u.c.}}$\\
         $\hat{W}^{(4)}_{\text{LR}}$ & $ -3.57(2)\times 10^{-6} N_{\text{u.c.}}$ & $-0.0309\times 10^{-6} N_{\text{u.c.}}$ & $ -5.35\times 10^{-6} N_{\text{u.c.}}$
    \end{tabular}
    \capt{The expectation values of the operators in the left column in ansatz wavefunctions $\ket{\Psi_{\text{RK}}}$, $\ket{\Psi_\text{IFM}}$ and $\ket{\Psi_\text{IAFM}}$ respectively. The operator $\hat{R}$ is defined as $\hat{R}=\sum_{\hexagon} \ket{\hexA} \bra{\hexB} + \text{H.c}$. In the RK wavefunction, $\mel{\Psi_{\text{RK}}}{\hat{R}}{\Psi_{\text{RK}}}=4\Bar{n}_{\text{flip}}N_{\text{u.c.}}$. To calculate expectation values in $\ket{\Psi_\text{RK}}$, we have used classical Monte Carlo sampling.
  }
    \label{tab:numerics}
    \end{ruledtabular}
\end{table*}
With these values at hand, we calculate the expectation value of $\hat{H}_{\text{eff}}$ using Eq.~\eqref{eq:Hpertlr} in $\ket{\Psi_{\text{RK}}}$, $\ket{\Psi_{\text{IFM}}}$, and $\ket{\Psi_{\text{IAFM}}}$, and the result is plotted in Fig.~\ref{fig:pertenergy}. As we turn on $\Omega$, the transition point $\Omega_C$ can be determined within our approximation as the $\Omega$ for which the energy of the ice ferromagnet becomes higher than that of the RK wavefunction, as calculated using Eq.~\eqref{eq:Hpertlr}.  We find
\begin{equation}
    \Omega_C=0.43927(1) V.
\end{equation}

There is an important question on whether our use of perturbation theory is justified. First, we argue that treating $\hat{H}_{\text{LR}}$ perturbatively is justified. $\left\{\hat{H}_{\text{LR}}\right\}$, $\left\{\hat{W}^{(2)}_{\text{LR}},\hat{L}^{(2)}_{\text{LR}},\hat{M}^{(2)}_{\text{LR}}\right\}$, $\left\{\hat{W}^{(3)}_{\text{LR}}\right\}$, and $\left\{\hat{W}^{(4)}_{\text{LR}}\right\}$ are sets of operators that are first, second, third, and fourth order respectively in $\hat{H}_{\text{LR}}$. As we can see from Table~\ref{tab:numerics}, the expectation values of these operators in $\ket{\Psi}_{\text{RK}}$ drops by an order of magnitude each time one goes one order higher in $\hat{H}_{\text{LR}}$. Next, is perturbation theory in $\hat{H}_{\Omega}$ justified, given that our calculated $\Omega_C$ is outside the $\Omega\ll V$ regime? We observe that the leading contribution to $J_{\text{ring}}$ that we dropped, $\frac{33833}{2592}\frac{(\Omega_C)^8}{V^7} = 0.018 V$ ~\cite{rochner2016spin}, is smaller than the one we kept, $\frac{63}{16}\frac{(\Omega_C)^6}{V^5}=0.028 V$. If we had kept higher order contributions to $J_{\text{ring}}$, it would only decrease the energy of the QSL relative to the ice ferromagnet and ice antiferromagnet. Further, the energy of the QSL that we present is a conservative estimate since we used the RK wavefunction which has higher energy than the true ground state of Hamiltonian~\eqref{eq:ringex}. This gives us hope that our result obtained using perturbation theory is qualitatively correct. 
\newtext{In Appendix~\ref{app:perturbation}, we further address the issue of convergence of the perturbation theory by calculating the Borel-Pad\'e approximants of the perturbative energies of the three ansatz states. We find that using the Borel-Pad\'e approximants for the ice FM and the ice antiferromagnet does not change the phase diagram qualitatively, while the Borel-Pad\'e approximant for the RK wavefunction does not capture the energy reduction coming from  quantum fluctuations.  However, rigorously ascertaining the convergence of our perturbative expansion is beyond the scope of this work.}

Within our approximation, for $\Omega<\Omega_C$, the ground state is an ice ferromagnet, an ordered state satisfying the ice rule. For $\Omega> \Omega_C$ but also close to $\Omega_C$, the ground state is in the QSL phase, i.e.,~the deconfined phase of a $U(1)$ gauge theory.  From the point of view of the QSL, the ordered ice ferromagnet state is obtained when monopole excitations of the spin liquid proliferate, and the monopole-antimonopole string operator, to be defined in Sec.~\ref{sec:monopole}, Eq.~\eqref{eq:monopolestring}, acquires an expectation value. As a consequence of this, the fractional ``electric charges", or spinons, get confined~\cite{polyakovCompactGaugeFields1975,polyakov1977}. The monopole creation operator (see Sec.~\ref{sec:monopole} and Ref.~\cite{hermele2004}) is diagonal in the $\hat{S}^z$ basis, and acts in the sector that obeys the ice rule. It is thus plausible that the confined phase is indeed the ice ferromagnet. While our calculation provides microscopic intuition for this transition, we emphasize that, to prove the existence of, locate and characterize this transition accurately, one needs to do a more careful quantum Monte Carlo calculation.

\subsection{Large $\Omega$---Higgs transition}\label{sec:Higgs}
\begin{figure}[t]
  \centering
  \includegraphics[width=0.9\columnwidth]{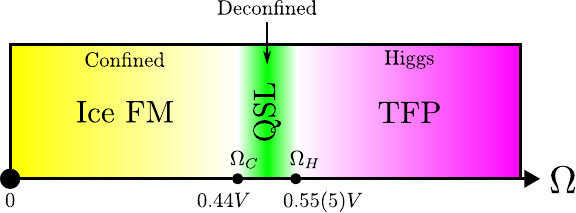}
  \capt{Approximate ground state phase diagram of the Hamiltonian in Eq.~\eqref{eq:hamsplit}. The ground state for $\Omega=0$ was calculated exactly to be an ice ferromagnet (ice FM)  in Sec.~\ref{sec:classical}.
  We assume that, as $\Omega$ is increased, no phase transition occurs to a different ordered state. The transition point from the ice ferromagnet (confined phase) to the QSL (deconfined phase) at $\Omega_C\approx 0.44 V$ is obtained by comparing energies of ansatz wavefunctions in the effective Hamiltonian obtained using perturbation theory in $\hat{H}_\Omega$ and $\hat{H}_{LR}$. For the Higgs transition to the TFP phase, 
  we make an approximation by dropping $\hat{H}_{LR}$, in which case $\Omega_H$ was calculated in Ref.~\cite{emonts2018monte} to be 0.55(5).}
  \label{fig:phase_diagram}
\end{figure}

From the Hamiltonian in Eq.~\eqref{eq:hamsplit}, it is clear that, in the limit $\Omega \gg V$, the ground state is a transverse-field-polarized (TFP) state, i.e.,~a product state of $(\ket{g} - \ket{r})_i$ at each site $i$. Thus, as $\Omega$ is increased away from $\Omega_C$, the system should eventually go through a phase transition from the putative QSL phase into the TFP phase. From the point of view of the QSL, this is a Higgs transition because the operator $\hat{S}^x$ that acquires expectation value in the TFP phase creates a pair of ``electric"-charge excitations in the spin liquid. 
The perturbation theory in $\Omega/V$ that we performed in Sec.~\ref{sec:mcpert} relies on the ability to go to a basis where the Hilbert space decouples into ice rule obeying and ice rule disobeying sectors separated by an energy gap of $V$. But the ground state in the $\Omega \gg V$ limit (TFP) straddles both of these sectors. So we do not expect perturbation theory in $\Omega/V$ to capture the phase transition into the TFP phase that contains the $\Omega \to \infty$ ground state. Hence, we will present an indirect reasoning below. 
In the $\Omega \ll V$ limit, $\hat{H}_{\text{LR}}$ was important, since it was the dominant term splitting the degeneracy in the ice manifold. 
On the other hand, in the vicinity of the putative Higgs transition, $\hat{H}_{\text{LR}}$ may not be as important since the largest term in $\hat{H}_{\text{LR}}$ has magnitude $V/27$, and as justified above using Table~\ref{tab:numerics}, the effect of $\hat{H}_{\text{LR}}$ is indeed perturbative. Therefore, we drop $\hat{H}_{\text{LR}}$ as a zeroth-order approximation for calculating the Higgs transition point. The resulting Hamiltonian is the transverse field Ising model on the pyrochlore lattice.  
Refs.~\cite{emonts2018monte} and~\cite{rochner2016spin} studied this model and found the transition point $\Omega_H$  to be at $\Omega_H=0.55(5)V$  and $0.6V$ respectively. This leads us to expect that, in the window $0.44 < \Omega < 0.55$, the ground state may be a QSL, leading us to sketch the phase diagram shown in Fig.~\ref{fig:phase_diagram}. 
Within our approximation, $\Omega_C<\Omega_H$ and there is a window where the QSL is the ground state. However, the introduction of $\hat{H}_{\text{LR}}$ may result in a lowering of the energy of the TFP state relative to the QSL. Calculating this effect and verifying that this does not bring down $\Omega_H$ far enough to destroy the QSL phase requires a more careful calculation which is beyond the scope of this work.
\newtext{We note that, to be certain about the existence of all the phases we found and about not missing any additional phases, a more detailed quantum Monte Carlo study is required, and we leave it for future work.}

In the remainder of this section, we provide some intuition for the Higgs transition by performing a gauge mean field theory (gMFT) calculation introduced in Ref.~\cite{savary2012coulombic}. 

\subsubsection{Gauge mean field theory---Higgs transition}\label{sec:gmft}
The main idea of this approach is to first recast the microscopic Hamiltonian as an exact gauge theory by introducing ancillary degrees of freedom followed by a mean-field decoupling of the interactions. This theory involves bosonic charges hopping in the presence of a fluctuating gauge field whose mean-field value is chosen self-consistently. If this mean-field gauge-field configuration is such that the hopping amplitudes of the bosonic charges is 0, then the theory is in a confined phase. If not, the theory is in the deconfined phase as long as the bosons do not condense. If the bosonic charges condense, then the theory is in a Higgs phase, which is adiabatically connected to the TFP state.

Concretely, the construction is as follows. For $\vb{r}\in A$, where $A$ is a sublattice of the diamond lattice,
\begin{equation}
\hat{S}^+_{\vb{r}\rightarrow\vb{r}+\vb{e}_\mu}=\hat{\Phi}^\dagger_{\vb{r}}\hat{\mathsf{s}}^+_{\vb{r}\rightarrow\vb{r}+\vb{e}_\mu}\hat{\Phi}_{\vb{r}+\vb{e}_\mu},
\end{equation}
where $\hat{S}^+_{\vb{r}\rightarrow\vb{r}+\vb{e}_\mu}\equiv \hat{S}^+_{\vb{r}+\vb{e}_\mu/2}=\hat{S}^+_{\vb{r},\mu}$ (and similarly $\hat{\mathsf{s}}^+_{\vb{r}\rightarrow\vb{r}+\vb{e}_\mu}\equiv \hat{\mathsf{s}}^+_{\vb{r}+\vb{e}_\mu/2}=\hat{\mathsf{s}}^+_{\vb{r},\mu}$) lives on a bond of the diamond lattice connecting sites $\vb{r}$ and $\vb{r}+\vb{e}_\mu$ (recall that centers of the bonds of the diamond lattice are sites of the pyrochlore lattice). $\hat{\mathsf{s}}^z$ is also a spin-1/2 operator and has eigenvalues $\pm 1/2$. Here, $\hat{\Phi}^\dagger_{\vb{r}}$ serves as a raising operator for $\hat{Q}_{\tet_{\mathbf{r}}}\equiv \eta_{\vb{r}} (\hat{n}_{\tet_{\mathbf{r}}} -2)$, where $\eta_{\vb{r}}=1$ for $\vb{r}\in A$ and $\eta_{\vb{r}}=-1$ for $\vb{r} \in B$. For convenience, we  drop the symbol $\tet$ from now on. $\hat{Q}_{\mathbf{r}}$ and $\hat{\Phi}^\dagger_{\vb{r}}$ satisfy the commutation relation: $\comm{\hat{Q}_{\mathbf{r}}}{\hat{\Phi}^\dagger_{\vb{r}}}=\hat{\Phi}^\dagger_{\vb{r}}$.
Note that $\hat{\Phi}_{\vb{r}}$ is not a canonical boson but a rotor satisfying
\begin{equation}\label{eq:constraint1}
    \hat{\Phi}_{\vb{r}}^\dagger \hat{\Phi}_{\vb{r}}=1.
\end{equation}
To recover the original spin Hilbert space, one imposes the constraint that the total gauge charge at $\vb{r}$ is 
\begin{equation}
\label{eq:constraint2}
    \hat{Q}_{\vb{r}}=\eta_{\vb{r}} \sum_{\mu}\hat{\mathsf{s}}^z_{\vb{r}+\eta \vb{e}_\mu/2}.
\end{equation}
 Rewriting the Hamiltonian \eqref{eq:hamsplit} in terms of the fictitious variables, $\hat{Q}_{\vb{r}}, \hat{\Phi}_{\vb{r}}$ and $\hat{\mathsf{s}}_{\vb{r},\mu}$ we get
\begin{equation}
\label{eq:hamfict}
\begin{aligned}
    \hat{H}=&  \frac{V}{2} \sum_{\vb{r}\in A,B}\hat{Q}^2_{\vb{r}} - \frac{\Omega}{2}\sum_{(\vb{r}\in A),\mu}\left(\hat{\Phi}_{\vb{r}}^\dagger \hat{\mathsf{s}}^+_{\vb{r}\rightarrow \vb{r}+\vb{e}_\mu}\hat{\Phi}_{\vb{r}+\vb{e}_\mu} + \text{ H.c.}\right)\\
    &+\frac{1}{2}\sum_{\vb{r},\vb{r}'\in A} \sum_{\mu,\nu } V_{\mu  \nu} (\vb{r}-\vb{r}')\hat{\mathsf{s}}^z_{\vb{r}\mu}\hat{\mathsf{s}}^z_{\vb{r}',\nu},
    \end{aligned}
\end{equation}
where $V_{\mu\nu}(\vb{r}-\vb{r}')=V\left(\frac{a}{\vb{r}-\vb{r}'+\vb{e}_\mu/2 - \vb{e}_\nu /2}\right)^6$ whenever $(\vb{r},\mu)$ and $(\vb{r}',\nu)$ are distinct and are not nearest-neighbors. $V_{\mu\nu}(\vb{r}-\vb{r}')$ is 0 otherwise.

Following Ref.~\cite{savary2012coulombic}, we perform the  zeroth-order mean-field decoupling: $\hat{\Phi}^\dagger \hat{\Phi} \hat{\mathsf{s}} \rightarrow \hat{\Phi}^\dagger \hat{\Phi} \expval{\hat{\mathsf{s}}}+\expval{\hat{\Phi}^\dagger \hat{\Phi}} \hat{\mathsf{s}} -\expval{\hat{\Phi}^\dagger \hat{\Phi}} \expval{\hat{\mathsf{s}}} $ and $\hat{\mathsf{s}} \hat{\mathsf{s}}\rightarrow \hat{\mathsf{s}}\expval{\hat{\mathsf{s}}}+\expval{\hat{\mathsf{s}}}\hat{\mathsf{s}}-\expval{\hat{\mathsf{s}}}\expval{\hat{\mathsf{s}}}$ (where $\hat{\mathsf{s}}$ could either be $\hat{\mathsf{s}}^+$, $\hat{\mathsf{s}}^-$, or $\hat{\mathsf{s}}^z$). Upon doing so, the Hamiltonian decouples into  a Hamiltonian of bosons hopping on the diamond lattice and a Hamiltonian of spins in an external field, which itself is set self-consistently by the Green's function of the bosons. Before solving the resulting theory, one needs to enforce the constraints \eqref{eq:constraint1} and \eqref{eq:constraint2} using Lagrange multipliers $\lambda_{\vb{r}}$ and $v_{\vb{r}}$, respectively. Within the mean-field theory, it is assumed that these Lagrange multipliers take a spatially homogeneous value at the saddle point. 
We then find  the minimum value of $\Omega^{\text{MF}}_H$ such that, for any $\Omega>\Omega^{\text{MF}}_H$, it is possible to self-consistently choose $\lambda$ only by macroscopically occupying a boson mode. 
This $\Omega^{\text{MF}}_H$ marks the location of the Bose-Einstein-condensation transition (or Higgs transition within the mean-field theory). We find $\Omega^{\text{MF}}_H\approx 0.7 V$. In Appendix~\ref{app:gmft}, we present more details of this calculation. An artifact of this technique is that, although we include long-range interactions in our calculation, they do not play any role at the saddle point near the Higgs transition. Therefore, the final steps and result of our calculation are identical to the ones carried out in~\cite{savary2017disorder}. 

In Appendix~\ref{app:gmft}, we also point out a major limitation of this technique in the small-$\Omega$ limit that may not have been appreciated in previous literature on gauge mean field theory.

\subsection{Comments on dynamical state preparation}\label{sec:dynamics}
So far, we have focused on the nature of the ground state of Hamiltonian~\eqref{eq:hamsplit} as a function of $\Omega/V$. However, what is often experimentally relevant is the nature of the state prepared by a ramping of parameters in a finite amount of time. In the context of the experiment in Ref.~\cite{semeghini2021}, it was shown in Ref.~\cite{giudici2022dynamical,sahay2022quantum} that a state in the same phase as the $\mathbb{Z}_2$ gauge theory can be prepared by a non-equilibrium time evolution starting from a trivial state.
\newtext{In the context of the experiment in Ref.~\cite{semeghini2021}, it was shown numerically in Ref.~\cite{giudici2022dynamical} that a state with a large overlap with the RVB state can be prepared by a non-equilibrium time evolution.
The question of dynamical state preparation was also studied in Ref.~\cite{sahay2022quantum}.}
Here, we will present an adaptation of the conclusions of Ref.~\cite{sahay2022quantum} to our setting. 

\begin{figure}[t]
  \centering
  \includegraphics[width = 0.9\columnwidth]{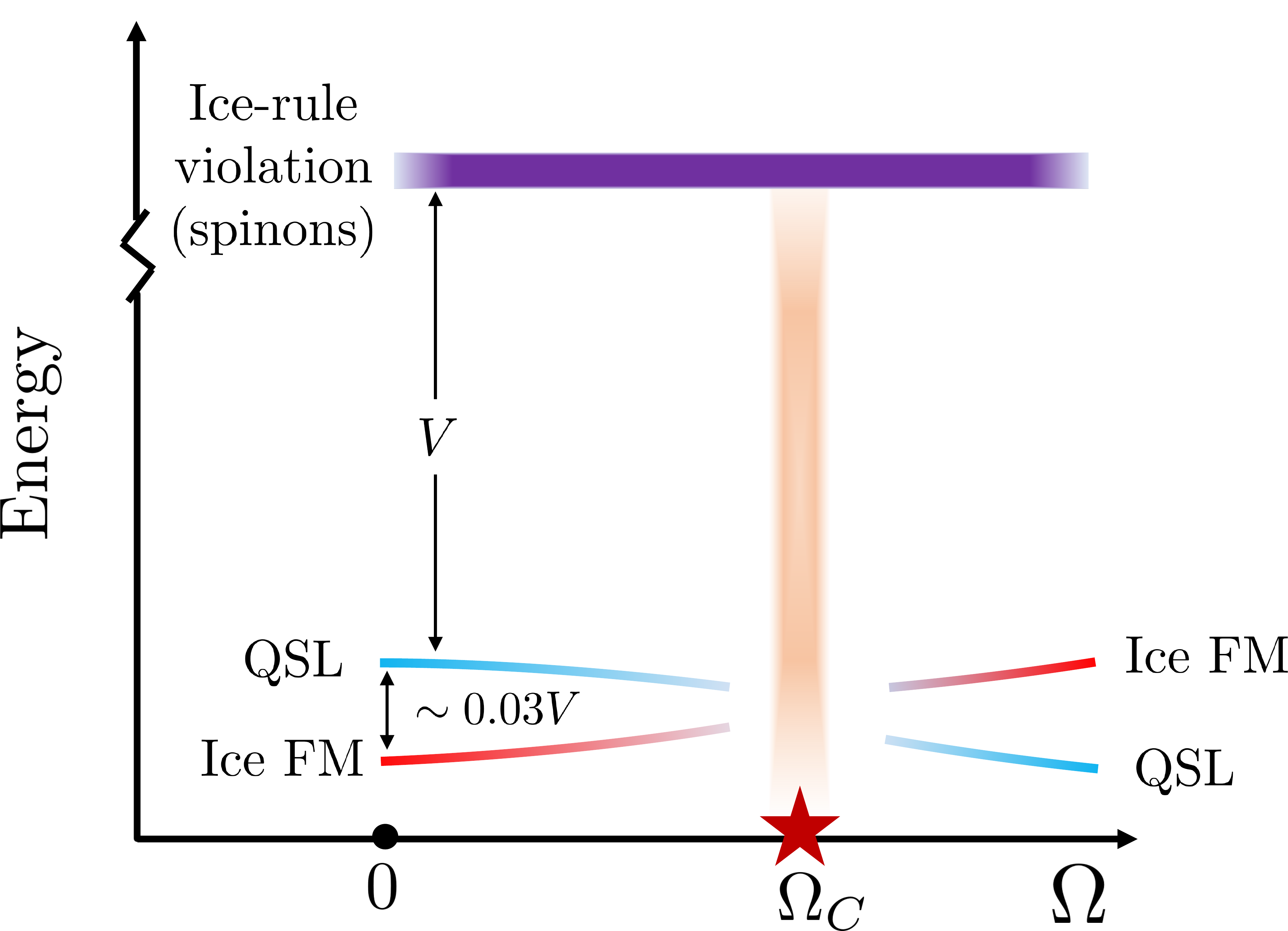}
  \capt{A qualitative sketch of the energy scales (per unit cell) in our problem. For $\Omega>\Omega_C$, the ground state is a $U(1)$ QSL. Ice ferromagnet is the ordered state obtained when monopoles proliferate, i.e.,~the ice ferromagnet has an extensive number of monopoles. We therefore use the energy difference per unit cell between the QSL and the ice ferromagnet at $\Omega=0$, obtained in Table~\ref{tab:numerics}, as a proxy for the monopole energy scale. This scale $\sim 0.03 V$ is much smaller than the spinon energy scale (``electric charge"), which is $\sim$ V.}
  \label{fig:energyscales}
\end{figure}

The excitations of a $U(1)$ QSL are gapless ``photons", magnetic monopoles, and ``electric charges" (spinons). The transition of a QSL to an ice ferromagnet is driven by the condensation of monopoles, while the transition to the TFP phase is driven by the condensation of spinons. The gapless ``photons" are not directly involved in these transitions. Also, a state with ``photon" modes excited on top of a QSL state is still in the deconfined phase of the $U(1)$ gauge theory. This allows us to ignore ``photons" in this section. Since the confined phase, ice ferromagnet has an extensive number of monopoles, we use the difference per unit cell between the energies of the QSL and ice ferromagnet states as a proxy for the monopole energy scale. At $\Omega=0$, this difference is $\left(\expval{\hat{H}_{\text{LR}}}_{\text{QSL}}-\expval{\hat{H}_{\text{LR}}}_{\text{IFM}}\right)/N_{\text{u.c.}}\approx 0.03 V$ (see Table~\ref{tab:numerics}), which is much smaller than the spinon energy scale (see Fig.~\ref{fig:energyscales} for a sketch).  
Suppose one starts with an initial state (for a small $\epsilon \sim \Omega/V$)
\begin{equation}
    \ket{\Psi_{(t=0)}}=\otimes_{i}\left(\ket{g}_i + \epsilon \ket{r}_i\right),
\end{equation}
which is the ground state in the limit of large negative $\delta/V$ and small $\Omega/V$. As shown in Sec.~\ref{sec:proposal}, the classical ground state lies in the ice manifold when $\delta \in \left(2.46 V, 4.46 V\right)$. Now suppose that $\delta$ is ramped up from its initial large negative value to a value in this range such that the ramp is adiabatic with respect to the spinon gap $V$, but is sudden with respect to the monopole scale $\sim 0.03 V$,  while keeping $\Omega/V \ll 1$. Using arguments in Ref.~\cite{sahay2022quantum}, this protocol will not prepare the ground state, which, from Fig.~\ref{fig:phase_diagram}, is an ice ferromagnet. Instead, it will (approximately) project out violations of the ice rule (due to adiabaticity with respect to the spinon scale) from the initial state $\ket{\Psi_{(t=0)}}$. The resulting final state is
\begin{equation}
   \ket{\Psi_{\text{final}}}\approx \hat{\mathcal{P}}\left\{\otimes_{i}\left(\ket{g}_i + \epsilon \ket{r}_i\right)\right\}=\ket{\Psi_{\textrm{RK}}},
\end{equation}
where $\hat{\mathcal{P}}$ is the projector onto the ice manifold. The projected wavefunction is an equal-weight superposition of all coverings, which is simply the RK wavefunction and which lies in the QSL phase~\cite{hermele2004}. There is one catch to the above argument---the spinon gap closes during the above ramp. So it is impossible to be sudden with respect to the monopole scale and yet be strictly adiabatic with respect to the spinon gap throughout the ramp. For a short duration (while the ramp is going through the spinon gap closing), adiabaticity with respect to the spinon gap will be violated. By the Kibble-Zurek mechanism, the resulting state is composed of finite-size puddles of QSL-like regions with a nonzero density of spinons interspersed~\cite{sahay2022quantum, zurek1985cosmological, kibble1976topology, del2014universality}. Thus, in summary, there are two different ways in which one can prepare a $U(1)$ QSL-like state in experiment and study a confinement-deconfinement transition\footnote{We note that the confinement-deconfinement transition of $U(1)$ gauge theory in 3+1D is strictly speaking, a ground state transition \cite{svetitsky1986symmetry,polyakovThermalPropertiesGauge1978}. Therefore, in this paper, when we use the phrase confinement-deconfinement transition, we mean signatures of this transition in a finite-size state prepared in finite time.}.
\begin{enumerate}
    \item \underline{$\Omega/V\ll 1$}: 
    Perform a ramp of $\delta$ starting from a large negative value and ending in the range $(2.46 V, 4.46 V)$ for a fixed $\Omega/V \ll 1$ such that the ramp is adiabatic with respect to $V$ (spinon gap) but sudden with respect to the monopole scale ($\sim 0.03 V$). Even though the ground state is not a QSL for these parameters, this procedure would create puddles of QSL-like regions by the argument in Ref.~\cite{sahay2022quantum}. To see a deconfinement-confinement transition, the ramp of $\delta$ should be slowed down and, once it is adiabatic with respect to the monopole gap, an ordered, i.e. confined state will be prepared.
    \item 
    \underline{Adiabatic:} Perform a ramp of $\delta$ starting from a large negative value and ending in the range $(2.46 V, 4.46 V)$ and a ramp in $\Omega$ starting from $\Omega/V \ll 1$ and ending in a final value $\Omega_f$, such that both ramps are adiabatic with respect to the monopole scale always. The two ramps can be performed simultaneously, or such that the ramp in $\delta$ precedes the ramp in $\Omega$. This would approximately create the ground state of Hamiltonian~\eqref{eq:hamsplit}. As the final value $\Omega_f$ goes through $\Omega_C$ ($\Omega_H$), the nature of the final state prepared this way goes through a confinement-deconfinement (Higgs) transition.
\end{enumerate} 
\newtext{We note that the first method above can prepare a state with a large overlap with the RK wavefunction even if the true ground state of the system is not in the QSL phase.}
Once a state is prepared by either of the above schemes, one needs to devise measurements that can tell whether the state is in the confined phase or in the deconfined phase. We address this in the following section.

\section{Diagnosis of the quantum spin liquid}\label{sec:diagnostics}
Access to wavefunction snapshots in the $\hat{S}^z$ basis, combined with access to unitary evolution, allows one to use the Rydberg-atom platform to measure non-local observables, a feature generally unavailable in traditional condensed matter systems. In this section, we describe some measurable correlators which can be used to observe the signatures of a quantum spin liquid state. In this section, we  assume that the detuning is chosen such that $\rho = 2$.

\subsection{Plaquette-plaquette correlators}

 \newtext{The plaquette operators are  off-diagonal in the $\hat{S}^z$ basis.} Thus they can distinguishing a coherent quantum superposition from a classical admixture of states. We define two plaquette operators $\hat{X}_P$ and $\hat{Y}_P$ for a hexagonal plaquette $P$ of the pyrochlore lattice as 
 \begin{equation} \label{eq:xp_yp_def}
  \begin{aligned}
    \hat{X}_{P} &=  \prod_{i =1}^{6} \left( 2\hat{S}_i^x \right) ,\\
    \hat{Y}_{P} & = \prod_{i = 1}^{3} \left( 2\hat{S}_{2i-1}^x \right) \left( 2\hat{S}_{2i}^y \right),
\end{aligned}
\end{equation}
where $1,2,\ldots,6$ denote the sites around a plaquette $P$. \newtext{We are interested in the following two connected correlators of the plaquette operators:
\begin{equation}
\begin{aligned}
\label{eq:conn_corr_x_y}
    \langle \hat{X}_{P} \hat{X}_{P'} \rangle_c = \langle \hat{X}_P \hat{X}_{P'} \rangle - \langle \hat{X}_P \rangle \langle \hat{X}_{P'} \rangle,\\
    \langle \hat{Y}_P \hat{Y}_{P'} \rangle_c = \langle \hat{Y}_P \hat{Y}_{P'} \rangle - \langle \hat{Y}_P  \rangle \langle  \hat{Y}_{P'} \rangle,
\end{aligned}
\end{equation}
where $P$ and $P'$ denote two plaquettes of the pyrochlore lattice (see Fig.~\ref{fig:b_b_correlator}).
}

Either of the two correlators, $\langle \hat{X}_P \hat{X}_{P'} \rangle_c $ and $\langle \hat{Y}_P \hat{Y}_{P'} \rangle_c $, can distinguish a QSL phase from other phases including a classical spin ice (see Table~\ref{tab:correlators}).

\begin{table*}[t]
\begin{ruledtabular}
  \begin{tabular}{ccccc}
 Correlator & Confined (Ice FM) & Deconfined (QSL) & Higgs (TFP) & Classical Spin Ice\\
 \hline 
 $\langle \hat{X}_P \hat{X}_{P'} \rangle_c $ & Exp. or faster decay & $ 1/\mathsf{R}^8$ & \newtext{$1/\mathsf{R}^{12}$} & Exp. or faster decay\\ 
 $\langle \hat{Y}_P \hat{Y}_{P'} \rangle_c $ & Exp. or faster decay & $ 1/\mathsf{R}^4$ & \newtext{$1/\mathsf{R}^{6}$} & Exp. or faster decay\\ 
 $\expval{\hat{\mathcal{M}}^\dagger \hat{\mathcal{M}} (\dual{r}_1 \rightarrow \dual{r}_2) }$ & Nonzero const. & Exp.~decay & Exp.~decay & Exp.~or faster decay\\ 
 $\chi_\mathcal{C}^E$ & Nonzero const.\footnote{\label{fn:FM}Distinguishing this nonzero constant from zero for $\chi_{\mathcal{C}}^E$ in the confined phase (ice FM) and for $\chi_{\mathcal{C}}^M$ in the Higgs phase (TFP) may be practically challenging.}  & Exp. or faster decay  & Nonzero const. & Exp. or faster decay \\ 
 $\chi_\mathcal{C}^M$ & Nonzero const. & Exp. or faster decay  & Nonzero const.\textsuperscript{\ref{fn:FM}}   
 & Exp.~or faster decay\\ 
 $  \langle \hat{S}^z_{\mathbf{r}i} \hat{S}^z_{\mathbf{r'}j} \rangle$  & Nonzero const. & $1/R^4$ & \newtext{$1/R^6$} & $1/R^4$ \\
\end{tabular}
\capt{Behavior of various correlators. $\hat{X}_P$ and $ \hat{Y}_P$ are plaquette operators defined in Eq.~\eqref{eq:xp_yp_def}. $\hat{\mathcal{M}}^\dagger \hat{\mathcal{M}} (\dual{r}_1 \rightarrow \dual{r}_2)$ is a monopole string operator defined in Eq.~\eqref{eq:monopolestring}. $\chi_\mathcal{C}^E$ and $\chi_\mathcal{C}^M$ are Fredenhagen-Marcu order parameters defined in Eq.~\eqref{eq:FM_e} and Eq.~\eqref{eq:FM_m}, respectively. In this table, we have omitted the form factors multiplying $1/R^4$ and $1/R^8$ that are provided in Eqs.~\eqref{eq:correlator_theoretical} and \eqref{eq:maxwellEE}. \label{tab:correlators}}
\end{ruledtabular}
\end{table*}

We compare the two correlators and provide protocols to measure them.
We assume throughout that the two plaquettes $P$ and $P'$ do not have any sites in common. 
\newtext{We now explain the behavior of these plaquette correlators in the ice FM, QSL, and TFP phases.}

\newtext{\subsubsection{Plaquette correlators in the ice FM phase}}

\newtext{We will determine the behavior of the correlators deep inside the ice FM phase, that is, for $\Omega \ll V$. In this limit, the ice FM phase is a product state in the $\hat{S}^z$ basis with perturbative corrections on top of it produced by $\hat{H}_\Omega$. Our ansatz for the ice FM state is given by $\ket{\Psi_{\text{ord}}}=\hat{U}_S^{\dagger}\ket{\Psi_{\text{IFM }}}$, where $\hat{U}_S^\dagger$ is the unitary that performs the Schrieffer-Wolff transformation [see Eq.~\eqref{eq:scwtrans}], and $\ket{\Psi_\text{IFM}}$ is a product state in the $\hat{S}^z$ basis described in Sec.~\ref{sec:classical} and shown in Fig.~\ref{fig:ice_fm}.}

\newtext{At zeroth order in $\Omega/V$, $\hat{U}_S = \hat{1}$ implying $\langle \hat{X}_P \hat{X}_{P'} \rangle_c = 0$ because the diagonal components of $\hat{X}_{P}$ in the $\hat{S}^z$ basis are zero. Similarly, $\langle \hat{Y}_P \hat{Y}_{P'} \rangle_c =0 $ at zeroth order. A nonzero contribution to the connected correlators is obtained only by terms in perturbation theory that are of an order proportional to $\mathsf{R}/a$. Thus, the plaquette correlators decay exponentially with distance in the ice FM phase. }

\subsubsection{Plaquette correlators in the QSL phase}

\newtext{Here we provide alternative plaquette correlators which agree with the plaquette correlators defined in Eq.~\eqref{eq:conn_corr_x_y} up to sixth order in $\Omega/V$. We then interpret them in terms of the gauge theory to understand their behavior in the QSL phase. }

\newtext{Let $\ket{\Psi_g}$ be the ground state of the system and let $\hat{U}_{S} = e^{\hat{S}}$ be the operator that implements the Schrieffer–Wolff transformation so that $\hat{U}_{S} \hat{H} \hat{U}_{S}^\dagger$ is the effective Hamiltonian in the ice manifold.
We use the same notation as Sec.~\ref{sec:expression_for_Heff} here.
Thus $\ket{\Psi_0} = \hat{U}_S \ket{\Psi_g}$ is the ground state of the effective Hamiltonian and lies in the ice manifold. 
Consider two new plaquette X and Y operators defined as
 \begin{equation} \label{eq:xp_yp_tilde_def}
  \begin{aligned}
    \hat{\tilde{X}}_{P} &=  \left( \hat{S}_1^+ \hat{S}_2^-\hat{S}_3^+\hat{S}_4^-\hat{S}_5^+\hat{S}_6^-   + \text{H.c.} \right) ,\\
      \hat{\tilde{Y}}_{P} & = -i \hat{S}_1^+ \hat{S}_2^-\hat{S}_3^+\hat{S}_4^-\hat{S}_5^+\hat{S}_6^-   + \text{H.c.},
\end{aligned}
\end{equation}
First, note that 
\begin{equation} \label{eq:xpxp_to_tildexp_tildexp}
\langle \Psi_0 | \hat{X}_P \hat{X}_{P'} | \Psi_0 \rangle = \langle \Psi_0  | \hat{\tilde{X}}_P \hat{\tilde{X}}_{P'} | \Psi_0 \rangle. 
\end{equation}}
This can be seen by writing $2\hat{S}^x_i  = \hat{S}^+_i + \hat{S}^-_i$ and noticing that the only terms that preserve the ice rule are ring exchanges over $P$ and $P'$.
When the remaining terms act on a state in the ice manifold, they either take it outside of the ice manifold or annihilate it. Thus the  expectation value of these remaining operators in $\ket{\Psi_0}$ is zero. 
For example, $\hat{S}_1^+ \hat{S}_2^+ \hat{S}_3^+ \hat{S}_4^- \hat{S}_5^- \hat{S}_6^- \hat{S}_{7}^+ \hat{S}_{8}^- \hat{S}_{9}^+ \hat{S}_{10}^- \hat{S}_{11}^+ \hat{S}_{12}^{-}$ acting on a state in the ice manifold would either annihilate this state or give a state that violates the ice rule on four of the tetrahedra surrounding $P$.
\newtext{An identity similar to Eq.~\eqref{eq:xpxp_to_tildexp_tildexp} also holds for the expectation value for a single plaquette X operator:
\begin{equation} \label{eq:xp_to_tildexp}
    \langle \Psi_0 | \hat{X}_P | \Psi_0 \rangle  =  \langle \Psi_0 | \hat{\tilde{X}}_P | \Psi_0 \rangle.
\end{equation}
Equations analogous to Eqs.~(\ref{eq:xpxp_to_tildexp_tildexp},\ref{eq:xp_to_tildexp})
also hold true for the plaquette Y operator. 
Now, $\ket{\Psi_g} = \ket{\Psi_0} + \mathcal{O} \left( \Omega/V \right)$, where the corrections of order $\Omega/V$ come from the perturbation $\hat{H}_\Omega$. Thus, one would expect $ \langle \hat{X}_P \hat{X}_{P'} \rangle_{c, \ket{\Psi_g}} = \langle \hat{\tilde{X}}_P \hat{\tilde{X}}_{P'} \rangle_{c, \ket{\Psi_g}} $ up to first order in $\Omega/V$ (The expectation values are calculated in $\ket{\Psi_g}$ here). However, in Appendix~\ref{app:corrections_plaq}, we show that this is true up to sixth order: 
\begin{equation}\label{eq:sixth_order_x}
    \langle \hat{X}_P \hat{X}_{P'} \rangle_c = \langle \hat{\tilde{X}}_P \hat{\tilde{X}}_{P'} \rangle_c + \Theta\left( \left( \Omega/V \right)^6 \right),
\end{equation}
\begin{equation}\label{eq:sixth_order_y}
    \langle \hat{Y}_P \hat{Y}_{P'} \rangle_c = \langle \hat{\tilde{Y}}_P \hat{\tilde{Y}}_{P'} \rangle_c + \Theta\left( \left( \Omega / V \right)^6 \right), 
\end{equation}
where the expectation values are again calculated in $\ket{\Psi_g}$. We will ignore these sixth-order corrections and now move on to understanding the behavior of $\langle \hat{\tilde{X}}_P \hat{\tilde{X}}_{P'} \rangle_c$ and $ \langle \hat{\tilde{Y}}_P \hat{\tilde{Y}}_{P'} \rangle_c$ in the QSL phase by mapping the operators $\hat{\tilde{X}}_{P}$ and $\hat{\tilde{X}}_{P'}$ to gauge fields. }

\begin{figure}[t]
  \centering
  \includegraphics[width = 0.8\columnwidth]{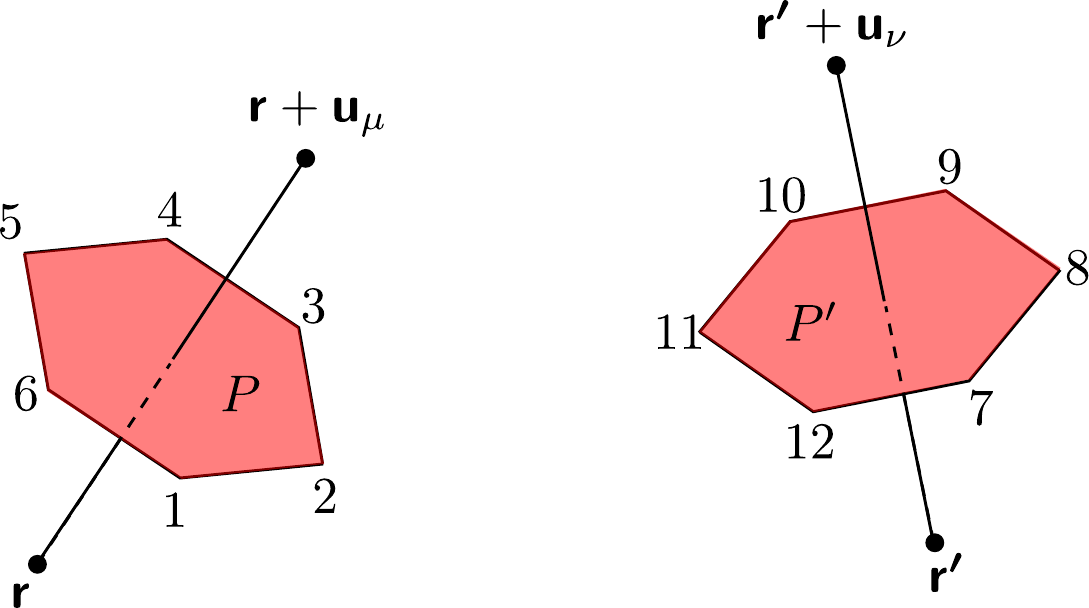}
  \capt{Notation for the plaquette correlators. $P$ and $P'$ are two hexagonal plaquettes of the pyrochlore lattice. $\dual{r}, \dual{r'}$, $\dual{r}+\dual{u}_\mu$, and $\dual{r'}+\dual{u}_\nu$ are the sites of the dual diamond lattice. $\dual{u}_\mu$ and $\dual{u}_\nu$ are vectors perpendicular to $P$ and $P'$.}
  \label{fig:b_b_correlator}
\end{figure}

 Using the mapping between the spins and the effective $U(1)$ gauge theory from Eq.~\eqref{eq:rotor_mapping}, we see that the operators $\hat{\tilde{X}}_P$ and $\hat{\tilde{Y}}_{P}$
 are equal to (twice) the cosine and the  sine of the magnetic field operator $\hat{B}_{\dual{r}, \mu}$, respectively:
\begin{equation}
  \begin{aligned}
  \hat{\tilde{X}}_P&= 2 \cos( \hat{\theta}_1 - \hat{\theta}_2 + \hat{\theta}_3 - \hat{\theta}_4 + \hat{\theta}_5 - \hat{\theta}_6 ) = 2 \cos(\hat{B}_{\dual{r},\mu}), \\
  \hat{\tilde{Y}}_P&= 2 \sin( \hat{\theta}_1 - \hat{\theta}_2 + \hat{\theta}_3 - \hat{\theta}_4 + \hat{\theta}_5 - \hat{\theta}_6 ) = 2 \sin(\hat{B}_{\dual{r},\mu}),
\end{aligned}
\end{equation}
where $\dual{r}$ belongs to the dual diamond lattice [see Fig.~\ref{fig:pyrochlore_diamond}(c)], and $\mu \in \{0,1,2,3\}$ labels the direction of magnetic field. $\hat{B}_{\dual{r},\mu}$ is along $\dual{u}_\mu$, which are vectors joining an A site of the dual diamond lattice to its neighboring B sites. These vectors are perpendicular to the plaquettes of the pyrochlore lattice, see Fig.~\ref{fig:b_b_correlator}. 
The effective theory in the deconfined phase (QSL) is Maxwell electromagnetism. \newtext{Thus the distance dependence of the plaquette correlators can be determined from the magnetic field correlator in the 3+1D continuum Maxwell electromagnetism.}

\newtext{Note that, for the plaquette correlators, we need the correlator of the magnetic field along the normal to the plaquettes, $\dual{u}_\mu$ and $\dual{u}_\nu$ (see Fig.~\ref{fig:b_b_correlator}). 
This can be calculated by first calculating the correlators of the Cartesian components of the magnetic field $\hat{B}_{\dual{r},i}$ for $i \in \{ x,y,z \}$ and appropriately projecting them on $\dual{u}_\mu$ and $\dual{u}_\nu$. }
In the $3+1$D continuum Maxwell electromagnetism, the correlator of the Cartesian components of the magnetic field can be evaluated analytically~\cite{hermele2004} and we explain it here for completeness. 

\newtext{We first express the magnetic field in terms of the gauge field $\hat{A}_\mu(\dual{r})$:
\begin{equation}
    \hat{B}_{\dual{r}, i} (t) = \sum_{j,k \in \{x,y,z\}} \epsilon_{ijk} \left( \partial_j \hat{A}_k(\dual{r},t) - \partial_k \hat{A}_j(\dual{r},t) \right),
\end{equation}
where $i \in \{ x,y,z\}$. Then we express the magnetic field in momentum space:
\begin{equation} \label{eq:b_in_momentum_space}
    \hat{B}_{\dual{k},i} (\mathsf{k}_0) = i\sum_{j,k \in \{x,y,z\}} \epsilon_{ijk} \left( \dual{k}_j \hat{A}_k(\dual{k}, \mathsf{k}_0) - \dual{k}_k \hat{A}_j(\dual{k}, \mathsf{k}_0) \right).
\end{equation}
Now the photon propagator in the Maxwell electrodynamics is given by
\begin{equation} \label{eq:photon_propagator}
    \langle \hat{A}_i (\dual{k}, \mathsf{k}_0) \hat{A}_j (-\dual{k}, -\mathsf{k}_0) \rangle_0 = \frac{1}{\mathsf{k}^2 + \mathsf{k}_0^2} \left( \delta_{i,j} - \frac{\mathsf{k}_i \mathsf{k}_j}{\mathsf{k}^2 + \mathsf{k}_0^2} \right),
\end{equation}
where  $\langle \cdot \rangle _0$ is the expectation value with respect to the Gaussian Maxwell action. Using Eqs.~\eqref{eq:b_in_momentum_space} and~\eqref{eq:photon_propagator}, the correlator of the magnetic fields in frequency-momentum space is
\begin{equation}
    \langle \hat{B}_{\dual{k},i} (\mathsf{k_0}) \hat{B}_{-\dual{k},i} (-\mathsf{k_0}) \rangle_0 = \frac{\mathsf{k}^2 \delta_{i,j} - \dual{k}_i\dual{k}_j}{\mathsf{k}^2 + \mathsf{k}_0^2}.
\end{equation}
Finally, the correlator in real space is obtained by performing a Fourier transform of the above momentum space correlator. The equal-time real-space magnetic-field correlator is given by~\cite{hermele2004}
}
\begin{equation}\label{eq:maxwellB2B2}
  \langle \hat{B}_{\boldsymbol{0},i} \hat{B}_{\dual{R},j} \rangle_0  \propto \frac{1}{\mathsf{R}^4} \left(2 \frac{\mathsf{R}_i \mathsf{R}_j}{\mathsf{R}^2} - \delta_{ij} \right)\equiv \mathcal{C}_{ij}^{B} (\dual{R}) .
\end{equation}

\newtext{Having obtained the correlator of the Cartesian components of the magnetic field, we now project the magnetic fields along the normals $\dual{u}_\mu$ and $\dual{u}_\nu$ to obtain the correlator of the magnetic fields along the plaquette normals. 
Thus the correlator of the magnetic field operators $\hat{B}_{\dual{r},\mu}$ for $\mu \in \{ 0,1,2,3 \}$ on the pyrochlore plaquettes is
\begin{equation}
    \langle \hat{B}_{\boldsymbol{0},\mu} \hat{B}_{\dual{R},\nu} \rangle_0  \propto \frac{1}{\mathsf{R}^4} \sum_{k,l \in \{x,y,z \} } ( \dual{u}_\mu )_k ( \dual{u}_\nu )_l  \left(  2 \frac{\mathsf{R}_l \mathsf{R}_k}{\mathsf{R}^2} -  \delta_{k,l} \right),
\end{equation}
where $\hat{B}_{\boldsymbol{0},\mu}$ is the magnetic field along the normal vector $\dual{u}_\mu$. }

\newtext{Now we return to the plaquette correlators and determine their behaviors in the QSL phase:
\begin{equation}
\begin{aligned}
    \langle \hat{\tilde{X}}_P \hat{\tilde{X}}_{P'} \rangle & = \expval{ \cos(\hat{B}_{\dual{r},\mu}) \cos(\hat{B}_{\dual{r'},\nu}) }_0\\
    & = e^{-\langle \hat{B}^2\rangle_0} \cosh{ \expval{\hat{B}_{\dual{r},\mu} \hat{B}_{\dual{r'},\nu}} _0 }.
\end{aligned}
\end{equation}
Similarly,
\begin{equation}
\begin{aligned}
    \langle \hat{\tilde{Y}}_P \hat{\tilde{Y}}_{P'} \rangle & = \expval{ \sin(\hat{B}_{\dual{r},\mu}) \sin(\hat{B}_{\dual{r'},\nu}) }_0\\
    & = e^{-\langle \hat{B}^2\rangle_0} \sinh{ \expval{\hat{B}_{\dual{r},\mu} \hat{B}_{\dual{r'},\nu}} _0 }.
\end{aligned}
\end{equation}
The connected correlators thus become
\begin{equation}
\begin{aligned}
    \langle \hat{\tilde{X}}_P \hat{\tilde{X}}_{P'} \rangle_c &= e^{-\langle \hat{B}^2\rangle_0} \left( \cosh{ \expval{\hat{B}_{\dual{r},\mu} \hat{B}_{\dual{r'},\nu}} _0 } - 1\right)\\
    & \approx \frac{e^{-\langle \hat{B}^2\rangle_0}}{2}   \expval{\hat{B}_{\dual{r},\mu} \hat{B}_{\dual{r'},\nu}}_0^2  
\end{aligned}
\end{equation}
and 
\begin{equation}
    \langle \hat{\tilde{Y}}_P \hat{\tilde{Y}}_{P'} \rangle_c  \approx \frac{e^{-\langle \hat{B}^2\rangle_0}}{2}   \expval{\hat{B}_{\dual{r},\mu} \hat{B}_{\dual{r'},\nu}}_0.  
\end{equation}
}
Thus the connected plaquette correlators in the QSL phase vary as
\begin{equation}\label{eq:correlator_theoretical}
  \begin{aligned}
    \langle \hat{\tilde{X}}_P \hat{\tilde{X}}_{P'} \rangle_c & \propto \frac{1}{\mathsf{R}^8} \left[ \sum_{k,l  } ( \dual{u}_\mu )_k ( \dual{u}_\nu )_l  \left(  2 \frac{\mathsf{R}_l \mathsf{R}_k}{\mathsf{R}^2} -  \delta_{k,l} \right) \right]^2 ,\\
    \langle \hat{\tilde{Y}}_P \hat{\tilde{Y}}_{P'} \rangle_c &  \propto \frac{1}{\mathsf{R}^4} \left[ \sum_{k,l  } ( \dual{u}_\mu )_k ( \dual{u}_\nu )_l  \left(  2 \frac{\mathsf{R}_l \mathsf{R}_k}{\mathsf{R}^2} -  \delta_{k,l} \right) \right],
\end{aligned}
\end{equation}
where the summation is over $k,l \in \{ x,y,z \}$, $\dual{R} = \dual{r} - \dual{r}'$, and $\mathsf{R}$ is assumed to be large compared to the monopole correlation length. The factors inside the square brackets are geometric factors, which depend on the direction of the vectors $\dual{u}_\mu$, $\dual{u}_\nu$, and $\dual{R}$, but are  independent of the distance $\mathsf{R}$ between the two plaquettes. Ref.~\cite{hermele2004} also separately studied the correlators precisely at the RK point (which sits at the phase boundary between deconfined and confined phases) where the effective field theory differs from the regular Maxwell theory. In the RK wavefunction, while the behavior of the plaquette correlators differs from Eq.~\eqref{eq:correlator_theoretical}, it is still a power law with a slower decay~\cite{hermele2004}. We note that, if the experimentally prepared state is close to an RK wavefunction (see discussion in Sec.~\ref{sec:dynamics}), then this distinction will be important.

\newtext{\subsubsection{Plaquette correlators in the TFP phase}}

\newtext{Now we calculate the dependence of the two-plaquette correlators deep inside the TFP phase, that is for $\Omega \gg V$. Our strategy is to treat the van der Waals interactions, which we denote in this section as $\hat{H}_V = \hat{H}_0 + \hat{H}_{\text{LR}}$ as a perturbation over $\hat{H}_\Omega$ using  perturbation theory. Recall that $\hat{H}_V$ is given by 
\begin{equation}
\label{eq:H_V}
    \hat{H}_V = \frac{V}{2} \sum_{i\neq j} \frac{\hat{S}_i^z \hat{S}_j^z}{|\bm{x}_i - \bm{x}_j|^6}.
\end{equation}
The unperturbed ground state is simply the product state
\begin{equation}
\label{eq:minus_state_def}
    \ket{-} =  \underset{i}{\otimes} \ket{\hat{S}^x_i = -1/2}.
\end{equation}
$\hat{H}_V$ flips two spins at $\bm{x}_i$ and $\bm{x}_j$ with an amplitude proportional to $V \left( a /|\bm{x}_i - \bm{x}_j | \right)^6$. The first-order correction from perturbation theory is
\begin{equation}
\label{eq:tfp_state_pert1}
    \ket{\chi_1} = -\frac{V}{8\Omega} \sum_{\text{pairs }i,j} \left( \frac{a}{|\bm{x}_i -\bm{x}_j|} \right)^6 \ket{i,j} ,
\end{equation}
where the summation is over all distinct pairs of sites $i, j$ and
\begin{equation}
    \ket{i,j} \equiv \ket{\hat{S}^x_i = 1/2}\ket{\hat{S}^x_j = 1/2} \underset{k \neq i,j}{\otimes}  \ket{\hat{S}^x_k = -1/2}.
\end{equation}
We find that the first-order terms in $\langle \hat{X}_P \hat{X}_{P'} \rangle_c$ are 0 and, up to second order in perturbation theory (see Appendix~\ref{app:plaquette_TFP} for derivation),
\begin{equation}
\label{eq:plaquette_x_conn_matrix_ele}
        \langle \hat{X}_P \hat{X}_{P'}  \rangle_{c} \propto \frac{V^2 a^{12}}{ \Omega^2}  \sum_{i\neq j} \frac{\langle i, j | \left( \hat{X}_P -1\right) \left(   \hat{X}_{P'} -1 \right) | i, j \rangle }{| \bm{x}_i - \bm{x}_j|^{12} }.
\end{equation}
The matrix element in Eq.~\eqref{eq:plaquette_x_conn_matrix_ele} is nonzero only if $i \in P$ and $j \in P'$ or $i \in P'$ and $j \in P$. 
If the distance between the plaquettes $\mathsf{R}$ is large, then we find
\begin{equation}
\label{eq:deep_tfp_state}
    \expval{ \hat{X}_P \hat{X}_{P'} }_{c} \propto \frac{V^2}{\Omega^2 } \left( \frac{a}{\mathsf{R}} \right)^{12}.
\end{equation}}

\newtext{Now consider the connected plaquette Y correlator. 
Note that the Hamiltonian Eq.~\eqref{eq:hamsplit} has a global $\mathbb{Z}_2$ symmetry: $\hat{S}^z \rightarrow -\hat{S}^z$, $\hat{S}^x \rightarrow \hat{S}^x$, and $\hat{S}^y \rightarrow -\hat{S}^y$ for $h=0$. Under this symmetry, $\hat{Y}_P \rightarrow -\hat{Y}_P$, implying  $\langle \hat{Y}_P \rangle = 0$. 
However, the product $\hat{Y}_P \hat{Y}_{P'}$ is $\mathbb{Z}_2$-symmetric, and its expectation value need not be zero. }

\newtext{Note that $\hat{Y}_P \hat{Y}_{P'}$ flips three spins of $P$ and three spins of $P'$, where the spins are assumed to be in the eigenbasis of  $\hat{S}^x$. On the other hand, the perturbation $\hat{H}_V$ flips two spins in $\hat{S}^x$ basis. Thus the first nonzero contribution in the perturbation series for $\langle \hat{Y}_P \hat{Y}_{P'} \rangle_c$ can only be obtained at third order or higher in perturbation theory. For a large distance between the plaquettes, the dominant contribution to the plaquette Y correlator will come from the process where two spins of $P$ are flipped by one application of $\hat{H}_V$, two spins of $P'$ are flipped by another application of $\hat{H}_V$, and one spin of $P$ and another of $P'$ are flipped by the third application of $\hat{H}_V$. Such a process will give a contribution that will fall off with distance as $\left(a / \mathsf{R} \right)^6$. Overall, in the TFP phase,
\begin{equation}
    \label{eq:Y_plaq_corr_TFP}
        \expval{ \hat{Y}_P \hat{Y}_{P'} }_{c} \propto \frac{V^3}{\Omega^3 } \left( \frac{a}{\mathsf{R}} \right)^{6}.
\end{equation}}

Since the plaquette correlators involve off-diagonal operators, they cannot be read out directly from the snapshots of a Rydberg-atom array. However, we show  that they can be measured by evolving the system under a modified Hamiltonian for a specific time duration followed by measurement of a diagonal operator~\cite{verresen2021, semeghini2021}. We describe the protocols to measure both plaquette X and plaquette Y correlators in the sections below.\\

\subsubsection{Measurement of the plaquette correlators} \label{sec:xy_plaq_corr}

\newtext{To measure the plaquette X correlator, one simply needs to change the basis from $\hat{S}^x$ to $\hat{S}^z$ on every site.} 
This can be accomplished by abruptly changing the phase and the amplitude of the Rabi frequency, so that the new Hamiltonian is $\hat{H}_\text{Y} \approx \Omega_Y \sum_i \hat{S}^y_i$ with $\Omega_Y \gg V$. 
(Achieving $\Omega_Y \gg V$ may require working with atom spacings that are sufficiently large and/or with Rydberg principal quantum numbers that are sufficiently low, but not low enough to make Rydberg lifetime a problem.) 
It is assumed that this change of the Hamiltonian is done sufficiently rapidly so that the sudden approximation is valid and the state of the system does not change.
Then evolve the system under $\hat{H}_Y$ for a time  $t_Y = \pi/(2 \Omega_Y)$, which amounts to a $\pi/2$ pulse about the $y$-axis,  \newtext{transforming $\hat{S}^x_i$ into $\hat{S}^z_i$}. 
Finally, measure all the atoms in the $\{\ket{g},\ket{r}\}$ basis \newtext{and get $\expval{ \hat{S}^z_i}$ in the final state, which is the same as the $\expval{ \hat{S}^x_i}$ of the state right before the sudden change of the Hamiltonian.
The connected plaquette X correlator can be calculated using these values of $\expval{\hat{S}_i^x}$.} 

The procedure to measure the connected plaquette Y correlator is similar to the procedure for measuring the plaquette X correlator, except that now the $\pi/2$ pulses on sites  $2i$ for $i= 1,2,\ldots,6$ are about the $x$-axis on the Bloch sphere while the $\pi/2$ pulses on  sites $2i-1$ for $i=1,2,\ldots,6$ are around the $y$-axis, where the sites $1$ to $6$ are on $P$ and those from $7$ to $12$ are on $P'$. 
\newtext{These pulses transform $\hat{S}_{2i}^{x} \rightarrow \hat{S}^{z}_{2i}$ and $\hat{S}_{2i-1}^{y} \rightarrow \hat{S}^{z}_{2i-1}$.}
After applying these $\pi/2$ pulses, $\expval{ \hat{S}_i^z}$ is measured by taking snapshots of the array and the connected plaquette Y correlator is calculated from it.

\newtext{We note that the power-law decays of the plaquette correlators in the QSL and the TFP phases are very rapid, and it might be difficult practically to distinguish them from an exponential decay.}
This connected plaquette Y correlator has an advantage over the connected plaquette X correlator with regards to this issue because the power law decays of the plaquette Y correlator are slower. The disadvantage of the of the plaquette Y correlator is that measuring it requires control over individual sites.

\subsection{Monopole-monopole correlator}\label{sec:monopole}
In the deconfined phase, monopoles are gapped. Therefore, the expectation value of an (equal-time) operator that creates a string with a monopole and antimonopole at its endpoints should decay exponentially with the length of the string. On the other hand, in the confined phase, monopoles are condensed, and hence the expectation value should approach a nonzero constant as the length of the string increases. In the continuum, the following operator inserts a string that creates a monopole at $\dual{r}_1$ and an antimonopole at $\dual{r}_2$ \cite{hermele2004}:
\begin{equation}\label{eq:monopolestring}
    \hat{\mathcal{M}}^\dagger \hat{\mathcal{M}} (\dual{r}_1 \rightarrow \dual{r}_2)\sim e^{i\int \dd[3]{\dual{r'}}\vb{\mathcal{A}}(\dual{r'})\cdot \hat{\boldsymbol{e}}(\dual{r'})}. 
\end{equation}
Here $\vb{\mathcal{A}}(\dual{r'})$ is a classical (non single-valued) vector potential such that the flux $\phi_{\Sigma}$ of $\vb{\mathcal{B}}=\grad \times \vb{\mathcal{A}}$ through a closed surface $\Sigma$ is 
\begin{equation}\label{eq:fluxcondition}
    \phi_{\Sigma}\equiv \oint_{\Sigma} \vb{\mathcal{B}}\cdot d\vb{S}=2\pi q Q_{\Sigma},
\end{equation}
where $Q_{\Sigma}=1$ when $\Sigma$ encloses $\dual{r}_1$ and not $\dual{r}_2$, $Q_{\Sigma}=-1$ when $\Sigma$ encloses $\dual{r}_2$ and not $\dual{r}_1$, and $Q_{\Sigma}=0$ otherwise. $q$ is an integer and denotes the ``charge" of the monopole string.  For simplicity, we will set $q=1$ in this section. We clarify that $\vb{\mathcal{B}}$ and $\phi_\Sigma$ are classical numbers and are different from $\hat{\vb{b}}$ and $\hat{\Phi}_{\Sigma}$ which are operators. $\hat{\vb{b}}\equiv \grad \times \hat{\vb{a}}$, for gauge-field (operator) $\hat{\vb{a}}$, and $\hat{\Phi}_{\Sigma}$ is defined as
\begin{equation}
    \hat{\Phi}_{\Sigma}\equiv \oint_{\Sigma} \hat{\vb{b}}\cdot d\vb{S}=2\pi \hat{m},
\end{equation}
where $\hat{m}$ takes integer eigenvalues.
The form of the monopole string operator is chosen so that it increases the flux through $\Sigma$ by $2 \pi Q_\Sigma$, i.e.,
\begin{equation}
    \comm{\hat{\Phi}_{\Sigma}}{\hat{\mathcal{M}}^\dagger \hat{\mathcal{M}} (\dual{r}_1 \rightarrow \dual{r}_2)}=2\pi Q_{\Sigma} \hat{\mathcal{M}}^\dagger \hat{\mathcal{M}}(\dual{r}_1 \rightarrow \dual{r}_2).
\end{equation}

\begin{figure}[t]
  \centering
  \includegraphics[width = 0.76\columnwidth]{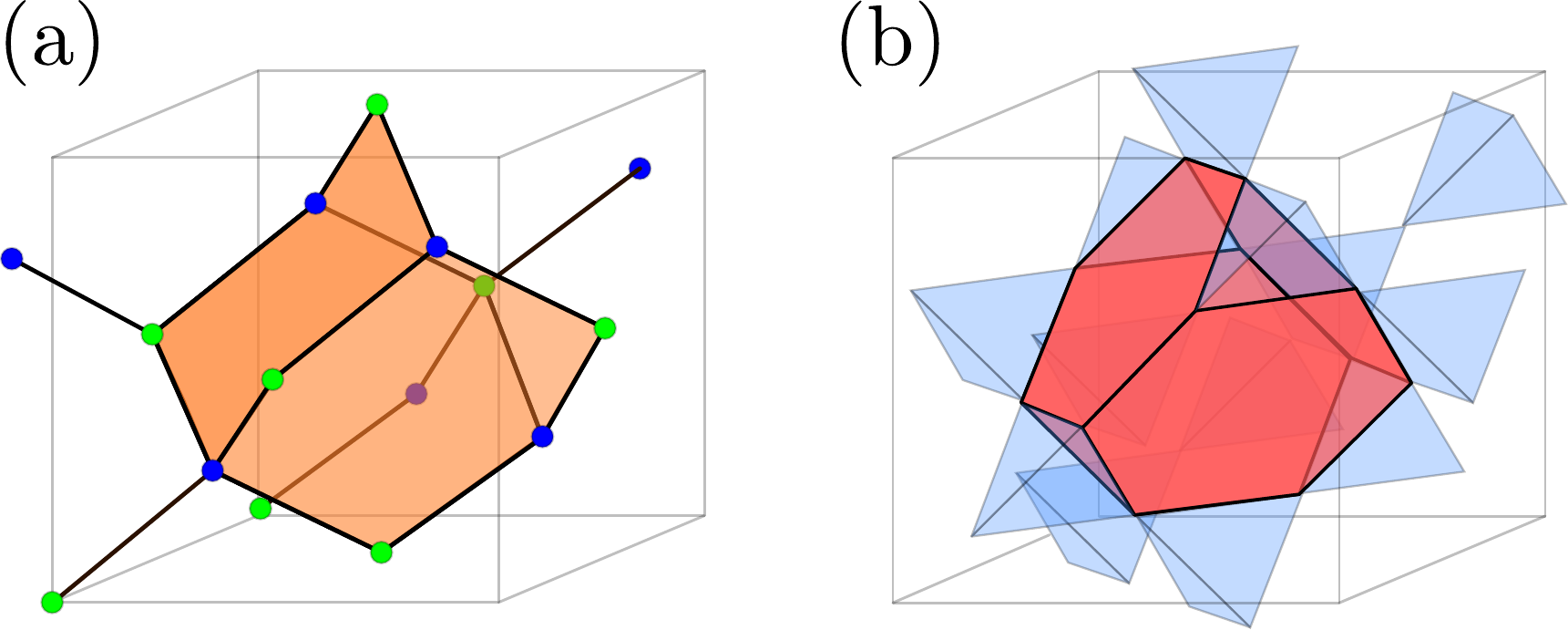}
  \capt{(a) The ``polyhedron" formed by four puckered hexagons of the diamond lattice is shown in orange. The centers of these ``polyhedra" form the dual diamond lattice. (b) The center of the ``polyhedron" in (a) is also the center of a truncated tetrahedron (shown in red) of the pyrochlore lattice.}
  \label{fig:polyhedra}
\end{figure}

We now adapt this operator to the Rydberg setting. Consider the diamond lattice formed by the centers of tetrahedra of the pyrochlore lattice, Fig.~\ref{fig:pyrochlore_diamond}(b). Unlike the continuum, it is now important to specify that the endpoints of the monopole string $\dual{r}_1$ and $\dual{r}_2$ belong to the  dual diamond lattice [see Fig.~\ref{fig:pyrochlore_diamond}(c)], whose sites are centers of ``polyhedra" made of four puckered-hexagonal ``plaquettes" of the diamond lattice\footnote{In terms of the original pyrochlore lattice, the vertices of the dual diamond lattice are centers of the truncated tetrahedra [see Fig.~\ref{fig:polyhedra}(b)] which fill the voids between the tetrahedra.}, see Fig.~\ref{fig:polyhedra}(a). Let $\bm{x} \equiv \vb{r}+\vb{e}_\mu/2$ be a site on the pyrochlore lattice, where $\vb{r}$ is an A-site of the diamond lattice. $\mathcal{A}_{\bm{x}}\equiv \mathcal{A}_{\vb{r},\vb{r}+\vb{e}_\mu}$ is the discrete version of $\mathcal{A}$ integrated (Fig.~\ref{fig:pyrochlore_diamond}(b) shows the vectors $\vb{e}_\mu$) along the line pointing from the center of an A tetrahedron (centred at $\vb{r}$) to the B tetrahedron (centred at $\vb{r}+\vb{e}_\mu$) such that the two tetrahedra touch at $\bm{x}$.  

 $\mathcal{A}_{\bm{x}}$ is required to satisfy the discrete version of Eq.~\eqref{eq:fluxcondition}, and hence depends on $\dual{r}_1$, $\dual{r}_2$, the ``magnetic field" configuration $\vb{\mathcal{B}}$ and the gauge choice for $\mathcal{A}_{\bm{x}}$. For the pyrochlore lattice, we have
\begin{equation}\label{eq:monopoledecay}
    \hat{\mathcal{M}}^\dagger \hat{\mathcal{M}} (\dual{r}_1\rightarrow \dual{r}_2)= e^{i\sum_{\bm{x}\in \text{pyrochlore}}\mathcal{A}_{\bm{x}}  \left(\hat{n}_{\bm{x}}-1/2\right)}.
\end{equation}
This operator is purely diagonal in the $\hat{n}_{\bm{x}}$ basis (i.e., in the $\hat{S}^z$-basis). So, experimentally, one can calculate this phase for each snapshot and average over shots. 

Theoretically, one expects
\begin{equation} \label{eq:monopole_correlator_behavior}
    \abs{\expval{\hat{\mathcal{M}}^\dagger \hat{\mathcal{M}} (\dual{r}_1 \rightarrow \dual{r}_2)}} \sim \begin{cases}
    e^{-\abs{\dual{r}_2-\dual{r}_1}/\lambda}, &\text{deconfined phase,}\\
    \text{constant,} &\text{confined phase}, 
    \end{cases}
\end{equation}
where $\lambda$ is a correlation length that depends on the monopole gap and the ``photon" velocity. 
 \begin{figure*}
    \centering
    \includegraphics[width=0.8\textwidth]{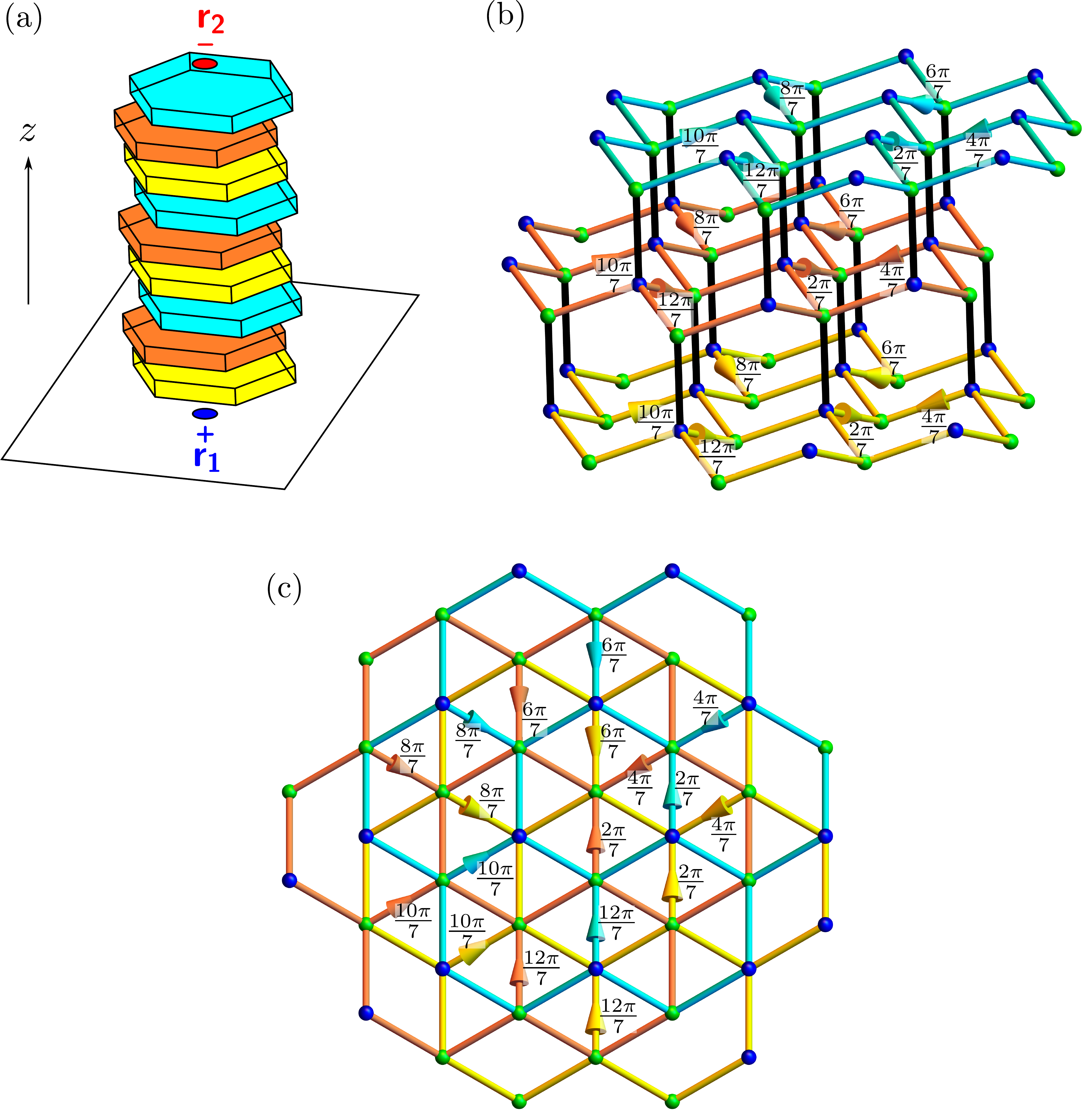}
    \capt{An example of the monopole string operator $\hat{\mathcal{M}}^\dagger \hat{\mathcal{M}} (\dual{r}_1\rightarrow \dual{r}_2)$ for which we provide $\mathcal{A}_{\bm{x}}$ explicitly. In our example, the string carries $2\pi$ flux through a tube with a width of 7 puckered hexagons of the diamond lattice. \newtext{The tube runs along the ${z}-$direction.}
    (a)~A schematic of the tube running along the $z$-direction. The diamond lattice (whose vertices are centers of tetrahedra of the pyrochlore lattice) can be seen as ABC stacking of layers of ``honeycomb" lattices made of chair-like puckered hexagons.  The tube consists of three types of layers shown in yellow, orange, and cyan. Each layer is made of 7 puckered hexagons. To convey a sketch, we depict such a layer by a big hexagon with some thickness. (b) A side view of the stack showing three of its layers, \newtext{where each layer is made of 7 puckered hexagons of the diamond lattice. The bonds within each of these layers are colored in yellow, orange, and cyan. The bonds (of the diamond lattice) connecting sites of two different layers are shown in black.} These layers are repeated in the $z$ direction to get the entire string. For bonds $\bm{x}$ with \newtext{(conical)} arrows, the value of $\mathcal{A}_{\bm{x}}$ is written next to the bond. For bonds $\bm{x}$ without arrows, $\mathcal{A}_{\bm{x}}=0$. \newtext{The two sub-lattices of the diamond lattice are represented by blue and green sites.} (c) Top view of three of the layers of the stack.  
    \newtext{See also Supplementary Material for an animation showing other points of views~\cite{supp}.}
    It can be seen from all three sub-figures (a)-(c) that the flux through any closed surface $\Sigma$ that completely encloses an integer number of layers, such that the bottom layer is included but not the top, is $2\pi$. However, if $\Sigma$ partially encloses a layer, then $\Phi_{\Sigma}$ is 0. This difficulty in defining arbitrary integer multiples of $2\pi$ flux through a volume enclosed by a finite number of plaquettes has been observed before \cite{hermele2004}. Therefore in our construction, $\dual{r}_1$ and $\dual{r}_2$ have to be seen as being smeared across 7 points of the dual diamond lattice below the bottom layer and above the top layer respectively, in order to be consistent with Eq.~\eqref{eq:fluxcondition}.}
    \label{fig:monopolecorr}
\end{figure*}
In Fig.~\ref{fig:monopolecorr}, we provide an example of one configuration of the classical numbers $\mathcal{A}_{\bm{x}}$ that defines a monopole string operator. Below, we comment on the freedom in choosing $\mathcal{A}_{\bm{x}}$. 

\subsubsection{Choice of $\mathcal{A}$}
The classical numbers $\mathcal{A}_{\bm{x}}$ should of course obey the constraint that the flux of $\curl \mathcal{A}$ through a closed surface $\Sigma$ is $2 \pi Q_{\Sigma}$, as mentioned above. However, one still has a freedom in the choice of $\mathcal{A}$ in the following two respects:
\begin{enumerate}
    \item Freedom in the arrangement of the field lines of $\curl \mathcal{A}$. For example, they can be confined to a thin tube connecting $\dual{r}_1$ and $\dual{r}_2$, or be spread out according to Coulomb's law, or be something in between. Different such arrangements, due to their different energy costs, would differ in sub-leading corrections to the exponentially decaying behavior, but the leading behavior would be unchanged. In Fig.~\ref{fig:monopolecorr}, 
    we provide a choice of $\mathcal{A}$, such that the monopole string is localized to a thin tube.
    \item For a fixed choice of field lines, we still have a gauge choice for $\mathcal{A}$. Consider a gauge transformation $\mathcal{A}_{\vb{r},\vb{r}+\vb{e}_\mu}\to \mathcal{A}_{\vb{r},\vb{r}+\vb{e}_\mu}+\lambda_{\vb{r}+\vb{e}_\mu}-\lambda_{\vb{r}}$, where $\lambda_{\vb{r}}$ is an $\vb{r}$-dependent real number. It results in
    \begin{equation}
        \hat{\mathcal{M}}^\dagger \hat{\mathcal{M}} (\dual{r}_1 \rightarrow \dual{r}_2) \to \hat{\mathcal{M}}^\dagger \hat{\mathcal{M}} (\dual{r}_1 \rightarrow \dual{r}_2) e^{-i\sum_{\mathbf{r}}\lambda_{\mathbf{r}} \eta_{\vb{r}} (\hat{n}_{\tet_{ \mathbf{r}}} -2)},
    \end{equation}
     where $\eta_{\vb{r}} = 1$ for $\mathbf{r} \in A$ and $\eta_{\vb{r}} = -1$ for $\mathbf{r} \in B$. In the $\Omega/V\ll1$ limit, we have $\hat{n}_{\tet_{\mathbf{r}}}=2$, so the expectation value is invariant under the gauge transformation.  
     Away from this limit, a gauge transformation on $\mathcal{A}_{\vb{r},\vb{r}+\vb{e}_\mu}$ generically results in a physical transformation on the monopole string operator. 
     However, as long as the external field $h=0$ [$h$ is defined in Eq.~\eqref{eq:h_def}], by particle-hole symmetry, we have $\expval{\hat{n}_{\tet_\mathbf{r}}}=2$. Since the variance $\expval{(\hat{n}_{\tet_\mathbf{r}}-2)^2}$ is bounded, we do not expect the gauge transformation on $\mathcal{A}_{\vb{r},\vb{r}+\vb{e}_\mu}$ to qualitatively change the behavior of Eq.~\eqref{eq:monopole_correlator_behavior}. But this question needs to be studied more closely in future work. 
\end{enumerate}

\newtext{\subsubsection{Monopole correlator in the ice FM phase}} \label{sec:monopole_corr_ice_fm}

\newtext{When $\Omega = 0$, the ground state is a product state in which each spin is in an eigenstate of $\hat{S}^z$, as discussed in Sec.~\ref{sec:classical}, and the monopole correlator, $\expval{\hat{\mathcal{M}}^\dagger \hat{\mathcal{M}} (\dual{r}_1 \rightarrow \dual{r}_2) }$, evaluates to a single phase (as opposed to a sum of phases for a state that is a superposition of the basis states). Thus $\abs{\expval{\hat{\mathcal{M}}^\dagger \hat{\mathcal{M}} (\dual{r}_1 \rightarrow \dual{r}_2) }} = 1$ and does not decay with the length of the string. For $\Omega \ll V$, $\abs{ \expval{\hat{\mathcal{M}}^\dagger \hat{\mathcal{M}} (\dual{r}_1 \rightarrow \dual{r}_2) }} $ will not be equal to 1, but we expect it to saturate to a nonzero constant for large strings  because the monopoles are condensed in the ice FM phase. }

\newtext{\subsubsection{Monopole correlator in the QSL phase}}
\label{sec:monopole_corr_qsl}

\newtext{The monopole correlator at the RK point and away from it in the QSL phase was calculated by Hermele et. al. in Ref.~\cite{hermele2004} using perturbation theory and field theory techniques. They showed that the correlator decays exponentially with the distance between the monopole and the antimonopole. They also verified the exponential decay numerically at the RK point.}

\newtext{\subsubsection{Monopole correlator in the TFP phase}}
\label{sec:monopole_corr_tfp}

\newtext{In this section, we show that the monopole correlator decays exponentially with the length of the string deep inside the TFP phase, that is for $\Omega \gg V$. We start by rewriting the monopole correlator from Eq.~\eqref{eq:monopoledecay} as
\begin{equation}
    \label{eq:monopoledecay2}
        \hat{\mathcal{M}}^\dagger \hat{\mathcal{M}} (\dual{r}_1\rightarrow \dual{r}_2)=  \underset{i \in \text{string}}{\otimes} \left[ \cos\left( \frac{\mathcal{A}_{i}}{2} \right) + 2i  \sin\left( \frac{\mathcal{A}_{i}}{2} \right) \hat{S}^z_{i} \right],
\end{equation}
where the tensor product is over the string between $\dual{r}_1$ and $\dual{r}_2$, and one choice of $\mathcal{A}_i$ is shown in Fig.~\ref{fig:monopolecorr}.}

\newtext{For $V=0$, the ground state is $\ket{-}$ (see Eq.~\eqref{eq:minus_state_def} for its definition), and it can be easily seen that each of the factors of the tensor product in Eq.~\eqref{eq:monopoledecay2} has an expectation value whose absolute value is less than 1.
Thus $\abs{\expval{\hat{\mathcal{M}}^\dagger \hat{\mathcal{M}} (\dual{r}_1\rightarrow \dual{r}_2)}}$ decays exponentially with the string length. For $V \ll \Omega$, at first order in perturbation theory, only two spins are flipped (in the $\hat{S}^x$ basis). Since the monopole correlator involves a product of a number of terms proportional to the length of the string, only two of which are altered by the perturbation, we expect that the monopole correlator will decay exponentially even at first order in perturbation theory. 
Thus the monopole correlator decays exponentially in the TFP phase. }\\

\subsection{Fredenhagen-Marcu order parameters} \label{sec:fm_order_param}

It is known that the confined and deconfined phases of a gauge theory without matter fields can be distinguished by the scaling of the Wilson loops $W_\mathcal{L} = \expval{ e^{i \oint_\mathcal{L} A_\mu dx^\mu} }$, where $A_\mu$ is the gauge field and $\mathcal{L}$ is a closed loop. In the deconfined phase, the Wilson loop follows the perimeter law, $W_{\mathcal{L}} \propto e^{-\text{Perimeter of }\mathcal{L}}$, while in the confined phase, it follows the area law, $W_{\mathcal{L}} \propto e^{-\text{Area of }\mathcal{L}}$. However, in the presence of matter fields (which are generically always present), the Wilson loop follows the perimeter law in both phases \cite{fradkin2013field,shankar2017quantum}, and it cannot be used to distinguish  them. The Fredenhagen-Marcu order parameter is useful in such cases and has  a different behavior in the two phases~\cite{fredenhagen1983charged,fredenhagen1986,fredenhagen1988dual,marcu1986uses,bricmont1983order,gregor2011diagnosing,verresen2021,semeghini2021}. 
\newtext{In its original formulation~\cite{fredenhagen1983charged}, the Fredenhagen-Marcu order parameters involved expectation values of operators in space-time. Subsequently, real-space versions were proposed and studied~\cite{fredenhagen1988dual,gregor2011diagnosing,verresen2021,xu2024critical}. We will be using the real-space version of the Fredenhagen-Marcu order parameters here.}
The Fredenhagen-Marcu order parameter, denoted here by $\chi^E_{\mathcal{C}}$, is defined as
\begin{equation} \label{eq:FM_e}
  \chi_{\mathcal{C}}^E = \frac{\abs{ \expval{ e^{i\sum_\mathcal{C} \hat{a}_{\vb{r}\vb{r}'}  } +\text{H.c.} } } }{\sqrt{ \abs{ \expval{ e^{i \sum_{\mathcal{L}} \hat{a}_{\vb{r} \vb{r}'} } + \text{H.c.} }  }}},
\end{equation}
where $\mathcal{C}$ is an open curve and $\mathcal{L}$ is the closed loop formed by combining $\mathcal{C}$ with its mirror image about a plane that intersects $\mathcal{C}$ only at its end points. This order parameter detects long-range order in the ``electric charge"-creation string. In the Higgs phase, ``electric charges" are condensed, and hence $\chi_{\mathcal{C}}^E$ approaches a nonzero constant. In the deconfined phase, the numerator in Eq.~\eqref{eq:FM_e} (calculated on an open curve) decays to zero faster than the denominator (calculated on a closed loop, giving the Wilson loop), as the length of $\mathcal{C}$ is increased. Therefore, in the deconfined phase, $\chi_{\mathcal{C}}^E$ goes to 0 as the length of $\mathcal{C}$ is increased. In the confined phase, it was argued in Ref.~\cite{fredenhagen1986} that while both the numerator and the denominator go to zero as the length of $\mathcal{C}$ is increased, the limit of their ratio approaches a constant. However distinguishing this constant from zero in finite systems for finite length of $\mathcal{C}$ may be difficult.
Below we explain how to measure $\chi_{\mathcal{C}}^E$. 

Using the mapping from spin operators to gauge fields, Eqs.~\eqref{eq:rotor_mapping} and~\eqref{eq:EM_mapping}, we see that
\begin{equation} 
   e^{i\sum_\mathcal{C} \hat{a}_{\vb{r} \vb{r}'}} \simeq \hat{S}_1^+\hat{S}_2^-\hat{S}_3^+\cdots,
\end{equation}
where the product of $\hat{S}^+$ and $ \hat{S}^-$ operators is over the sites on the curve $\mathcal{C}$.
The denominator in $\chi_{\mathcal{C}}^E$ has a similar expression in terms of spin operators. \newtext{Thus, $\chi_{\mathcal{C}}^{E}$ is given by}
\begin{equation}\label{eq:FM_e_plus_minus}
      \chi_{\mathcal{C}}^E = \frac{\abs{ \expval{ \hat{S}_1^+\hat{S}_2^-\hat{S}_3^+\cdots +\text{H.c.} } } }{\sqrt{ \abs{ \expval{ \hat{S}_1^+\hat{S}_2^-\hat{S}_3^+\cdots + \text{H.c.} }  }}},
\end{equation}
where the product in the numerator is along the open curve $\mathcal{C}$ and the one in the denominator is along the closed loop $\mathcal{L}$.

From the point of view of measurement, it is more convenient to consider another quantity,  which has the same behavior as $\chi_\mathcal{C}^E$ in the three phases, defined as:
\begin{equation}
\tilde{\chi}_\mathcal{C}^E \equiv \frac{\abs{ \expval{ \prod_{i\in \mathcal{C}} \hat{S}_i^x  } } }{\sqrt{ \abs{ \expval{ \prod_{i\in \mathcal{L}} \hat{S}_i^x  }  }}},
\end{equation}
In the transverse-field-polarized (Higgs) phase, $\tilde{\chi}_\mathcal{C}^E$ approaches a nonzero constant, just like $\chi_\mathcal{C}^E$. Now, we argue that even in the QSL and confined phases, $\tilde{\chi}_\mathcal{C}^E$ and $\chi_\mathcal{C}^E$ have the same behavior.  
For a state $\ket{\Psi} $ that dominantly lies in the ice manifold, with corrections from outside the ice manifold being of order $\Omega/V$ (such as the ground state $\ket{\Psi_g}$),  we have
\begin{multline}
  \langle \Psi | \hat{S}_1^+ \hat{S}_2^- \hat{S}_3^+ \cdots + \text{H.c.} | \Psi \rangle \\ = \langle \Psi | (2\hat{S}_1^x) (2\hat{S}_2^x) (2\hat{S}_3^x) \cdots |\Psi \rangle + \Theta \left( (\Omega/V)^L \right),
\end{multline}
where $L$ is the number of sites on $\mathcal{C}$. The correction is of order $( \Omega/V )^L$ by an argument similar to the one used to show that the error is sixth order in the protocol to measure the plaquette X correlator (see Appendix~\ref{app:corrections_plaq}). Thus, for small $\Omega/V$, $\chi_\mathcal{C}^E$ and $\tilde{\chi}_\mathcal{C}^E$ are equal up to order $(\Omega/V)^L$. 

The numerator and the denominator of $\tilde{\chi}_\mathcal{C}^E$ can be measured by applying $\pi/2$ pulses about the $y$-axis and measuring, from the snapshots, products of $\hat{S}^z$ along $\mathcal{C}$ and $\mathcal{L}$. This procedure is similar to the protocol to measure the plaquette X correlator, described in Sec.~\ref{sec:xy_plaq_corr}. 

The operator $e^{i \sum_{\mathcal{C}} \hat{a}_{\vb{r} \vb{r}'}}$ creates two opposite ``electric charges" at the endpoints of $\mathcal{C}$. So a magnetic analogue of $\chi^E_\mathcal{C}$ can also be defined, where the  numerator is the expectation value of the operator that creates a monopole and an antimonopole at the endpoints of $\mathcal{C}$. Such an order parameter, $\chi_{\mathcal{C}}^M$, detects long-range order in the monopole string operator and is given by
\newtext{
\begin{equation}\label{eq:FM_m}
  \chi_{\mathcal{C}}^M = \frac{\sqrt{\expval{ \hat{\mathcal{M}}^{\dagger} \hat{\mathcal{M}}\left(\dual{r}_1 \xrightarrow{\mathcal{C}_a} \dual{r}_2\right) }\expval{ \hat{\mathcal{M}}^{\dagger} \hat{\mathcal{M}}\left(\dual{r}_1 \xrightarrow{\mathcal{C}_b} \dual{r}_2\right) }} }{  \sqrt{\expval{ \hat{\mathcal{M}}^{\dagger} \hat{\mathcal{M}}\left(\dual{r}_1 \xrightarrow{\mathcal{L}} \dual{r}_1\right)}}  },
\end{equation}
where $\hat{\mathcal{M}}^\dagger \hat{\mathcal{M}} (\dual{r}_1 \xrightarrow{\mathcal{C}_a} \dual{r}_2)$ inserts a monopole-antimonopole string along $\mathcal{C}_a$ and was defined in Eq.~\eqref{eq:monopoledecay}. The open strings $\mathcal{C}_a$, $\mathcal{C}_b$ and the closed loop $\mathcal{L}$ are chosen so that $\mathcal{L}$ is obtained upon joining $\mathcal{C}_a$ and $\mathcal{C}_b$.} In this section, we use the notation where the path of the monopole-antimonopole string is explicitly written in the argument of $\hat{\mathcal{M}}^\dagger \hat{\mathcal{M}}$. 
Since this operator is diagonal in the $\hat{S}^z$ basis, it can be measured straightforwardly from the snapshots of the Rydberg-atom array.

In the confined phase, monopoles are condensed, so $\chi_{\mathcal{C}}^M$ should be a nonzero constant. In the deconfined phase, by the argument of Ref.~\cite{fredenhagen1986}, the numerator of Eq.~\eqref{eq:FM_m} decays to zero faster than the denominator as the length of $\mathcal{C}$ increases. Therefore, in the deconfined phase, $\chi_{\mathcal{C}}^M$ goes to zero as the length of $\mathcal{C}$ increases. In the Higgs phase, even though there is no long-range order in the monopole string and both the numerator and denominator go to zero, by the argument in Ref.~\cite{fredenhagen1986}, the ratio (i.e.~$\chi_{\mathcal{C}}^M$) approaches a nonzero constant as the length of $\mathcal{C}$ increases. But distinguishing this nonzero constant from zero in finite-size numerics and experiment may be challenging (similar to the situation for $\chi_{\mathcal{C}}^E$ in the confined phase). The behavior of the Fredenhagen-Marcu order parameters in various phases is summarized in Table~\ref{tab:correlators}.

Before proceeding, we note that our protocols to measure the plaquette correlators and the Fredenhagen-Marcu order parameter $\chi_\mathcal{C}^E$ work in the limit $\Omega/V \ll 1$, which is outside the window in which the ground state of Hamiltonian \eqref{eq:hamsplit} is a QSL. However, we explained in Sec.~\ref{sec:dynamics} that it is possible to dynamically prepare finite puddles of QSL regions even in the $\Omega /V \ll 1$ limit when the ground state is not a QSL. Our protocols can then be applicable. 

\newtext{\subsubsection{Fredenhagen-Marcu order parameters in the ice FM phase}}
\label{sec:FM_ice_fm}

\newtext{We argued in Sec.~\ref{sec:monopole_corr_ice_fm} that $\abs{ \expval{ \hat{\mathcal{M}}^\dagger \hat{\mathcal{M}} \left(\dual{r}_1 \xrightarrow{\mathcal{C}} \dual{r}_2 \right)} }$ approaches a nonzero constant for large open curves $\mathcal{C}$ in the ice FM phase. By the same reasoning, we expect  $\abs{ \expval{ \hat{\mathcal{M}}^\dagger \hat{\mathcal{M}} \left(\dual{r}_1 \xrightarrow{\mathcal{L}} \dual{r}_2 \right) } }$ to approach a nonzero constant for large closed loops $\mathcal{L}$, implying that $\chi_\mathcal{C}^M$ approaches a nonzero constant for large loops.} 

\newtext{Now we consider the behavior of $\chi_{\mathcal{C}}^E$ in the ice FM phase. 
For $\Omega \ll V$, the ground state will be $\ket{\Psi_\text{IFM}}$ plus perturbative corrections in $\Omega/V$ on top of it coming from $\hat{H}_\Omega$. 
The ground state in the ice FM phase can be written as $\ket{\Psi_{\text{ord}}} = \hat{U}_S^\dagger \ket{\Psi_{\text{IFM}} }$ [see Eq.~\eqref{eq:scwtrans}]. 
Also, call the operator in the numerator of $\chi_{\mathcal{C}}^E$ as $\hat{\chi}_{\mathcal{C}, \text{num}}^E \equiv \hat{S}_1^+ \hat{S}_2^- \hat{S}_3^+\cdots + \text{H.c.}$, where the product is over the $\hat{S}^\pm$ operators of sites on $\mathcal{C}$.
The factor in the numerator of $\chi_\mathcal{C}^E$ in the ice FM phase can thus be written as $\abs{\langle \Psi_{\text{IFM}} | \hat{U}_S \hat{\chi}_{\mathcal{C}, \text{num}}^E \hat{U}_S^\dagger | \Psi_{\text{IFM}} \rangle} $. Let $\abs{\mathcal{C}}$ be the length of $\mathcal{C}$. Now, acting on a basis state in which spins along $\mathcal{C}$ are alternating, $\hat{\chi}_{\mathcal{C}, \text{num}}^E$ flips these $\abs{\mathcal{C}}$ spins along $\mathcal{C}$.
To compensate, the same number of flips must come from $\hat{U}_S$ and $\hat{U}_S^\dagger$ combined. 
This happens at order $\abs{\mathcal{C}}$ in perturbation theory.
Thus for a fixed and small $\Omega$ (as compared to $V$), the numerator of $\chi_\mathcal{C}^E$ will be proportional to $\left( \Omega/V \right) ^ {\abs{\mathcal{C}}  }$.
By a similar argument, we conclude that the denominator of $\chi_\mathcal{C}^E$ will be proportional to $\left( \Omega / V \right) ^ {\abs{\mathcal{L}} /2 }$.
Since the loop $\mathcal{L}$ is formed by joining $\mathcal{C}$ and its mirror image, we have $\abs{\mathcal{L}} = 2\abs{\mathcal{C}}$, and the two exponential decays cancel out. 
Thus $\chi_{\mathcal{C}}^E$ approaches a nonzero constant in the ice FM phase.
}

\newtext{\subsubsection{Fredenhagen-Marcu order parameters in the QSL phase}} \label{sec:FM_qsl}

\newtext{Our ansatz for the QSL phase is $\ket{\Psi_{\text{QSL}}} = \hat{U}_{S}^\dagger \ket{\Psi_{\text{RK}}}$ [see Eq.~\eqref{eq:qsl_ansatz}]. 
The numerator of $\chi_{\mathcal{C}}^E$ is $\abs{\langle \Psi_{\text{RK}} | \hat{U}_S \hat{\chi}_{\mathcal{C}, \text{num}}^E \hat{U}_S^\dagger | \Psi_{\text{RK}} \rangle} $.
By an argument similar to the one in  Sec.~\ref{sec:FM_ice_fm}, we expect that the numerator is $\propto (\Omega/V)^{\abs{\mathcal{C}}}$.
However, unlike the case of Sec.~\ref{sec:FM_ice_fm}, the denominator of $\chi_\mathcal{C}^E$ for the QSL phase has a nonzero contribution even at zeroth order in $\Omega/V$. 
We can estimate the size of the denominator of $\chi_{\mathcal{C}}^E$ in $\ket{\Psi_{\text{RK}}}$ by a simple argument.}

\newtext{Let us call the operator in the denominator of $\chi_\mathcal{C}^E$ as $\hat{\chi}_{\mathcal{C}, \text{den}}^E \equiv \hat{S}_1^+ \hat{S}_2^- \hat{S}_3^+\cdots + \text{H.c.}$, where the product is over the $\hat{S}^\pm$ operators of sites on $\mathcal{L}$. Now we know that the number of dimer configurations on a lattice with $N$ lattice sites grows exponentially with $N$. Say this number is $\kappa^N$. (We know from Pauling's estimate for the residual entropy of water-ice that $\kappa\approx \sqrt{3/2}$~\cite{pauling1935}). Now $\hat{\chi}_{\mathcal{C}, \text{den}}^E$ has a nonzero expectation value in a basis state only if the loop $\mathcal{L}$ is flippable. If we fix the spins on the loop to be in a flippable configuration, the number of dimer coverings with the remaining $N-\abs{\mathcal{L}}$ spins will be approximately $\kappa^{N-\abs{\mathcal{L}}}$. Thus the expectation value of $\hat{\chi}_{\mathcal{C}, \text{den}}^E$ in the RK wavefunction will be approximately proportional to $\kappa^{-\abs{\mathcal{L}}}$. If we include the perturbative corrections, then the denominator of $\chi_\mathcal{C}$ will be $\sqrt{(\text{const.}) \kappa^{-\abs{\mathcal{L}}} + \mathcal{O} \left( \Omega/V\right)}$. }

\newtext{Combining the numerator and the denominator, we have
\begin{equation}
    \chi_\mathcal{C}^E \propto \frac{\left(\Omega/V \right)^{-\abs{\mathcal{C}}}}{\sqrt{ \kappa^{-\abs{\mathcal{L}}} + \mathcal{O}(\Omega/V)}} .
\end{equation}
Since $\abs{\mathcal{L}} = 2 \abs{\mathcal{C}}$, for small enough $\Omega/V$, $\chi_\mathcal{C}^E$ decays exponentially with the length of $\mathcal{C}$. Note that this is consistent with our expectation from field theory---the Fredenhagen-Marcu order parameter is supposed to go to zero as the loop size is increased in the deconfined phase of a gauge theory~\cite{fredenhagen1983charged, fredenhagen1988dual}.}

\newtext{For the Fredenhagen-Marcu order parameter corresponding to the monopoles, we do not have an argument based on the microscopics of our model which shows that the order parameter decays exponentially with loop length. However, we expect this is the case based on the result that the Fredenhagen-Marcu order parameter goes to zero in the deconfined phase of a gauge theory~\cite{fredenhagen1983charged, fredenhagen1986, fredenhagen1988dual}. Verifying this within the field theory and numerically for the microscopic model is an open problem.}

\newtext{\subsubsection{Fredenhagen-Marcu order parameters in the TFP phase}} \label{sec:FM_tfp}

\newtext{We first calculate the two Fredenhagen-Marcu order parameters for the ground state when $V = 0$, which is $\ket{-}$ defined in Eq.~\eqref{eq:minus_state_def}, and later we will consider the perturbative corrections coming from a small, but nonzero, $V$.} 

\newtext{Using the expression from Eq.~\eqref{eq:FM_e_plus_minus}, using $\abs{ \expval{ \hat{S}^{\pm}   }{\hat{S}^x = -1/2} } = 1/2$, and calculating the expectation value in the $\ket{-}$ state, we find 
\begin{equation}
    \chi_\mathcal{C}^E = \frac{2 \times (1/2)^{|\mathcal{C}|}}{\sqrt{ 2 \times (1/2)^{|\mathcal{L}|} }} = \sqrt{2},
\end{equation}
where we have used the fact that $\abs{\mathcal{L}} = 2 \abs{\mathcal{C}}$. Similarly, for the Fredenhagen-Marcu order parameter corresponding to the monopole-antimonopole string, $\chi_\mathcal{C}^M$., we have
\begin{equation}
    \chi_\mathcal{C}^M = \frac{\sqrt{\abs{\prod_{i \in \mathcal{C}_a} \cos(\frac{\mathcal{A}_i}{2})}\abs{\prod_{i \in \mathcal{C}_b} \cos(\frac{\mathcal{A}_i}{2})}   }}{\sqrt{\abs{\prod_{i \in \mathcal{L}} \cos(\frac{\mathcal{A}_i}{2}) } }}=1,
\end{equation}
i.e., the exponential decay of the numerator cancels the exponential decay of the denominator to give 1. 
For a small but nonzero value of $V$, the ground state up to first order in perturbation theory is $\ket{-} + \ket{\chi_1}$, where $\ket{\chi_1}$ is given in Eq.~\eqref{eq:tfp_state_pert1}. Using perturbation theory, the first-order correction to the numerator of Eq.~\eqref{eq:FM_e_plus_minus} is $\langle - | \hat{\mathcal{M}}^\dagger \hat{\mathcal{M}} (\dual{r}_1 \xrightarrow{\mathcal{C}} \dual{r}_2) | \chi_1 \rangle \propto (1/2)^{\abs{\mathcal{C}}} \mathcal{O} (V/\Omega)$. An analogous expression is true for the denominator with $\mathcal{C}$ replaced by $\mathcal{L}$. Thus we have
\begin{equation}
    \chi_\mathcal{C}^E = \frac{ 2 (1/2)^{\abs{\mathcal{C}}} \left( 1+\mathcal{O} \left( \frac{V}{\Omega} \right)\right)}{\sqrt{2 (1/2)^{\abs{\mathcal{L}}} \left( 1+\mathcal{O} \left( \frac{V}{\Omega} \right)\right)}},
\end{equation}
and $\chi_\mathcal{C}^E$ approaches a nonzero constant for large loops. Similarly, for the $\chi_\mathcal{C}^M$ correlator for a nonzero  but small $V$, we have
\begin{equation}
    \chi_\mathcal{C}^M = \frac{\sqrt{\prod_{\alpha=a,b}\abs{\prod_{i \in \mathcal{C}_\alpha} \cos(\frac{\mathcal{A}_i}{2}) \left(1+ \mathcal{O} \left(\frac{V}{\Omega} \right) \right) }  }}{\sqrt{\abs{\prod_{i \in \mathcal{L}} \cos(\frac{\mathcal{A}_i}{2})\left(1+ \mathcal{O} \left(\frac{V}{\Omega} \right) \right) } }}, 
\end{equation}
and $\chi_\mathcal{C}^M$ also remains a nonzero constant for large loops. This completes our discussion of the Fredenhagen-Marcu correlators in the TFP phase.}\\

\subsection{Two-point $\hat{S}^z$ correlator} \label{sec:sz_sz_correlator}

\begin{figure}[t]
  \centering
  \includegraphics[width=0.65\columnwidth]{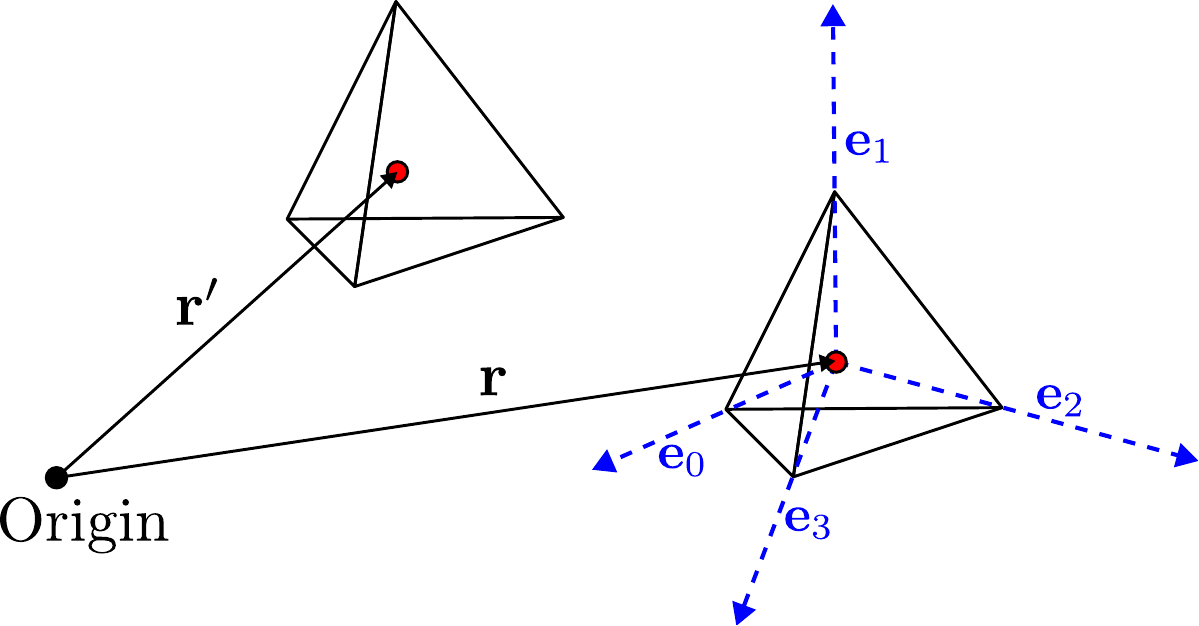}
  \capt{Notation for the two-point $\hat{S}^z$ correlator. $\mathbf{r}$ and $\mathbf{r}'$ are the positions of the centers of the tetrahedra. $\mathbf{e}_\mu$ are the  vectors joining the center of an up-pointing tetrahedron to the centers of its neighboring down-pointing tetrahedra.}
  \label{fig:sz_sz_correlator}
\end{figure}

Consider  two spins $\hat{S}^z_{\mathbf{r},\mu}$ and $\hat{S}^z_{\mathbf{r}',\nu}$ located on the sites $\mathbf{r} + \mathbf{e}_\mu/2$ and $\mathbf{r}' + \mathbf{e}_\nu/2$, where $\mathbf{r}$ and $\mathbf{r}'$ are the centers of two up-pointing tetrahedra and $\mu, \nu \in \{ 0,1,2,3 \}$ label the sites of the tetrahedra (see Fig.~\ref{fig:sz_sz_correlator}). From the mapping of spins to gauge theory, Eqs.~\eqref{eq:rotor_mapping} and~\eqref{eq:EM_mapping}, it can be seen that the two-point  correlator of these two spins $\langle \hat{S}^z_{\mathbf{r},\mu} \hat{S}^z_{\mathbf{r}',\nu}\rangle $ is the same as the two-point correlator of the electric field. Since $\hat{S}^z_{\mathbf{r},\mu} \hat{S}^z_{\mathbf{r'},\nu} $ is a diagonal operator, its correlator can be measured experimentally by capturing snapshots of the Rydberg-atom array and averaging over them.

\newtext{\subsubsection{Two-point $\hat{S}^z$ correlator in the ice FM phase}}
\label{sec:sz_sz_corr_ice_fm}

\newtext{
For $\Omega \ll V$, the ground state of the system is $\hat{U}_S^\dagger \ket{\Psi_\text{IFM}}$, and the two-point $\hat{S}^z$ correlator is $\langle \Psi_\text{IFM} | \hat{U}_S \hat{S}^z_{\mathbf{r},\mu} \hat{S}^z_{\mathbf{r}',\nu} \hat{U}_S^\dagger | \Psi_\text{IFM} \rangle $. 
Up to zeroth order in $\Omega/V$, $\hat{U}_S =\hat{1}$. 
Since $\ket{\Psi_\text{IFM}}$ is a  product state in the $\hat{S}^z$ basis, $\abs{\langle \Psi_\text{IFM} | \hat{S}^z_{\mathbf{r},\mu} \hat{S}^z_{\mathbf{r}',\nu} | \Psi_\text{IFM} \rangle } = (1/2)^2$. After taking into account corrections in $\Omega/V$, we still expect that $\langle \hat{S}^z_{\mathbf{r},\mu} \hat{S}^z_{\mathbf{r}',\nu}\rangle $ will approach a nonzero constant for large separation $\abs{ \mathbf{r} - \mathbf{r}'} $. }

\subsubsection{Two-point $\hat{S}^z$ correlator in the QSL phase}
\label{sec:sz_sz_corr_qsl}

The effective theory in the deconfined phase is the Maxwell electromagnetism. \newtext{By a derivation analogous to the derivation of Eq.~\eqref{eq:maxwellB2B2}}, one can show that in $3+1$D continuum Maxwell electromagnetism, the correlator of the Cartesian components of the electric field $\hat{e}_{\mathbf{r},i}$ for $i \in \{ x,y,z \}$ is given by~\cite{hermele2004}
\begin{equation}\label{eq:maxwellEE}
  \langle \hat{e}_{\boldsymbol{0},i} \hat{e}_{\mathbf{R},j} \rangle_0 \propto \frac{1}{R^4} \left( 2 \frac{R_i R_j}{R^2} - \delta_{ij} \right) ,
\end{equation}
where 
$\langle \cdot \rangle_0$ denotes expectation value with respect to the Maxwell action. Eq~\eqref{eq:maxwellEE} is the electric analogue of Eq~\eqref{eq:maxwellB2B2}. Now the correlator of the electric field operators $\hat{e}_{\mathbf{r},\mu}$ for $\mu \in \{ 0,1,2,3 \}$ along the links of the diamond lattice are obtained from Eq.~\eqref{eq:maxwellEE} by taking components of the Cartesian electric field along the vectors $\mathbf{e}_\mu$. Thus
\begin{equation} \label{eq:two_pt_sz}
  \langle \hat{S}^z_{\mathbf{r},\mu} \hat{S}^z_{\mathbf{r'},\nu} \rangle = \sum_{k,l \in \{x,y,z\}} (\mathbf{e}_\mu)_k (\mathbf{e}_\nu)_l  \langle \hat{e}_{\mathbf{r},k} \hat{e}_{\mathbf{r}',l} \rangle_0,
\end{equation} 

\newtext{\subsubsection{Two-point $\hat{S}^z$ correlator in the TFP phase}}
\label{sec:sz_sz_corr_tfp}
\newtext{For $V = 0$, the ground state is $\ket{-}$ and $\langle - | \hat{S}^z_{\mathbf{r},\mu}\hat{S}^z_{\mathbf{r'},\nu} | - \rangle = 0 $. The first-order correction to the ground-state wavefunction from the perturbation $\hat{H}_V$ is given by $\ket{\chi_1}$ defined in Eq.~\eqref{eq:tfp_state_pert1}. The first-order correction to the two-point $\hat{S}^z$ correlator is 
\begin{equation}
\langle - | \hat{S}^z_{\mathbf{r},\mu}\hat{S}^z_{\mathbf{r'},\nu} | \chi_1 \rangle + \text{H.c.} \propto \frac{V}{\Omega} \left( \frac{a}{R} \right)^6.
\end{equation}
Thus, in the TFP phase, the two-point $\hat{S}^z$ correlator is proportional to $\frac{V}{\Omega} \left( \frac{a}{R} \right)^6$.}\\

\section{Discussion}\label{sec:discussion}

In this work, we have presented a proposal to prepare and detect the deconfined phase of the $U(1)$ gauge theory in 3+1 dimensions on a Rydberg atom simulator. We first showed that laser-driven  neutral atoms trapped in a pyrochlore lattice using optical tweezer arrays naturally realise a $U(1)$ quantum spin liquid as the ground state when the laser detuning lies in a specified window and the interactions between Rydberg atoms are restricted to nearest-neighbor. We then studied the effect of van der Waals interactions beyond nearest-neighbor. In the classical limit obtained by dropping the Rabi frequency term, we showed that long-range interactions break the degeneracy to select an ice ferromagnet as the ground state. We then studied the competition between the long-ranged interactions that prefer an ordered state and quantum fluctuations that prefer a QSL state, by calculating the energies in ansatz wavefunctions using perturbation theory. We found that, for Rabi frequencies greater than $\Omega_C \approx 0.44 V$, the ground state is a QSL within our approximation. When $\Omega$ is increased further, we argued that the QSL goes into a transverse-field-polarized state via a Higgs transition. While we have focused on the ground state, we also commented on the effect of dynamical state preparation in deciding the nature of the prepared state. We then provided experimental protocols for measuring the plaquette correlators, Bricmont-Fr{\"o}lich-Fredenhagen-Marcu order parameters, the monopole-monopole correlator, and the ``electric field" correlator that can distinguish a QSL phase from ordered phases.

Our ground state phase diagram is the result of an approximate calculation. While it is possible that the true phase diagram differs from what we found, we note that there are other knobs one can tune to get a desired phase diagram. Dressed states created from multiple Rydberg and possibly ground levels can be used to customize the interactions away from the isotropic $1/r^6$ form we considered in this paper \cite{de2017Optical,glaetzle2015designing,van2015quantum,petrosyan2014binding,grass2018fractional,young2021asymmetric}. \newtext{ It is also possible to engineer interactions that are strongly peaked in distance~\cite{steinert2023spatially, hollerith2022realizing} which could allow the nearest-neighbor interactions to be much stronger than the interactions at other distances, and potentially make the QSL more stable.} Designing a dressing scheme compatible with the symmetries of the pyrochlore lattice and exploring the resulting phase diagrams is an interesting direction for future work. 
\newtext{It is known that dipolar-like interactions can preserve the degeneracy of the ice manifold \cite{isakov2005why}. The QSL region can potentially be extended to smaller Rabi frequencies by making the Rydberg atoms interact via dipolar interactions either by applying a DC electric field or microwave-dressing a Rydberg $s$ state with one or more Rydberg $p$ states~\cite{young2021asymmetric}.}
We also note that our proposal requires two Rydberg excitations per tetrahedron, meaning that it lies outside of the Rydberg-blockade regime and is therefore sensitive to imperfections and thermal fluctuations in nearest-neighbor spacing. It will therefore be useful to extend our proposal to the blockade regime of one excitation per tetrahedron. While previous numerical work on dimer models have required a nonzero RK potential (6-body term) to achieve this, it will be worthwhile to study if one can engineer long-range Rydberg interactions that stabilize a spin-liquid in the blockade regime.

\begin{figure}[t]
  \centering
  \includegraphics[width=0.7\columnwidth]{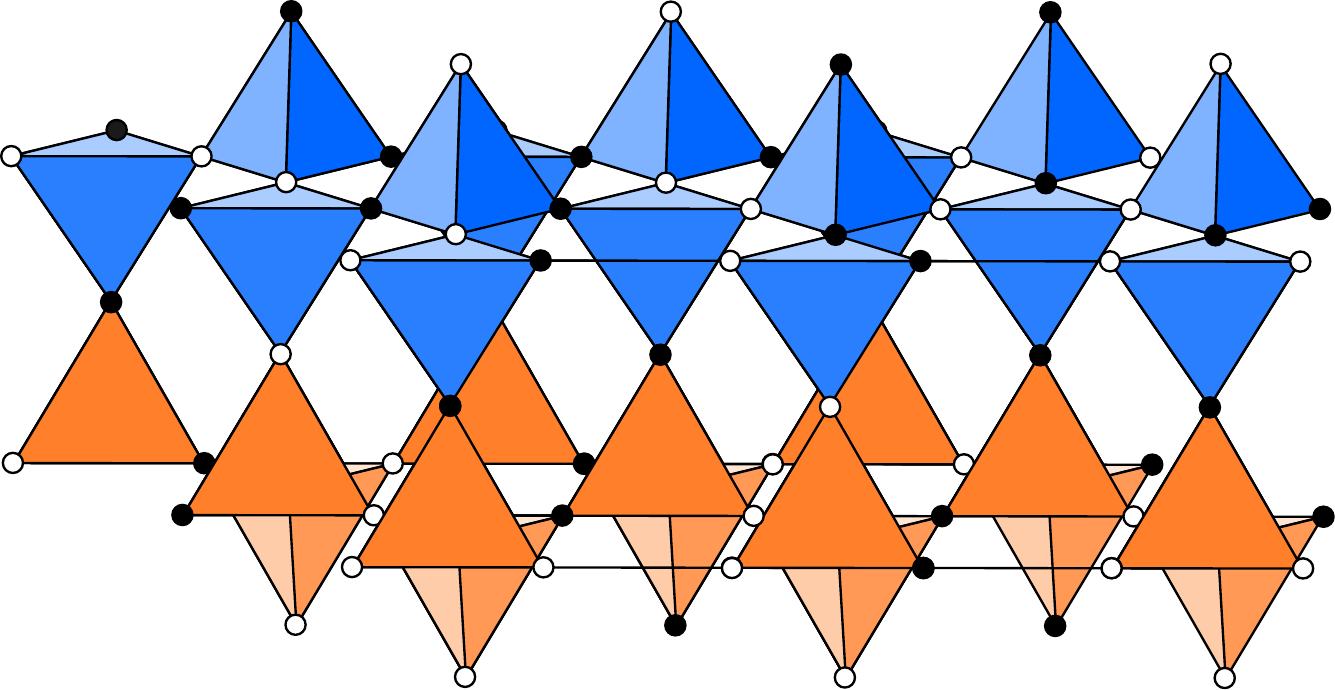}
  \capt{A lattice made of corner-sharing tetrahedra different from the pyrochlore lattice. The lattice consists of $ABAB\ldots$ stacking of the blue ($A$) and the orange ($B$) layers. A configuration satisfying $n_{\tet} = 2$ is shown here.}
  \label{fig:water_ice}
\end{figure}

One can also look for other lattices that could realize a $U(1)$ QSL ground state. One such possibility is a lattice of corner-sharing tetrahedra where all up-pointing tetrahedra (and separately all down-pointing tetrahedra) form a hexagonal close-packed lattice shown in Fig.~\ref{fig:water_ice}. If only nearest-neighbor interactions are considered between atoms positioned on the sites of this lattice, then, by perturbation theory in $\Omega/V$ for a particular range of detunings, one gets ring exchange terms similar to the ones obtained in Sec.~\ref{sec:perturbation_theory}, and the system maps onto a dimer model. It is not known if this dimer model is in the QSL phase when the RK potential is zero and long range van der Waals interactions are included. Another open problem is to construct lattices where a dimer model can be realized within the blockade regime without the RK potential.

Next, we note that, formally, a distinction between the confined and deconfined phases exists only in the thermodynamic limit. Experimentally, there are two finiteness effects that can be important. First, a realistic three dimensional Rydberg array will likely have a relatively small linear dimension.  
Some of the correlators presented in Sec.~\ref{sec:diagnostics} require asymptotic behavior in distance to distinguish different phases. Second, as found in Ref.~\cite{sahay2022quantum} and mentioned in Sec.~\ref{sec:dynamics}, a finite-time state preparation scheme would generically prepare puddles of spin-liquid regions as opposed to an entire spin liquid. It is therefore necessary to quantitatively study how the behavior of the correlators is modified under these conditions. \newtext{One must also estimate the size of the puddles of the QSL and compare them to the length scale at which the asymptotic behavior of the correlators is observed. We leave this for future work.} 

We also note that, to translate field-theory observables into microscopic variables, we relied on the perturbative limit of small $\Omega/V$. However in the phase diagram that we found, the region where the spin liquid is a ground state does not satisfy $\Omega/V\ll1$. Understanding how the field-theory operators (e.g.~plaquette, monopole, and electric-field operators) get renormalized away from the perturbative limit is important both from fundamental and practical standpoints.

Our work is a proposal to prepare a gapless $U(1)$ spin liquid using unitary evolution. An interesting research direction would be to come up with schemes that also use projective measurements to expedite the state preparation along the lines of~Refs.~\cite{tantivasadakarn2022shortest,lavasani2022monitored}. One can also explore how other exotic phases of matter such as fractons and 3+1D topological order can potentially be realized on a Rydberg simulator.

\begin{acknowledgments}
We thank Nikita Astrakhantsev, Peter Lunts, Nathan Schine, Alexander Schuckert, Dayal Singh, and Ashvin Vishwanath for discussions. J.S., G.N., and V.G.~were  supported by NSF DMR-2037158, US-ARO Contract No.~W911NF1310172, and Simons Foundation. J.S.~and A.V.G.~were supported in part by AFOSR, NSF QLCI (award No.~OMA-2120757), DoE ASCR Accelerated Research in Quantum Computing program (award No.~DE-SC0020312), the DoE ASCR Quantum Testbed Pathfinder program (award No.~DE-SC0019040), NSF PFCQC program, ARO MURI, AFOSR MURI, and DARPA SAVaNT ADVENT. Support is also acknowledged from the U.S.~Department of Energy, Office of Science, National Quantum Information Science Research Centers, Quantum Systems Accelerator.  

\end{acknowledgments}

\bibliography{rydbergs}

\appendix

\section{\newtext{Convergence of the perturbation theory}}\label{app:perturbation}

\newtext{In this appendix, we examine the issue of the convergence of the Taylor expansion of the perturbational energies of the ice FM, ice antiferromagnet and the RK ansatz wavefunctions. 
We find that the Pad\'e approximants for the perturbational energies of the ice FM, ice antiferromagnet, and the RK ansatz wavefunctions  have spurious singularities in the range $0 < \Omega/V < 0.6$ because of the vanishing of the denominators of the Pad\'e approximants. 
It is known that such singularities can appear in Pad\'e approximants and can be avoided by the Borel-Pad\'e analysis, and the Borel-Pad\'e approximants obtained from it do not have these spurious singularities. 
We determine the $[m/n]$ Borel-Pad\'e approximant of a series $f(x)$ by the procedure described in Section 3 of Ref.~\cite{deeb2016comparison} and we explain it briefly here: first, we perform a Borel transform on the series $f(x)$ giving a new series $\mathcal{B}f(x)$. Then, we calculate the $[m/n]$ Pad\'e approximant of $\mathcal{B}f(x)$ which we denote by $P_{[m/n]}(x)$. 
Finally, we obtain the $[m/n]$ Borel-Pad\'e approximant by calculating the Laplace transform of $P_{[m/n]}(x)$.
Here, $m+n$ should be equal to the degree of the truncated Taylor series. }

\begin{figure*}[t]
     \centering
     \begin{subfigure}[b]{0.45\textwidth}
         \centering
         \includegraphics[width=\textwidth]{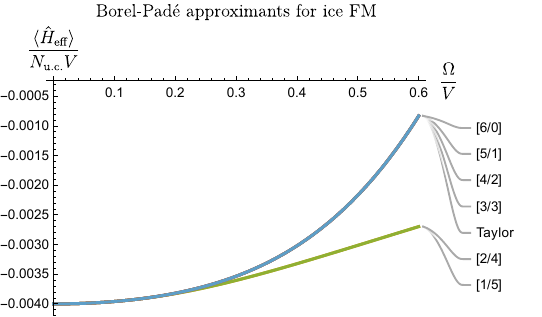}
         \caption{Ice ferromagnet}
         \label{fig:borel_pade_IFM}
     \end{subfigure}
     \hfill
     \begin{subfigure}[b]{0.45\textwidth}
         \centering
         \includegraphics[width=\textwidth]{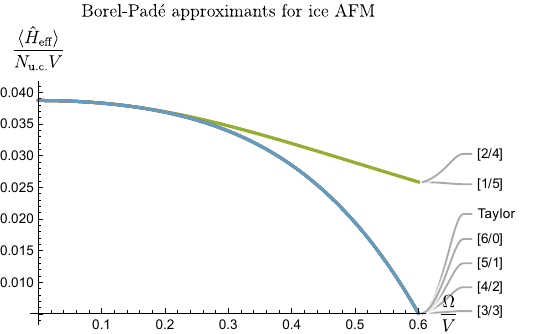}
         \caption{Ice antiferromagnet}
         \label{fig:borel_pade_IAFM}
     \end{subfigure}
     \hfill
     \begin{subfigure}[b]{0.45\textwidth}
         \centering
         \includegraphics[width=\textwidth]{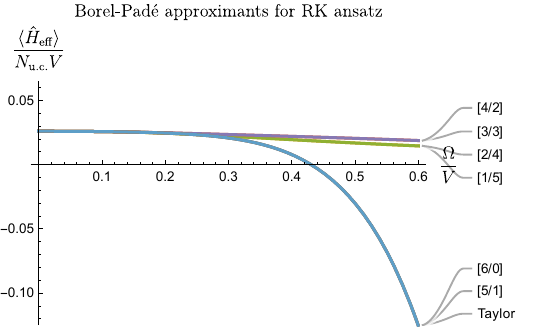}
         \caption{RK wavefunction}
         \label{fig:borel_pade_RK}
     \end{subfigure}
     \begin{subfigure}[b]{0.45\textwidth}
         \centering
         \includegraphics[width=\textwidth]{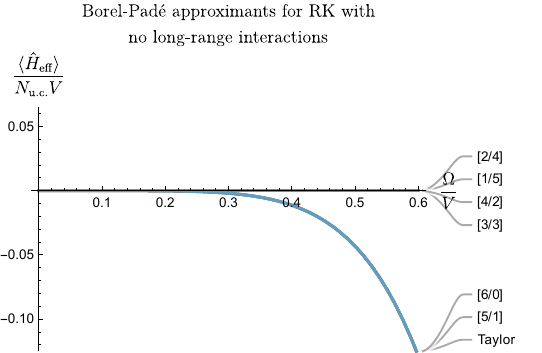}
         \caption{RK wavefunction with long-range interactions turned off}
         \label{fig:borel_pade_RK_noLR}
     \end{subfigure}
        \caption{\newtext{Sub-figures (a), (b) and (c) show the various Borel-Pad\'e approximants and the Taylor series for the three ansatz states: ice ferromagnet, ice antiferromagnet and the RK wavefunction. Sub-figure (d) shows the Borel-Pad\'e approximants and the Taylor series for the RK wavefunction without the long-range interactions. The curves labelled ``Taylor" are the energies of the ansatz states obtained from perturbation theory. The curves labelled by ``$[m/n]$" where $m, n \in \{0,1,  ..., 6\}$ such that $m+n=6$ are the $[m/n]$ Borel-Pad\'e approximants.}}
        \label{fig:borel_pade}
\end{figure*}

\newtext{From the perturbation theory calculation of Sec.~\ref{sec:numerics}, we have the Taylor series up to sixth order in $\Omega/V$ for the energies of the three ansatz states -- ice ferromagnet, ice antiferromagnet, and the RK wavefunction. Thus we have $m+n=6$. We have computed the various $[m/n]$ Borel-Pad\'e approximants and plotted them in Figs.~\ref{fig:borel_pade} (a)--(c). Based on these plots, we make the following comments:}
\begin{itemize}
    \item \newtext{Regarding the ice ferromagnet [Fig.~\ref{fig:borel_pade}(a)]: We find that the [6/0], [5/1], [4/2],  and [3/3] Borel-Pad\'e approximants are equal to the Taylor series while the [2/4] and [1/5] Borel-Pad\'e approximants have a lower energy than the Taylor series.
    At the transition point, $\Omega_C = 0.43927$, the [2/4] approximant differs from the Taylor series by about 17\%.
    If we use the [2/4] approximant instead of the Taylor series for the ice FM to determine the transition point between ice FM and QSL, it shifts from $\Omega_C = 0.43927 V$ to $0.44067 V$. 
    This change in the location of the transition point is very small, and using the Borel-Pad\'e approximants instead of the Taylor series does not change the phase diagram qualitatively.}
    \item \newtext{Regarding the ice antiferromagnet [Fig.~\ref{fig:borel_pade}(b)]: We again find that the [2/4] and [1/5] approximants are equal to each other and are different from the Taylor series. The other Borel-Pad\'e approximants, namely the [6/0], [5/1], [4/2], and [3/3] approximants are equal to the Taylor series. The [2/4] Borel-Pad\'e approximant differs from the Taylor series at the transition point, $\Omega_C = 0.43927 V$, by about 20\%. This is not a small amount, but even if we assume that the true energy is lower than the perturbation-theory energy (Taylor series) by 20\%, ice antiferromagnet continues to remain an excited state and the phase diagram does not change. This is under the assumption that the energies of the ice ferromagnet and the RK wavefunction are given by their Taylor series.}

    \item \newtext{Regarding the RK wavefunction [Fig.~\ref{fig:borel_pade}(c)]: We find that the [4/2], [3/3], [2/4], and [1/5] Borel-Pad\'e approximants are positive for all values of $\Omega/V > 0$, and the phase diagram would not have a QSL if we used these approximants as the energy of the RK wavefunction. 
    However, we believe this is an artifact of the Borel-Pad\'e approximants and is not representative of the underlying physics. To understand our claim, consider the Hamiltonian without $\hat{H}_{\text{LR}}$, i.e., the transverse-field Ising model. We know from Ref.~\cite{emonts2018monte} that the ground state is a QSL for $\Omega < 0.55(5) V$. For $\Omega =0$, all states in the ice manifold including the RK wavefunction are the ground states. 
    For a nonzero but small $\Omega/V$, the quantum fluctuations are present, and we expect them to decrease the energy of the ground state. In 
    Fig.~\ref{fig:borel_pade}(d), we show the Taylor series obtained from sixth-order perturbation theory and its Borel-Pad\'e approximants for the Hamiltonian without $\hat{H}_{\text{LR}}$. 
    We see that the Taylor series decreases as $\Omega/V$ is increased and captures the energy reduction from quantum fluctuations, however the [2/4], [1/5], [4/2], and [3/3] Borel-Pad\'e approximants remain equal to 0.
    Thus, the [2/4], [1/5], [4/2], and [3/3] Borel-Pad\'e approximants do not capture the physics. This could be because of the structure of the Taylor series---the sixth-order term has a large coefficient as compared to the zeroth-, second-, and fourth-order terms. (The Taylor series for the RK wavefunction with $\hat{H}_{\text{LR}}$ is $0.026 - 0.027 (\Omega/V)^2 - 0.098 (\Omega/V)^4 - 2.77 (\Omega/V)^6$). However, we are not certain about why the [2/4], [1/5], [4/2], and [3/3] approximants do not capture the energy decrease.
    Thus the only Borel-Pad\'e approximants we may be able to reliably use with the given data are [6/0] and [5/1], which are the same as the Taylor series. We would obtain the same phase diagram if we were to use the [6/0] or [5/1] approximants.}
\end{itemize}
\newtext{In summary, we find that using the $[6/0]$ and $[5/1]$ Borel-Pad\'e approximants only changes the critical coupling of the transition, but does not change the phase diagram qualitatively.}

\section{Gauge mean field theory}\label{app:gmft}
In this appendix, we first provide details of the gauge mean field theory calculation sketched in Sec.~\ref{sec:gmft}, with a focus on capturing the Higgs transition. Then, we attempt to use the same technique in the small-$\Omega$ limit to obtain the confinement-deconfinement transition. We find that, in this limit, the technique is fraught with a serious limitation stemming from neglecting gauge fluctuations. 

Starting from Eq.~\eqref{eq:hamfict} of Sec.~\ref{sec:gmft} and performing the mean-field decoupling, we get
\begin{equation}\label{eq:Hmf}
\begin{aligned}
    \hat{H}_{\text{MF}}&=\hat{H}_{\Phi}+ \hat{H}_{\mathsf{s}} + \hat{H}_{\text{c}}, \text{ where}\\
    \hat{H}_{\Phi}=&\frac{V}{2} \sum_{\vb{r}\in A,B} \hat{Q}^2_{\vb{r}}- \frac{\Omega}{2}\sum_{(\vb{r}\in A),\mu}\left(\hat{\Phi}_{\vb{r}}^\dagger \hat{\Phi}_{\vb{r}+\vb{e}_\mu} \expval{\hat{\mathsf{s}}^+_{\vb{r},\mu }}+ \text{ H.c.}\right),\\
    \hat{H}_{\mathsf{s}}=&-\frac{\Omega}{2}\sum_{(\vb{r}\in A), \mu}\left(\expval{\hat{\Phi}_{\vb{r}}^\dagger \hat{\Phi}_{\vb{r}+\vb{e}_\mu}} \hat{\mathsf{s}}^+_{\vb{r},\mu }+ \text{ H.c.}\right) \\&+ \sum_{(\vb{r}\in A), \mu} \hat{\mathsf{s}}^z_{\vb{r},\mu}\sum_{(\vb{r}'\in A), \nu}\left(V_{\mu\nu}(\vb{r}-\vb{r}')\expval{\hat{\mathsf{s}}^z_{\vb{r}',\nu}}\right),\\
    \hat{H}_{\text{c}}=&\frac{\Omega}{2}\sum_{(\vb{r}\in A),\mu}\left(\expval{\hat{\Phi}_{\vb{r}}^\dagger \hat{\Phi}_{\vb{r}+\vb{e}_\mu}}\expval{ \hat{\mathsf{s}}^+_{\vb{r},\mu }}+ \text{ H.c.}\right)\\&-\frac{1}{2}\sum_{(\vb{r}\in A), \mu} \hat{\mathsf{s}}^z_{\vb{r},\mu}\sum_{(\vb{r}'\in A), \nu}\left(V_{\mu\nu}(\vb{r}-\vb{r}')\expval{\hat{\mathsf{s}}^z_{\vb{r}',\nu}}\right).
\end{aligned}
\end{equation}
$\hat{H}_{\text{c}}$ is a constant, and $V_{\mu\nu}(\mathbf{r} - \mathbf{r'})$ was defined in Sec.~\ref{sec:gmft}. $\hat{H}_{\mathsf{s}}$ above is of the form $-\sum_{(\vb{r}\in A),\mu}\left(h^x_{\vb{r},\mu} \hat{\mathsf{s}}^x_{\vb{r},\mu}+h^z_{\vb{r},\mu} \hat{\mathsf{s}}^z_{\vb{r},\mu}\right)$, where 
\begin{equation}\label{eq:fieldexp}
    \begin{aligned}
    h^x_{\vb{r},\mu}&=\Omega \expval{\hat{\Phi}_{\vb{r}}^\dagger \hat{\Phi}_{\vb{r}+\vb{e}_\mu}},\\
    h^z_{\vb{r},\mu}&=-\sum_{(\vb{r}'\in A), \nu}\left(V_{\mu\nu}(\vb{r}-\vb{r}')\expval{\hat{\mathsf{s}}^z_{\vb{r}',\nu}}\right),
    \end{aligned}
\end{equation}
and $\expval{\hat{\Phi}_{\vb{r}}^\dagger \hat{\Phi}_{\vb{r}+\vb{e}_\mu}}$ is calculated in the ground state of $\hat{H}_{\Phi}$, which in turn depends on $\expval{\hat{\mathsf{s}}^+}$. (We have implicitly assumed here that $\expval{\hat{\Phi}_{\vb{r}}^\dagger \hat{\Phi}_{\vb{r}+\vb{e}_\mu}}$ is real, which we will show can be assumed self-consistently.) This implies that, in the ground state, 
\begin{equation}\label{eq:magfield}
    \expval{\hat{\mathsf{s}}^i_{\vb{r},\mu}}=\frac{h^i_{\vb{r},\mu}}{2\abs{\vb{h}_{\vb{r},\mu}}} \text{ for } i=x,z.
\end{equation}
Our goal is to self-consistently minimize the ground-state energy of the mean-field Hamiltonian subject to the constraints in Eqs.~\eqref{eq:constraint1} and ~\eqref{eq:constraint2}. We showed in Sec.~\ref{sec:classical} that the ordered ground state at $\Omega=0$ has momentum $\vb{k}=\boldsymbol{0}$. Also, the TFP state in the large-$\Omega$ limit is a $\vb{k}=\boldsymbol{0}$ state. So we start with a mean-field ansatz with full translation symmetry (similar to Ref.~\cite{savary2012coulombic}):
\begin{equation}
\begin{aligned}
    \expval{\mathsf{s}^+_{\vb{r},\mu}}&=\frac{1}{2}\cos\theta,\\
    \expval{\mathsf{s}^z_{\vb{r},\mu}}&=\frac{1}{2}\varepsilon_\mu \sin \theta, 
\end{aligned}
\end{equation}
where $\varepsilon_\mu=1,1,-1,-1$ for $\mu=0,1,2,3$, respectively. To solve the matter sector, it is convenient to deal with the Lagrangian instead of the Hamiltonian. The imaginary-time Lagrangian for the matter sector is
\begin{equation}
    \begin{aligned}
    \mathcal{L}& = \frac{1}{2V}\sum_{\vb{r} \in A,B}\abs{(\partial_\tau - i v_{\vb{r}}) \Phi_{\vb{r}}}^2\\ &-\frac{\Omega \cos \theta}{4}\sum_{(\vb{r}\in A), \mu}(\Phi_{\vb{r}}^* \Phi_{\vb{r}+\vb{e}_\mu}e^{ia_{\vb{r},\mu}}+\text{ c.c.})\\
    &-i \sum_{\mathbf{r} \in A,B} \left[ \eta_{\vb{r}} v_{\vb{r}} \left( \sum_{\mu}\mathsf{s}^z_{\vb{r}+\eta_{\vb{r}} \vb{e}_\mu/2}\right)+ \Tilde{\lambda}_{\vb{r}}(\abs{\Phi_{\vb{r}}}^2-1)\right] ,
    \end{aligned}
\end{equation}
where the Lagrange multiplier $\tilde{\lambda}_{\vb{r}}$ (which gets integrated over) enforces the constraint $\abs{\Phi_{\vb{r}}}^2=1$. The Lagrange multiplier $v_{\vb{r}}$ enforces the constraint \eqref{eq:constraint2}. To zeroth order, we ignore the gauge fluctuation $a_{\vb{r},\mu}$. The matter Lagrangian alone, despite being quadratic in the rotor variables, is nevertheless interacting  because a quadratic term in rotor operators is nonlinear in terms of canonical bosons (in other words, it is a cosine term in the phase of the rotor.) In order to make progress, Ref.~\cite{savary2012coulombic} assumes that, at the saddle point, $\Tilde{\lambda}_{\vb{r}}$ takes on a spatially uniform and purely imaginary value $i\lambda$, and also implicitly assumes that $v_{\vb{r}}$ is 0 at the saddle point. Here, we will follow suit while acknowledging that these approximations are uncontrolled. Making these simplifications, we obtain
\begin{equation}
\label{eq:lagsimp}
\begin{aligned}
    \mathcal{L}=&\frac{1}{2V}\sum_{\vb{r}}\abs{\partial_\tau \Phi_{\vb{r}}}^2 - \frac{\Omega \cos \theta}{4} \sum_{(\vb{r}\in A),\mu}\left(\Phi_{\vb{r}}^* \Phi_{\vb{r}+\vb{e}_\mu}+\text{ c.c.}\right)\\& + \lambda\sum_{\vb{r}} (\abs{\Phi_{\vb{r}}}^2-1).
    \end{aligned}
\end{equation}
The constraints now simplify to
 \begin{gather}
     \expval{\Phi^\dagger_{\vb{r}}\Phi_{\vb{r}}}=1 \label{eq:constrsimp1},  \\
     h^x=\Omega\expval{\Phi^\dagger_{\vb{r}}\Phi_{\vb{r}+\vb{e}_\mu}}. \label{eq:constrsimp2}
 \end{gather}
 Now, we have a quadratic Lagrangian, which we solve by Fourier transformation. Our Fourier transformation convention is (for $\alpha \in \{A,B\}$)
 \begin{equation}
     \Phi_{\vb{r},\alpha}(\tau)=T\sum_{\omega_n}\sum_{\vb{k}\in BZ}\Phi_{\vb{k},\alpha}(\omega_n)e^{i(\vb{k}\cdot \vb{r}-\omega_n \tau)},
 \end{equation}
where $T$ is the temperature, $\omega_n$ are Matsubara frequencies and we eventually take the limit $T \to 0$. Eq.~\eqref{eq:lagsimp} becomes
\begin{equation}
   \mathcal{L}= T \sum_{\vb{k},\omega_n}\begin{pmatrix}\Phi^*_{\vb{k},A}(\omega_n) & \Phi^*_{\vb{k},B}(\omega_n)\end{pmatrix} \mathcal{G}^{-1}_{\vb{k}}(\omega_n) \begin{pmatrix}\Phi_{\vb{k},A}(\omega_n) \\ \Phi_{\vb{k},B}(\omega_n)\end{pmatrix},
\end{equation}
where 
\begin{equation}
    \mathcal{G}^{-1}_{\vb{k}}(\omega_n)=\begin{pmatrix}\frac{\omega_n^2}{2V}+\lambda & -\frac{\Omega \cos \theta}{4}f_{\vb{k}}\\-\frac{\Omega \cos \theta}{4}f^*_{\vb{k}} & \frac{\omega_n^2}{2V}+\lambda\end{pmatrix}.
\end{equation}
Here,
\begin{equation}
    f_{\vb{k}}=1+e^{-i k_1} +e^{-i k_2} + e^{-i k_3},
\end{equation}
where $\vb{k}\equiv k_1 \vb{b}_1 + k_2 \vb{b}_2 + k_3 \vb{b}_3$, and $\vb{b}_1$, $\vb{b}_2$ and $\vb{b}_3$ are reciprocal lattice vectors of the FCC lattice satisfying $\vb{a}_i \cdot \vb{b}_j =\delta_{ij}$.

Upon inverting the above matrix, we find that the eigenvalues of $\mathcal{G}_{\vb{k}}(\omega_n)$ are $\frac{2 V}{\omega_n^2 + \left(\omega^{\pm}_{\vb{k}}(\lambda,\theta)\right)^2}$, where the dispersion of the two bosonic bands is
\begin{equation}
    \omega^{\pm}_{\vb{k}}(\lambda,\theta)=\sqrt{2V \left(\lambda \pm \frac{\Omega \cos \theta}{4} \abs{f_{\vb{k}}}\right)}.
\end{equation}
As long as the spinon dispersion is gapped, spinons will not condense. 
From the dispersion above, we see that the dispersion becomes gapless when $\lambda=\Omega \cos \theta$. However, as we will see below, for fixed $\theta$ and $\Omega$, $\lambda$ is determined by the constraint in Eq.~\eqref{eq:constrsimp1}. Therefore the condition $\lambda=\Omega \cos \theta$ is met for a specific $\Omega=\Omega^{\text{MF}}_{H}$, which we will calculate below. Before that, will go through a few intermediate steps. First, the matrix form of $\mathcal{G}_{\vb{k}}(\omega_n)$ is (assuming $\Omega>0$)
\begin{widetext}
\begin{equation}
    \begin{aligned}
        \mathcal{G}_{\vb{k}}(\omega_n)=V \begin{pmatrix}\frac{1}{\omega_n^2 + (\omega^+_{\vb{k}})^2}+\frac{1}{\omega_n^2 + (\omega^-_{\vb{k}})^2}& g_{\vb{k}}\left(\frac{1}{\omega_n^2 + (\omega^+_{\vb{k}})^2}-\frac{1}{\omega_n^2 + (\omega^-_{\vb{k}})^2}\right)\\ g_{\vb{k}}^*\left(\frac{1}{\omega_n^2 + (\omega^+_{\vb{k}})^2}-\frac{1}{\omega_n^2 + (\omega^-_{\vb{k}})^2}\right)&\frac{1}{\omega_n^2 + (\omega^+_{\vb{k}})^2}+\frac{1}{\omega_n^2 + (\omega^-_{\vb{k}})^2}\end{pmatrix}
    \end{aligned},
\end{equation}
\end{widetext}
where
\begin{equation}
g_{\vb{k}}=
\begin{cases}
-\frac{f_{\vb{k}}}{\abs{f_{\vb{k}}}} & \text{ when }0\leq\theta < \pi/2,\\
0 & \text{ when }\theta=\pi/2.
\end{cases}
\end{equation}

With the Green's function in hand, we are now ready to impose the constraints,  Eq.~\eqref{eq:constrsimp1} and Eq.~\eqref{eq:constrsimp2}. First, we calculate equal-time correlation functions of $\Phi$ (by performing the Matsubara sum on the Green's function). Using these, the constraints in Eq.~\eqref{eq:constrsimp1} and Eq.~\eqref{eq:constrsimp2}  become, respectively, 
\begin{align}
    F_1(\lambda,\theta)&\equiv \frac{V}{2N_{\text{u.c.}}}\sum_{\vb{k}}\left(\frac{1}{\abs{\omega^+_{\vb{k}}}}+ \frac{1}{\abs{\omega^-_{\vb{k}}}}\right)=1 \label{eq:constrfin1},\\
    \Omega F_2(\lambda,\theta)&\equiv \Omega \frac{V}{2N_{\text{u.c.}}}\sum_{\vb{k}}g_{\vb{k}}\left(\frac{1}{\abs{\omega^-_{\vb{k}}}}- \frac{1}{\abs{\omega^+_{\vb{k}}}}\right)=h^x.\label{eq:constrfin2}
\end{align}
Next, by imposing Eq.~\eqref{eq:magfield} with the help of Eq.~\eqref{eq:fieldexp}, we get 
\begin{equation}
    h^z=-\frac{\mathcal{B}\sin \theta}{2}, \text{ where }\mathcal{B}=\frac{\sin \theta}{2}\sum_{(\vb{r}'\in A), \nu} V_{0,\nu}(-\vb{r}')\varepsilon_\nu.
\end{equation}
For a given $\theta$, Eq.~\eqref{eq:constrfin1} determines $\lambda$. We see that there are three self-consistent solutions for $\theta$: 
\begin{equation}
\theta=
    \begin{cases}
    0,\\
    \pi/2,\\
    \cos^{-1}\left(\frac{2\Omega F_2(\lambda,\theta)}{\mathcal{B}}\right).
    \end{cases}
\end{equation}
Within gMFT (gauge mean field theory), these three solutions correspond to a QSL, a ``Coulomb ferromagnet" (spin liquid with nonzero ice ferromagnetic order parameter), and an ice ferromagnet, respectively~\cite{savary2012coulombic}. 
For a fixed parameter $\Omega$, the true solution depends on which of the three solutions above has lower energy with respect to the mean-field Hamiltonian \eqref{eq:Hmf}. Suppose that, for large enough $\Omega$, one is in the QSL phase, i.e.,~$\theta=0$ and $\expval{\hat{\Phi}_{\vb{r}}}=0$. Now, the bosons will condense when their dispersion becomes gapless, i.e.,~$\lambda=\Omega$. Using constraint \eqref{eq:constrfin1}, we find that this transition point is $\Omega^H_{\text{MF}}\approx 0.7 V$, as also found in Ref.~\cite{savary2017disorder}. For $\Omega>\Omega^H_{\text{MF}}$, the ground state is in the TFP phase.

\begin{figure}[t]
    \centering
    \includegraphics[width=0.47\textwidth]{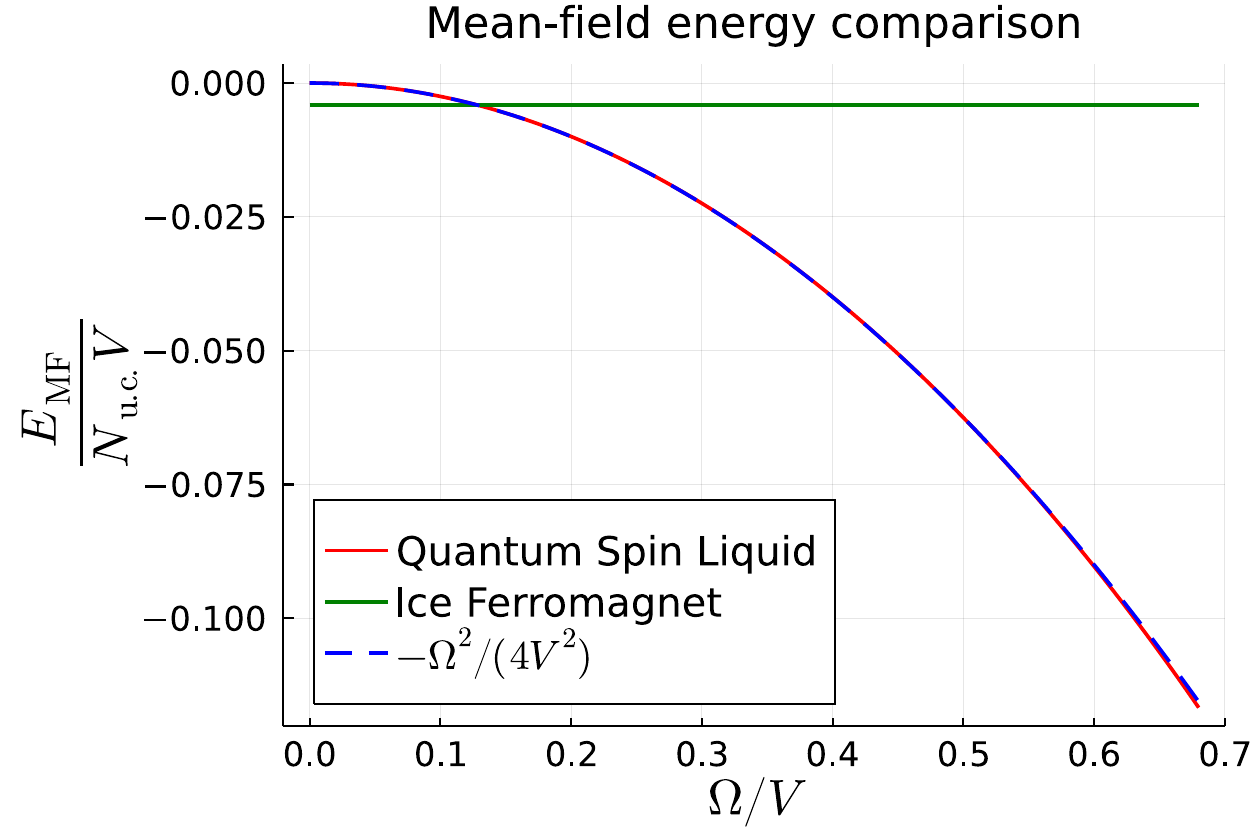}
    \capt{The energy per unit cell (in units of $V$) of saddle points $\theta=0$ (QSL) and $\theta=\pi/2$ (ice ferromagnet) given by Eq.~\eqref{eq:Emf} up to an overall additive constant that is the same for $\theta=0$ and $\theta=\pi/2$. We also plot $-\frac{\Omega^2}{4V^2}$ arising from trivial spin-flip pairs: this plot almost overlaps with the energy of the $\theta=0$ state.}
    \label{fig:meanfieldenergy}
\end{figure}

Having identified the Higgs transition point, we now attempt to identify the confinement-deconfinement transition for low $\Omega$, i.e.,~find $\Omega$ at which $\theta=0$ becomes the lowest-energy saddle-point. Using Eq.~\eqref{eq:Hmf}, we get the following expression for the mean-field energy:
\begin{equation}\label{eq:Emf}
    E_{\text{MF}}=K-N_{\text{u.c.}}\left(2\Omega F_2(\lambda,\theta)\cos \theta + \frac{\mathcal{B}}{2}\sin^2 \theta\right),
\end{equation}
where $K$ is the total kinetic energy of the bosons and can be calculated to be
\begin{equation}
    K=\frac{1}{2}\sum_{\vb{k}}\left(\omega^+_{\vb{k}}+\omega^-_{\vb{k}}\right).
\end{equation}
In Fig.~\ref{fig:meanfieldenergy}, 
we plot the energy $E_{\text{MF}}$ for $\theta=0$ (QSL) and $\theta=\pi/2$ (ice ferromagnet), and find a transition at $\Omega \approx 0.13 V$. (The third solution for $\theta$ becomes the lowest-energy solution only in a minuscule window around $\Omega\approx 0.13 V$, so we ignore it.) However, we will now argue that this result is misleading. 

In gMFT, the energy reduction in the QSL phase with respect to the ordered phase (ice ferromagnet) arises from the minimization of kinetic energy of the bosonic charges $\hat{\Phi}_{\vb{r}}$ that are allowed to hop. When $\theta=0$, the hopping coefficient is maximized, while, for $\theta=\pi/2$, the hopping coefficient is 0. However, microscopically, this hopping corresponds to a single spin-flip. A pair of spin-flips at the same site leads to a \textit{constant} reduction of energy coming from second order perturbation theory, given by $-\Omega^2 N_{\text{u.c.}}/V$. It is constant in the sense that this reduction is obtained for \textit{any} state including the QSL and the ice ferromagnet. The mean-field calculation, however, unfairly assigns this reduction to the QSL but not to the ordered state. In fact, in Fig.~\ref{fig:meanfieldenergy}, we have also plotted $-\Omega^2/(4V)$ (the factor of 1/4 can perhaps be attributed to using spin-$1/2$ and classical spins at the same time). As can be seen, this plot almost completely overlaps with the energy of the QSL calculated within gMFT. So it is clear that, within gMFT, the difference between the energies of the QSL and the confined phase is quadratic in $\Omega$ to leading order even though we know from perturbation theory that the leading order term should be proportional to $\Omega^6$. Hence, gMFT cannot be used in the vicinity of the confinement-deconfinement transition unless gauge-fluctuations are properly taken into consideration.

\section{Difference between $\langle \hat{X}_P \hat{X}_{P'} \rangle_c $ and $ \langle \hat{\tilde{X}}_P \hat{\tilde{X}}_{P'} \rangle_c$}\label{app:corrections_plaq}

\newtext{In this appendix, we show that the difference between $\langle \hat{X}_P \hat{X}_{P'} \rangle_c $ and $ \langle \hat{\tilde{X}}_P \hat{\tilde{X}}_{P'} \rangle_c$ evaluated in the ground state is of sixth order in $\Omega/V$, that is, derive Eq.~\eqref{eq:sixth_order_x}.}

\newtext{Let $\ket{\Psi_g}$ be the ground state of the system. Thus $\ket{\Psi_0} = \hat{U}_S \ket{\Psi_g}$ is in the ice manifold, where $\hat{U}_S$ is the unitary operator that implements the Schrieffer-Wolff transformation (see Sec.~\ref{sec:expression_for_Heff}).  
We have
\begin{equation}
\label{eq:xp_to_uxpu}
    \langle \Psi_g | \hat{X}_P  | \Psi_g\rangle =     \langle \Psi_0 | \hat{U}_S \hat{X}_P \hat{U}_S^\dagger| \Psi_0\rangle.
\end{equation}
At zeroth order in $\Omega/V$, the right-hand side of the above equation is $\langle \Psi_0 |  \hat{X}_P | \Psi_0\rangle$, which we know is equal to $ \langle \Psi_0 |\hat{\tilde{X}}_P |\Psi_0 \rangle$ since $\ket{\Psi_0}$ is in the ice manifold [see Eq.~\eqref{eq:xp_to_tildexp}]. 
Note that $\hat{U}_S = 1+\hat{S} + \hat{S}^2/2!+ \cdots$. The terms that are of order $(\Omega/V)^i$ flip $i$ spins.
When $\hat{U}_S$ and $\hat{U}_S^\dagger$ in Eq.~\eqref{eq:xp_to_uxpu} are expanded as a power series, the first term whose expectation value is nonzero (other than the zeroth order term) appears at sixth order in $\Omega/V$.
This is because $\hat{X}_P $ flips six spins which need to be compensated from another six spin flips coming from six powers of $\hat{S}$.
Thus, we have 
\begin{equation}
    \langle\Psi_0 | \hat{U}_S \hat{X}_P \hat{U}_S^\dagger| \Psi_0\rangle = \langle\Psi_0 |  \hat{X}_P | \Psi_0\rangle + \Theta \left((\Omega/V)^6\right).
\end{equation}
A similar argument applied to $ \langle \hat{\tilde{X}}_P \rangle$ shows that 
\begin{equation}
    \langle\Psi_0 | \hat{U}_S \hat{\tilde{X}}_P \hat{U}_S^\dagger| \Psi_0\rangle = \langle\Psi_0 |  \hat{\tilde{X}}_P | \Psi_0\rangle + \Theta \left((\Omega/V)^6\right).
\end{equation}
Using Eq.~\eqref{eq:xp_to_tildexp}, we find that $ \langle  \hat{\tilde{X}}_P \rangle = \langle  \hat{X}_P \rangle + \Theta \left((\Omega/V)^6\right)$. 
An analogous argument applies to show  $ \langle  \hat{\tilde{X}}_P \hat{\tilde{X}}_{P'}\rangle = \langle  \hat{X}_P\hat{X}_{P'} \rangle + \Theta \left((\Omega/V)^{12}\right)$. Finally, putting together all the pieces, we obtain Eq.~\eqref{eq:sixth_order_x}. 
By similar arguments, Eq.~\eqref{eq:sixth_order_y} can also be derived.}

\section{\newtext{Plaquette correlators in TFP phase}}
\label{app:plaquette_TFP}

\newtext{In this Appendix, we derive  the plaquette X correlator deep inside the TFP phase at second order in perturbation theory, treating the van der Waals interaction as the perturbation. That is, we derive Eq.~\eqref{eq:plaquette_x_conn_matrix_ele}. }

\newtext{For $\Omega \gg V$, the ground state up to  first order in $V/\Omega$ is $\ket{\xi} = \ket{-} +\ket{\chi_1} $ [see Eqs.~\eqref{eq:minus_state_def} and ~\eqref{eq:tfp_state_pert1} for the definitions of $\ket{-}$ and $\ket{\chi_1}$ respectively]. Here $\ket{-}$ is of zeroth order, and $\ket{\chi_1}$ is of first order in $\Omega/V$.  The connected plaquette X correlator is
\begin{equation}
    \frac{\langle \xi | \hat{X}_P \hat{X}_{P'}| \xi \rangle}{\langle \xi | \xi \rangle} -    \frac{\langle \xi | \hat{X}_P | \xi \rangle\langle \xi | \hat{X}_{P'} | \xi \rangle}{\langle \xi | \xi \rangle^2}.    
\end{equation}
Since $\langle - | \chi_1 \rangle = 0$ and $\hat{X}_P \ket{-} = \ket{-}$, the first-order term in the plaquette X correlator above will be zero. Keeping only terms up to the second order, the plaquette X correlator becomes
\begin{equation}
\begin{aligned}
    (&1+\langle \chi_1 | \hat{X}_{P}\hat{X}_{P'} | \chi_1 \rangle )( 1- \langle \chi_1 | \chi_1 \rangle )\\
    & - (1+\langle \chi_1 | \hat{X}_P | \chi_1\rangle ) (1+\langle \chi_1 | \hat{X}_{P'} | \chi_1\rangle ) (1-2\langle \chi_1 | \chi_1\rangle ). 
    \end{aligned}
\end{equation}
Simplifying this expression and keeping only terms that are second-order in $\Omega/V$ gives
\begin{equation}
\begin{aligned}
    \langle &\chi_1 | \hat{X}_{P}\hat{X}_{P'} | \chi_1 \rangle - \langle \chi_1 | \hat{X}_{P}| \chi_1 \rangle - \langle \chi_1 | \hat{X}_{P'} | \chi_1 \rangle + \langle \chi_1 | \chi_1 \rangle\\
    &= \langle \chi_1 | (\hat{X}_P - 1) (\hat{X}_{P'} - 1) | \chi_1 \rangle.
\end{aligned}
\end{equation}
Substituting the definition of $\ket{\chi_1}$ from Eq.~\eqref{eq:tfp_state_pert1}, we obtain the desired Eq.~\eqref{eq:plaquette_x_conn_matrix_ele}.
}

\end{document}